\ifx\pdfminorversion\undefined\else\pdfminorversion=7\fi
\ifx\texorpdfstring\undefined\let\texorpdfstring\relax\fi
\ifx\paperheight\undefined\else\paperheight=11in\fi
\ifx\pdfsuppresswarningpagegroup\undefined\else\pdfsuppresswarningpagegroup=1\fi
\documentclass[aps,prd,10pt,twocolumn,showpacs,showkeys,preprintnumbers,floatfix,nofootinbib,superscriptaddress,unicode,hyperfootnotes=false]{revtex4}
\usepackage{amsfonts} 
\usepackage{amssymb} 
\usepackage{amsmath} 
\usepackage{graphicx} 
\usepackage{xcolor,cancel}

\usepackage{subfigure} 
\usepackage{array} 
\usepackage{dcolumn} 
\usepackage{bm} 
\usepackage{latexsym} 
\usepackage{longtable} 
\usepackage{hyperref} 
\usepackage{bm}
\usepackage{slashed}
\usepackage{bbold}
\usepackage{color}
\usepackage{multirow}
\usepackage{rotating}
\usepackage{comment}

\usepackage{textgreek}
\graphicspath{{./FIGS/}{./Figs_Q/}}

\allowdisplaybreaks

\newcommand\pslash{\slashed p}

\newcommand{\CPV}{$\cancel{\text{CP}}$}
\renewcommand\Re{\mathop{\rm Re}}
\renewcommand\Im{\mathop{\rm Im}}

\DeclareMathOperator{\Tr}{Tr}

\definecolor{green}{rgb}{0.1, 0.8, 0.1}

\newcolumntype{.}[1]{D{.}{.}{#1}}

\begin{document}


\title{Contribution of the QCD \texorpdfstring{$\Theta$}{\texttheta}-term to nucleon electric dipole moment}
%
%


\author{Tanmoy Bhattacharya}
\email{tanmoy@lanl.gov}
\affiliation{Los Alamos National Laboratory, Theoretical Division T-2, Los Alamos, NM 87545}

\author{Vincenzo Cirigliano}
\email{cirigliano@lanl.gov}
\affiliation{Los Alamos National Laboratory, Theoretical Division T-2, Los Alamos, NM 87545}

\author{Rajan Gupta}
\email{rajan@lanl.gov}
\affiliation{Los Alamos National Laboratory, Theoretical Division T-2, Los Alamos, NM 87545}

\author{Emanuele Mereghetti}
\email{emereghetti@lanl.gov}
\affiliation{Los Alamos National Laboratory, Theoretical Division T-2, Los Alamos, NM 87545}

\author{Boram Yoon}
\email{boram@lanl.gov}
\affiliation{Los Alamos National Laboratory, Computer Computational and Statistical Sciences, CCS-7, Los Alamos, NM 87545}

%
%
%
\preprint{LA-UR-20-30515}
\date{\today}
\begin{abstract}
We present a calculation of the contribution of the $\Theta$-term to
the neutron and proton electric dipole moments using seven
2+1+1-flavor HISQ ensembles. We also estimate the topological
susceptibility for the 2+1+1 theory to be $\chi_Q = (66(9)(4)~{\rm
MeV})^4$ in the continuum limit at $M_\pi = 135$~MeV. The calculation
of the nucleon three-point function is done using Wilson-clover
valence quarks. The \CPV\ form factor $F_3$ is calculated by expanding
in small $\Theta$. We show that lattice artifacts introduce
a term proportional to $a$ that does not vanish in the chiral limit,
and we include this in our chiral-continuum fits. A chiral
perturbation theory analysis shows that the $N(\bm 0) \pi (\bm 0)$
state should provide the leading excited state contribution, and we
study the effect of such a state. Detailed analysis of the
contributions to the neutron and proton electric dipole moment using
two strategies for removing excited state contamination are
presented. Using the excited state spectrum from fits to the two-point function, 
we find  $d_n^\Theta$ is small, 
$|d_n^\Theta| \lesssim 0.01\ {\overline \Theta}\ {\rm e \cdot fm}$, 
whereas for the proton we get $|d_p^\Theta| \sim 0.02\ {\overline \Theta}\ {\rm e \cdot fm} $. 
On the other hand, if the dominant excited-state contribution is
from the $N \pi$ state, then $|d_n^\Theta|$ could be as large as 
$0.05\ {\overline \Theta}\ {\rm e \cdot fm}$ and  
$|d_p^\Theta| \sim 0.07\ {\overline \Theta}\ {\rm e \cdot fm} $. Our overall conclusion is that
present lattice QCD calculations do not provide a reliable estimate of
the contribution of the $\Theta$-term to the nucleon electric dipole
moments, and a factor of ten higher statistics data are needed to get 
better control over the systematics and possibly a $3\sigma$ result.
\end{abstract}
\maketitle
%
%
%
%
\section{Introduction}

The permanent electric dipole moments (EDMs) of nondegenerate 
states of elementary particles,
atoms and molecules are very sensitive probes of CP violation
(\CPV). Since the EDMs are necessarily proportional to their spin, and
under time-reversal the direction of spin reverses but the electric
dipole moment does not, a nonzero measurement confirms CP violation
assuming CPT is conserved.  Of the elementary particles, atoms and
nuclei that are being investigated, the electric dipole moments of the
neutron (nEDM) and the proton (pEDM) are the simplest quantities for which lattice QCD can provide the
theoretical part of the calculation needed to connect the
experimental bound or value to the strength of \CPV\ in a
given theory~\cite{Pospelov:2005pr,Engel:2013lsa}.

EDMs can shed light on one of the deepest mysteries of the observed universe,  the origin of the baryon
asymmetry: the universe has \(6.1^{+0.3}_{-0.2}\times 10^{-10}\)
baryons for every black body photon~\cite{Bennett:2003bz}, whereas in
a baryon symmetric universe, we expect no more than about \(10^{-20}\)
baryons and anti-baryons for every photon~\cite{Kolb:1990vq}.  It is difficult to
include such a large excess of baryons as an initial condition in an
inflationary cosmological scenario~\cite{Coppi:2004za}.  The way out
of the impasse lies in generating the baryon excess dynamically during
the evolution of the universe.  But, if the matter-antimatter
symmetry was broken post inflation and reheating, then one is faced
with Sakharov's three necessary conditions~\cite{Sakharov:1967dj} on the dynamics: the
process has to violate baryon number, evolution has to occur out of
equilibrium, and charge-conjugation and CP invariance have to be
violated. 


CP violation  exists in the electroweak sector of the standard model (SM) of particle interactions due to a phase in the Cabibbo-Kobayashi-Maskawa (CKM)
quark mixing matrix~\cite{Kobayashi:1973fv}, and possibly due to a similar
phase in the Ponte\-corvo--Maki--Nakagawa--Sakata (PNMS) matrix in the leptonic sector~\cite{Maki:1962mu,Nunokawa:2007qh}.  The effect of these on nEDM and pEDM is, however, small: that arising from the CKM
matrix is about $O(10^{-32})$ $e$~cm \cite{Khriplovich:1981ca,Czarnecki:1997bu,Seng:2014lea}, much smaller than
the current  90\% confidence level (CL) experimental bound $d_n < 1.8 \times 10^{-26}$ $e$~cm~\cite{Abel:2020gbr}\rlap,
\footnote{The slightly stronger 95\% CL bounds $d_n<1.6 \times 10^{-26}$ $e$~cm and $d_p<2.0\times 10^{-25}$ $e$~cm  can be obtained 
from the experimental limit  on the $^{199}$Hg~\protect\cite{Graner:2016ses} EDM, assuming that  nucleon EDMs are the dominant contributions to the nuclear EDM.}
and than the reach of  ongoing experiments,  $d_n < 3.4 \times 10^{-28}$ $e$~cm at 90\% confidence~\cite{Ito2019}.


In principle, the SM has an additional source of CP violation arising
from the effect of QCD instantons.  The presence of these localized finite
action nonperturbative configurations in a non-Abelian theory 
leads to inequivalent quantum theories defined over various
`$\Theta$'-vacua~\cite{Jackiw:1976pf,Callan:1976je}. Because of asymptotic freedom, all
nonperturbative configurations including instantons are strongly
suppressed at high temperatures~\cite{Gross:1980br,Dolgov:1991fr} 
where baryon number violating
processes occur. Because of this, CP violation due to such vacuum
effects does not lead to appreciable baryon number production~\cite{Kuzmin:1992up}. 
Nonetheless, understanding the contribution of such a term 
to the nucleon EDM is very important for two reasons.
First, the $\Theta$ term  constitutes a `background' contribution to all hadronic EDMs that needs to be understood before one can claim discovery of new sources of CP violation through nucleon or hadronic EDM measurements; and second,
besides generating higher-dimensional CP-odd operators, new sources of CP-violation beyond the Standard Model (BSM) also generate a so-called `induced $\Theta$ term'~\cite{Pospelov:2005pr,Bigi:1990kz,Pospelov:1999ha} if one assumes  that the Peccei-Quinn mechanism is at work~\cite{PhysRevLett.38.1440}. 
Therefore, in the large class of viable models of CP violation that incorporate the Peccei-Quinn mechanism, quantifying the contribution of the induced  $\Theta$ to the nucleon EDM (operationally, the calculation is the same as in the first case) is essential to bound or establish such sources of CP violation.


Until recently, the calculation of hadronic matrix elements needed to connect 
nucleon EDMs to SM and  BSM sources of CP violation 
relied on chiral symmetry supplemented by dimensional
analysis~\cite{Crewther:1979pi,Pich:1991fq,Cho:1992rv,Borasoy:2000pq,Hockings:2005cn,Narison:2008jp,Ottnad:2009jw,deVries:2010ah,Mereghetti:2010kp} or  QCD sum rules~\cite{Pospelov:1999ha,Pospelov:2000bw,Lebedev:2004va,Pospelov:2005pr,Fuyuto:2012yf,Haisch:2019bml},  
both entailing  large theoretical errors.  Large-scale simulations of lattice QCD  provide a first-principles  method for calculating these matrix elements with controlled uncertainties.  Several groups have reported results of lattice QCD calculations of the neutron EDM induced by  the  QCD $\Theta$ term~\cite{Shintani:2005xg,Berruto:2005hg,Shindler:2014oha,Guo:2015tla,Shindler:2015aqa,Alexandrou:2015spa,Shintani:2015vsx,Dragos:2019oxn}  and by higher-dimensional operators, 
such as the quark EDM~\cite{Bhattacharya:2015esa,Gupta:2018lvp} 
and at a more exploratory level the quark chromo-EDM~\cite{Abramczyk:2017oxr,Bhattacharya:2018qat,Kim:2018rce}.
In this paper, we present a new calculation of the contribution of the $\Theta$-term  to the nEDM and pEDM and show that the statistical and systematic uncertainties are still too large to extract reliable estimates.

This paper is organized as follows: In Section~\ref{sec:lagrangian},
we describe our notation by introducing the Lagrangian with \CPV\ and
the needed matrix elements. In Section~\ref{sec:FF}, we describe the
decomposition of the matrix elements into the electromagnetic form
factors. Section~\ref{sec:lattparm} provides the lattice parameters
used in the calculations. In Section~\ref{sec:charge}, we present the
implementation of the gradient flow scheme, and in Sec.\ref{sec:ChiQ}
the calculation of the topological
susceptibility. Section~\ref{sec:alpha} describes the methodology for
extracting the \CPV\ phase $\alpha$ for the ground state created by the nucleon
interpolating operator used, from the two-point function. This phase  
controls the CP transformation of the asymptotic
nucleon state. Section~\ref{sec:3pt} describes the calculation
strategy for obtaining the form factors when this phase $\alpha$ is
nonzero and gives the  formulae used to extract the \CPV\ form
factor \(F_3\) from the matrix elements. In Section~\ref{sec:ESC}, we
discuss the extraction of $F_3(q^2)$ and the removal of the excited
states contamination. The extrapolation of $F_3(q^2)$ to $q^2 = 0$ is
presented in Sec.~\ref{sec:qsq}.  Section~\ref{sec:Oa} discusses the
lattice-spacing artifacts. Our results with the excited state spectrum
taken from the two-point function are presented in
Sec.~\ref{sec:Results} and those with an $N \pi$ excited state in
Sec.~\ref{sec:Npi}.  These results are compared to previous
calculations in Section~\ref{sec:comp}. Conclusions are presented in
Section~\ref{sec:Conclusions}. Further details on the connection 
between Minkowski and Euclidean notation, the extraction of the 
form factors, the chiral extrapolation, excited-state contamination, and 
the $O(a)$ corrections in the Wilson-clover theory  are presented
in five appendices.

\section{The QCD \texorpdfstring{$\Theta$}{\texttheta}-term}
\label{sec:lagrangian}

QCD allows for the existence of a P and T (and \CPV\ if CPT is
conserved) violating dimension-four operator, i.e., the $\Theta$-term.  In
its presence, the QCD Lagrangian density in Euclidean notation becomes
{\begin{equation}
{\cal L}_{\rm QCD}  \mathbin{{\longrightarrow}} {\cal L}_{\rm QCD}^{\cancel{\text{CP}}} =  {\cal L}_{\rm QCD}  + 
   i \Theta
   \frac{G^a_{\mu\nu}  {\widetilde G^a_{\mu\nu}} }{32\pi^2}
\label{eq:Lcpv}
\end{equation}}%
where $G^a_{\mu \nu}$ is the chromo-field strength tensor, \(\widetilde G^a_{\mu\nu} =\frac12 \epsilon_{\mu\nu\lambda\delta} G^{a\lambda\delta}\) is its dual, and $\Theta$
is the coupling.\footnote{Throughout the paper, we work in Euclidean space, 
using $q$ for the Euclidean 4-momentum and  $Q$ for the topological charge. The gauge field includes a factor of the strong coupling, \(g\), so that the kinetic term is \(G^a_{\mu\nu}G^a_{\mu\nu}/4g^2\).
Also, our conventions for connecting the Euclidean and Minkowski metrics 
are given in  Appendix~\ref{sec:appendix0}.} \(G_{\mu\nu}  {\widetilde G_{\mu\nu}}\) is a total derivative
of a gauge-variant current and its space-time integral gives the
 topological charge 
 \begin{equation}
     Q = \int d^4x \, \frac{G^a_{\mu\nu}  {\widetilde G^a_{\mu\nu}} }{32\pi^2}~.
 \end{equation}
Non-zero values of $Q$ are tied to the topological structure of
QCD and the $U(1)$ axial anomaly. In addition, higher
dimension operators that arise due to novel \CPV\ couplings
at the TeV scale generate this term under renormalization in a hard cutoff scheme like lattice regularization or gradient flow~\cite{Bhattacharya:2015esa}. Also, BSM models in which the Peccei-Quinn mechanism is operative induce such a term~\cite{Pospelov:2005pr}. 

Under a
chiral transformation, one can rotate $\Theta$ into a complex phase of
the quark matrix and vice versa. It is, therefore, necessary to
work with the convention independent ${\overline \Theta} = \Theta + {\rm Arg\ Det}M_q$, which
includes both, $\Theta$ from all sources and the overall phase of the
quark matrix $M_q$. Since, the argument of the determinant is ill-defined when it is zero, all physical effects of \(\overline\Theta\) vanish in the presence of even a single massless quark flavor.

If the
overall $\overline \Theta$ is nonzero, then this operator would induce an nEDM $d_n$
of size 
\begin{eqnarray}
            d_n &=& {\overline \Theta} \ X 
            \\
X &\equiv&            \lim_{q^2\to0} \frac{F_3(q^2)}{2 M_N \overline\Theta} \,.
\label{eq:dndef}
\end{eqnarray}
Here $X$ is obtained from the \CPV\ part of the matrix
element of the electromagnetic vector current within the neutron
state in the presence of the $\Theta$-term and $F_3$ is the \CPV\ violating form factor defined in 
Eq.~\eqref{eq:FFdef}. This is obtained, at the leading order, from the \CPV\ part of the matrix element
\begin{align}
&  \left.\langle N \mid  J^{\rm EM}_\mu \mid N \rangle\right|
^{{\overline \Theta}} \approx \left.\langle N \mid J^{\rm EM}_\mu \mid N \rangle\right|^{{\overline \Theta}=0} \nonumber \\
     & \qquad\qquad -i {\overline \Theta}    \left\langle N \left| J^{\rm EM}_\mu \ \int d^4 x\   
       \frac{G^a_{\mu\nu} \widetilde G^a_{\mu\nu}}{32\pi^2} \right| N \right\rangle \,,
\label{eq:ThetaEDM}
\end{align}
where we have assumed that the \({\overline \Theta}\)-term is the only source of \CPV.
In other words, $X$ provides the connection  between the 
\CPV\ coupling (${\overline \Theta}$) and the nEDM ($d_n$).

\begin{table*}[tbhp]
\centering
\renewcommand{\arraystretch}{1.2}
\begin{ruledtabular}
\begin{tabular}{ c|c|c|c|c|c|c|c|c|c|c|c }
Ensemble&$a$&$M_\pi^{val}$& $L^3\times T$&$M_\pi^{\rm val}L$&$\tau/a$      &$aM_N$      &$N_{conf}$&Confs.&$N_{HP}$ &$N_{LP}$    &$\chi_Q^{1/4}$    \\
ID &[fm]&[MeV]&&&&&&Per Bin&&&[MeV]\\                                                                                                                 
\hline                                                       

\hline                                                                                                                                         
$a12m310$ &$0.1207(11)$&$310.2(2.8)$&$24^3\times64$&$4.55$&$\{8,10,12\}$   &$0.6660(27)$&$1013$    & 8  &$4,052$   &$64,832$    &  145.9(2.7)   \\
$a12m220$ &$0.1184(09)$&$227.9(1.9)$&$32^3\times64$&$4.38$&$\{8,10,12\}$   &$0.6122(25)$&$1000$    & 8  &$4,000$   &$64,000$    &   145.3(2.4)  \\
$a12m220L$&$0.1189(09)$&$227.6(1.7)$&$40^3\times64$&$5.49$&$\{8,10,12\}$   &$0.6125(21)$&$1000$    & 8  &$4,000$   &$128,000$   &   141.3(2.5)  \\
                                                                                                                                                
\hline                                                                                                                                          
$a09m310$&$0.0888(08)$&$313.0(2.8)$&$32^3\times96$&$4.51$&$\{10,12,14\}$   &$0.4951(13)$&$2196$    & 18 &$8,784$   &$140,544$   &   129.5(2.3) \\
$a09m220$&$0.0872(07)$&$225.9(1.8)$&$48^3\times96$&$4.79$&$\{10,12,14\}$   &$0.4496(18)$&$961$     & 8  &$3,844$   &$123,008$   &   115.0(2.2) \\
$a09m130$&$0.0871(06)$&$138.1(1.0)$&$64^3\times96$&$3.90$&$\{10,12,14\}$   &$0.4204(23)$&$1289$    & 11 &$5,156$   &$164,992$   &   106.8(1.7) \\
\hline                                                                                                                                          
$a06m310$&$0.0582(04)$&$319.3(5)$  &$48^3\times144$&$4.5$&                 &            &$970$     &    &          &            &   127.0(5.5)  \\
$a06m220$&$0.0578(04)$&$229.2(4)$  &$64^3\times144$&$4.4$&                 &            &$1014$    &    &          &            &   103.0(4.2)  \\
$a06m135$&$0.0570(01)$&$135.6(1.4)$&$96^3\times192$&$3.7$&$\{16,18,20,22\}$&$0.2704(32)$&$453$     & 9  &$1,812$   &$28,992$    &    89.3(2.8)  \\
\end{tabular}
\end{ruledtabular}
\caption{Lattice parameters, nucleon mass $M_N$, number of
  configurations analyzed, and the total number of high precision (HP)
  and low precision (LP) measurements made. We also give the bin size
  (Confs. per bin) used in the statistical analysis of two- and three-point functions. The last column gives
  the topological susceptibility $\chi_Q$ calculated at flow time
  $\tau_{gf} = 0.68$~fm and with a bin size of 20 configurations.  The
  ensembles $a06m310$ and $a06m220$ have been used only for the
  calculation of $\chi_Q$, and 861 configurations 
  were used to calculate $\chi_Q$ on the $a06m135$ ensemble. }
\label{tab:ensembles}
\end{table*}

At present, the upper bound on the nEDM, $|d_n| \allowbreak < \allowbreak 1.8 \times
10^{-26}$~$e$~cm (90\% CL)~\cite{Abel:2020gbr}, is used along with an
estimate \(X\sim(2.50 \pm 1.25)\times10^{-16}\)~$e$ cm~\cite{Pospelov:2005pr}
to set a limit on the size of ${\overline \Theta} \lesssim
10^{-10}$. This is an unnaturally small number! One solution to this
unnaturalness is the dynamical tuning of ${\overline \Theta} = 0$
using the Peccei-Quinn mechanism\footnote{The Peccei-Quinn mechanism
relaxes \(\overline\Theta\) dynamically to \(\Theta_{\rm ind}\), the point
where the effective potential achieves its minimum.  In the absence of other
sources of CP violation in the theory, \(\Theta_{\rm ind}=0\).}~\cite{PhysRevLett.38.1440}.

{Our goal is to calculate $X$ using lattice QCD, which multiplied
by the cumulative value, ${\overline \Theta}$, from all sources (SM or
BSM), gives the full contribution to nEDM 
from the dimension-4 $G \tilde G$ operator in Eq.~(\ref{eq:Lcpv}). Knowing $X$ will
allow current and future bounds on (or measured value of) $d_n$ 
to more stringently constrain or pin down ${\overline \Theta}$.

In the rest of the paper,} all the analyses are carried out assuming that the only \CPV\ coupling
arises from the \(\Theta\)-term, whose strength is \(\overline\Theta\).
Results are presented for  \(\overline\Theta=0.2\), which we have checked 
is small enough {so that \(O(\overline\Theta^2)\) corrections are negligible for all quantities of interest
($\alpha$ and $F_3$ defined later)}.  \looseness-1

\begin{figure}[b]
  \subfigure{
    \includegraphics[width=0.4\linewidth]{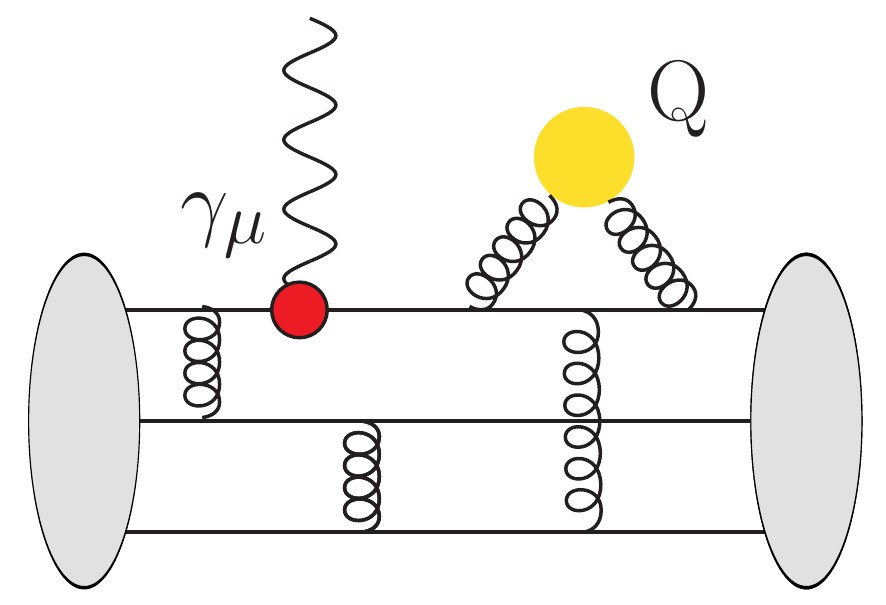}
  }
  \hspace{0.04\linewidth}
  \subfigure{
    \includegraphics[width=0.4\linewidth]{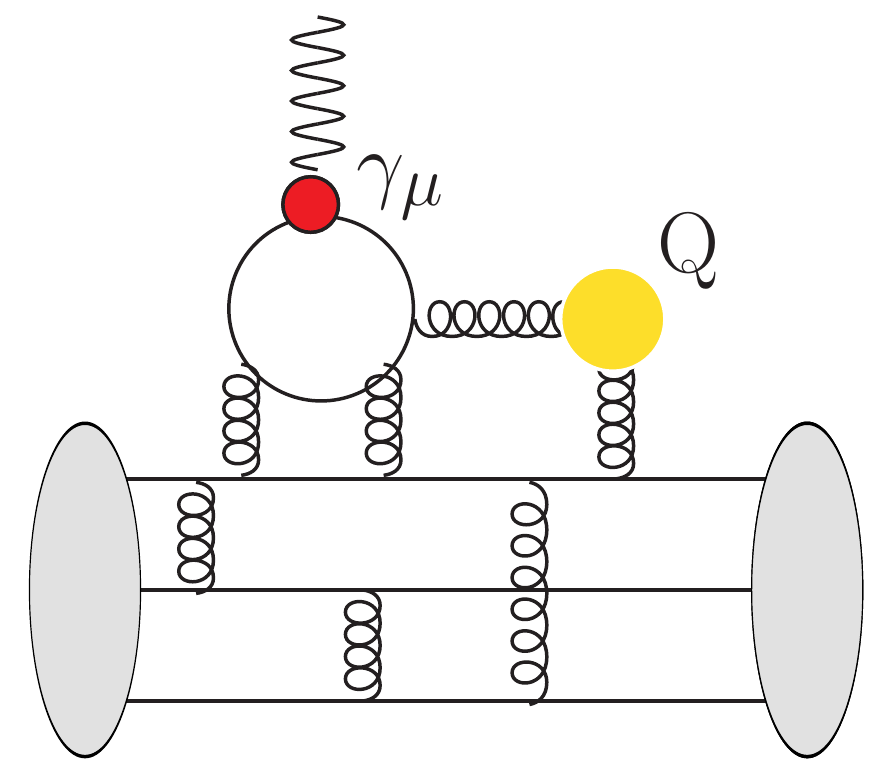}
  }
\caption{The
  connected (left) and disconnected (right) diagrams with
  the insertion of the bilinear vector current (red filled circle) in
  the nucleon two-point function. The signal is given by the
  correlation between this 3-point function and the topological charge
  shown by the filled yellow circle.
  \label{fig:con_disc}}
\end{figure}

The lattice
calculation consists of the evaluation of the connected
and disconnected diagrams shown in Fig.~\ref{fig:con_disc}. The
disconnected diagram gets contributions from all quark flavors in the loop---but their contributions to the CP-conserving form-factors of the vector current are small~\cite{Alexandrou:2018sjm}. In this work, we assume the same holds for the CP-violating ones and neglect the contribution to the electric dipole moment coming from these diagrams.

%

\section{Form Factor of the Electromagnetic Current}
\label{sec:FF}

The parameterization of the matrix element of the electromagnetic
current, $J^{\rm EM}_\mu(q)$, defined in Eq.~\eqref{eq:ThetaEDM}, within the
  nucleon state in terms of the most general set of form factors
  consistent with the symmetries of the theory is
\begin{align}
& \langle N (p^\prime,s') \mid J^{\rm EM}_\mu \mid N (p,s) \rangle_{\not{\rm CP}}^{{\overline \Theta}} 
\ = \   \overline {u}_N (p',s') \Bigg[   \gamma_\mu   F_1 (q^2)
\notag \\
& \qquad\qquad\qquad {} + \frac{1}{2 M_N} \sigma_{\mu \nu} q_\nu \Big( F_2(q^2)  - i F_3 (q^2) \gamma_5 \Big) 
\notag \\
& \qquad\qquad\qquad {} +  \frac{F_A(q^2)}{M_N^2}  (\slashed q   q_\mu  - q^2 \gamma_\mu) \gamma_5~    \Bigg] u_N (p,s)\,,
\label{eq:FFdef}
\end{align}
where \(M_N\) is the nucleon mass, \(q=p^\prime -p\) is the Euclidean 4-momentum transferred 
by the electromagnetic current, $\sigma_{\mu \nu} = (i/2) [\gamma_\mu, \gamma_\nu]$, 
and \(u_N(p,s)\) represents the free
neutron spinor of momentum \(p\) and spin \(s\) obeying \(( i \pslash + M_N) u_N (p,s)=0\),
with \(\gamma_4\) implementing the asymptotic (i.e., free) parity operation.
Throughout, we work in Euclidean space  and refer the reader to Appendix~\ref{sec:appendix0} 
for details on our conventions. 
\(F_1\) and \(F_2\) are the Dirac and Pauli form factors, in terms of
which the Sachs electric and magnetic form factors are
\(G_E = F_1 - (q^2/4M_N^2) F_2\) 
and \(G_M = F_1 + F_2\), respectively.\footnote{We emphasize that
we use \(q^2\) for the \emph{Euclidean} four-momentum-squared that is
denoted by \(Q^2\) in our previous work and throughout the literature.  As noted in the Appendix~\ref{sec:appendix0},
it is the negative of the Minkowski four-momentum-squared.}
The anapole form factor \(F_A\) and the electric dipole form factor
\(F_3\) violate parity P; and \(F_3\) violates CP as well.  
The zero momentum limit of these form
factors gives the charges and dipole moments: the
electric charge is \(G_E(0) = F_1(0)\), the magnetic
dipole moment is \(G_M(0)/2 M_N = (F_1 (0) + F_2(0)) / 2 M_N\), and the 
EDM is defined in Eq.~\eqref{eq:dndef}.

In all the discussions in this paper, the current $J_\mu^{\rm EM}$ used is the  
renormalized local vector current $Z_V \sum_i e_i {\overline \psi_i} \gamma_\mu \psi_i$, where 
$e_i$ is the electric charge of a quark with flavor $i$. 
The renormalization is carried out by taking ratios of all 
three-point fermion correlators with the lattice estimate of the vector charge, $g_V \equiv 1/Z_V$, which is given by the forward matrix element of $ {\overline \psi_i} \gamma_4 \psi_i$.  
These ratios are constructed with identical source, sink, and current insertion positions and within the single jackknife loop used for the 
statistical analysis of the data to take advantage of error reduction due to correlated fluctuations.\footnote{This forward matrix element has very small excited state contamination and, therefore, does not affect our excited state fits at this level of precision.}

%
\section{Lattice Parameters}
\label{sec:lattparm}

We present results on seven ensembles, whose parameters are defined in 
Table~\ref{tab:ensembles}.   These were generated by the MILC
collaboration~\cite{Bazavov:2012xda} using 2+1+1-flavors of highly
improved staggered quarks (HISQ) action.  For the construction of the
nucleon correlation functions we use the clover-on-HISQ formulation
that has been used extensively by us in the calculation of the nucleon
charges and form factors as described in
Refs.~\cite{Gupta:2018qil,Jang:2019jkn}. These ensembles cover three
values of the lattice spacing, $a \approx 0.12$, $0.09$  and $0.06$~fm and
three values of the pion mass $M_\pi \approx 315, 220$ and $
130$~MeV. Further details of the lattice parameters and methodology,
statistics, and the interpolating operator used to construct the nucleon
2- and 3-point correlation functions can be found in
Refs.~\cite{Jang:2019jkn,Gupta:2018qil}.


%
\begin{figure*}[htpb]
  \begin{minipage}[t]{0.45\linewidth}
    \includegraphics[width=0.48\linewidth]{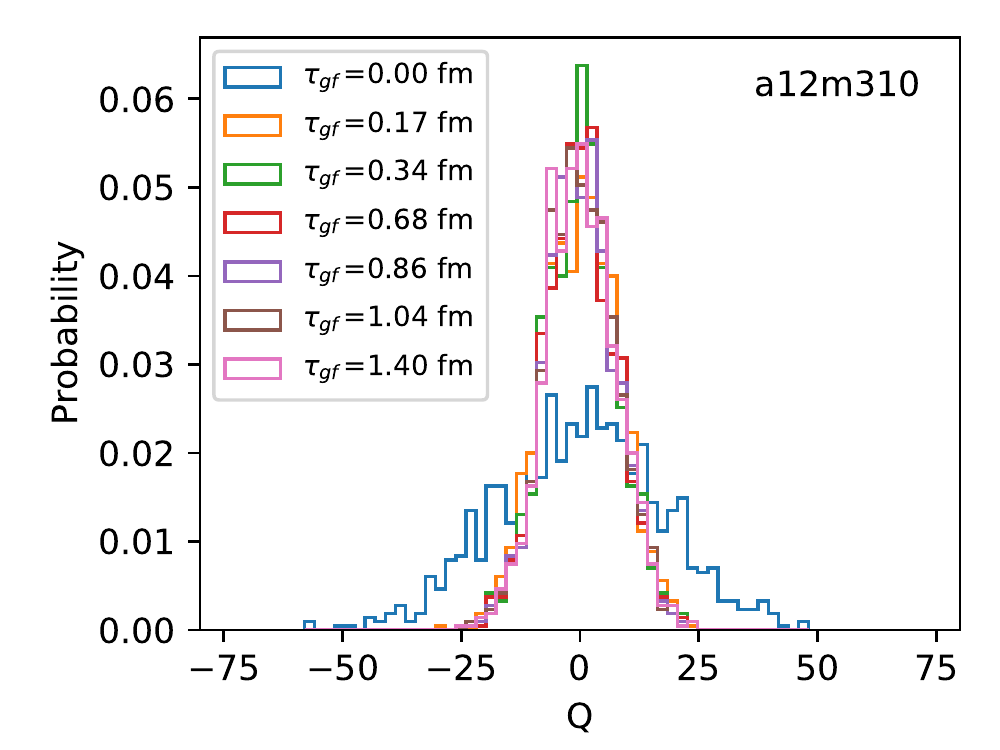}
    \includegraphics[width=0.48\linewidth]{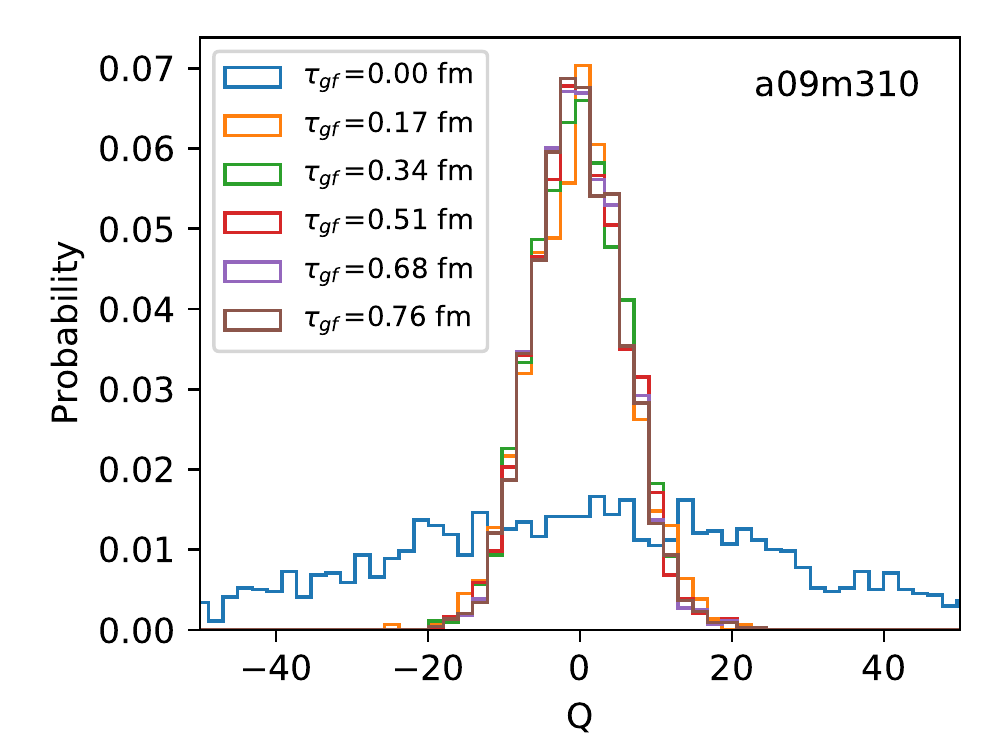}\\
    \includegraphics[width=0.48\linewidth]{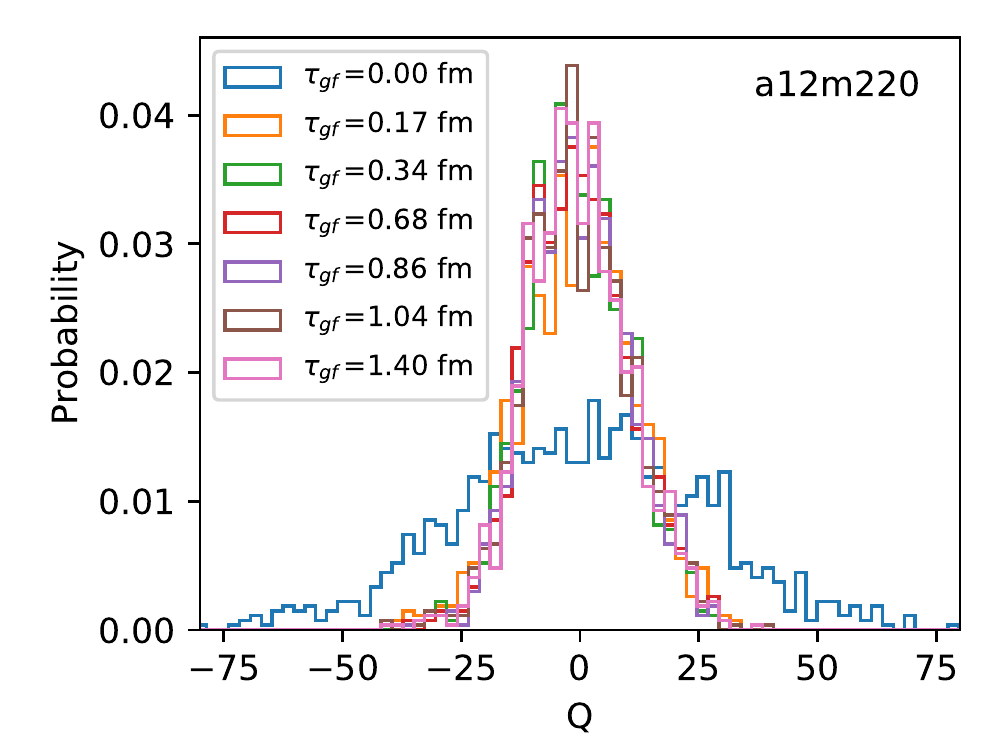}
    \includegraphics[width=0.48\linewidth]{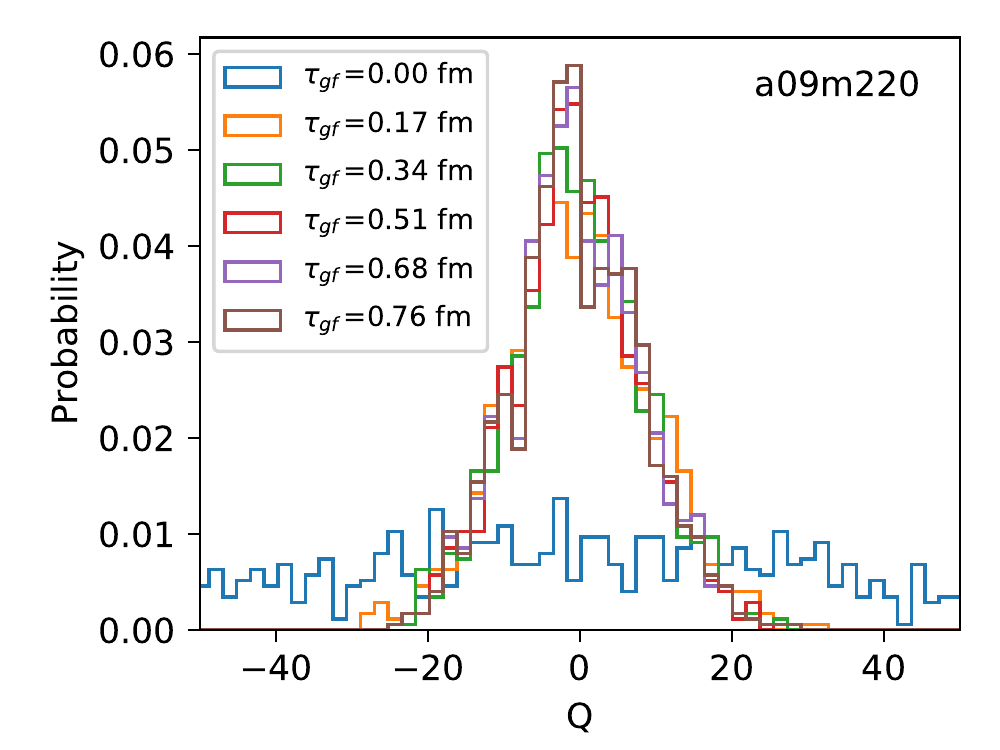}\\
    \includegraphics[width=0.48\linewidth]{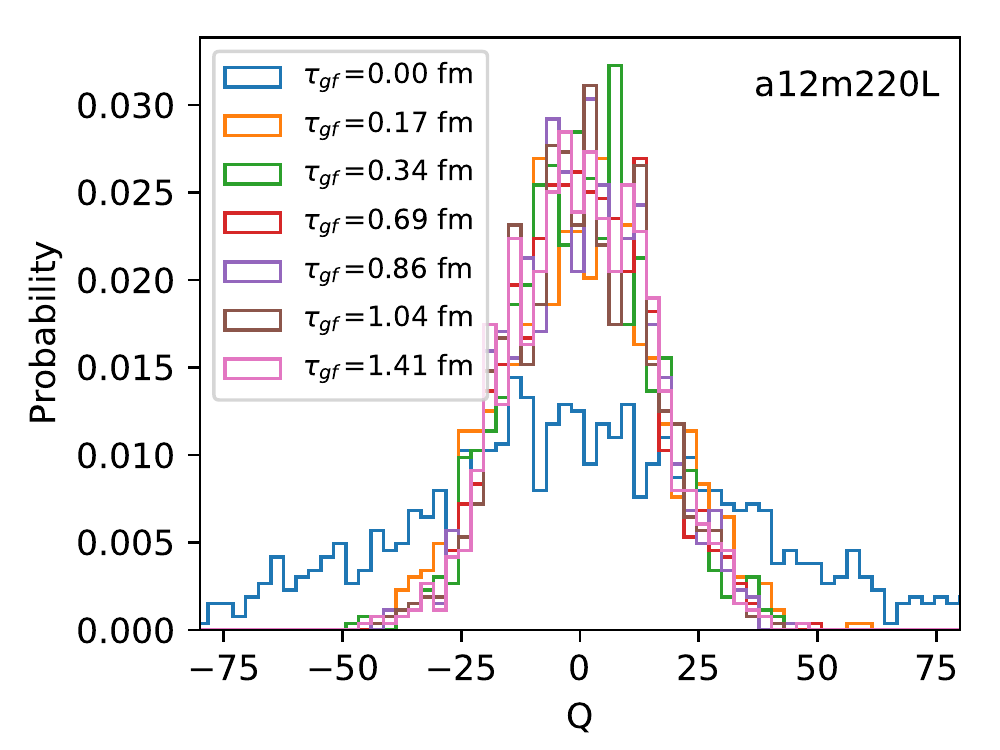}
    \includegraphics[width=0.48\linewidth]{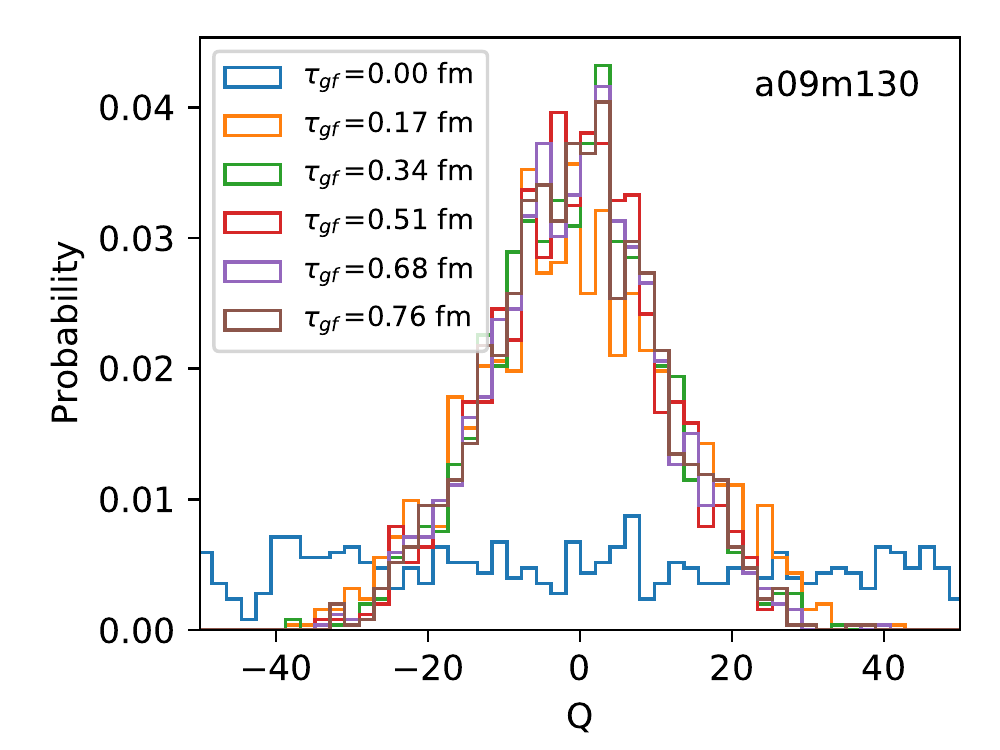}\\
    \includegraphics[width=0.48\linewidth]{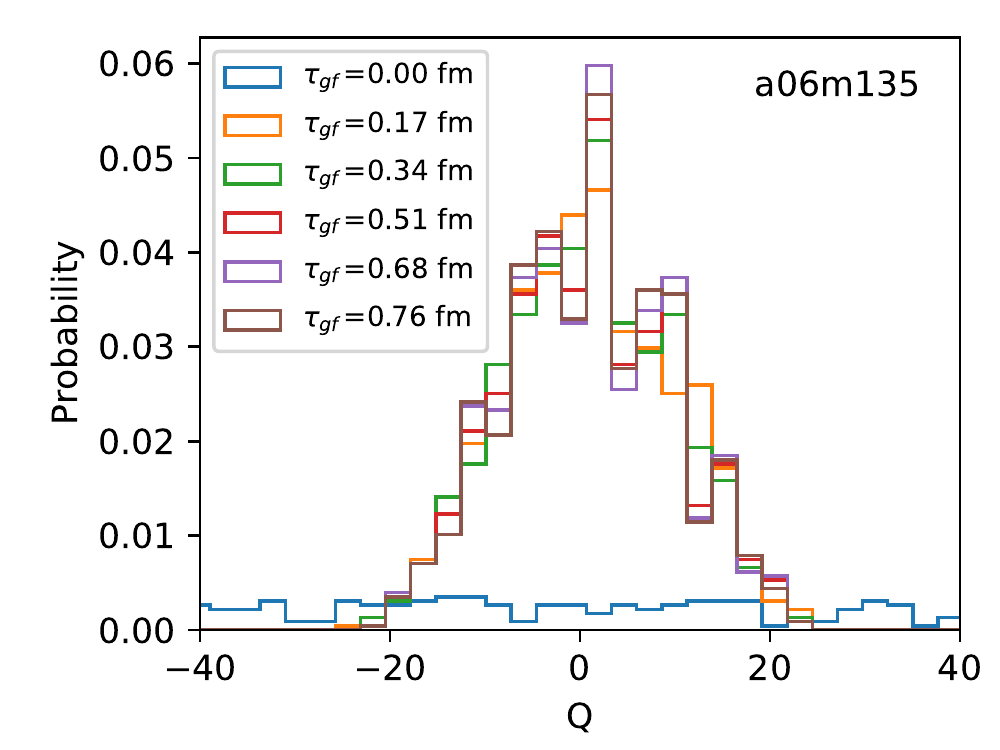}
\caption{The distribution of the topological charge $Q$ as a function of
  the flow time $\tau_{\rm gf}$. The panels on the left (right) show data for the $a=0.12$~fm ($a=0.09$~fm) ensembles 
\label{fig:Qdist}}
  \end{minipage}
\hspace{0.05\linewidth}
  \begin{minipage}[t]{0.45\linewidth}
    \includegraphics[width=0.48\linewidth]{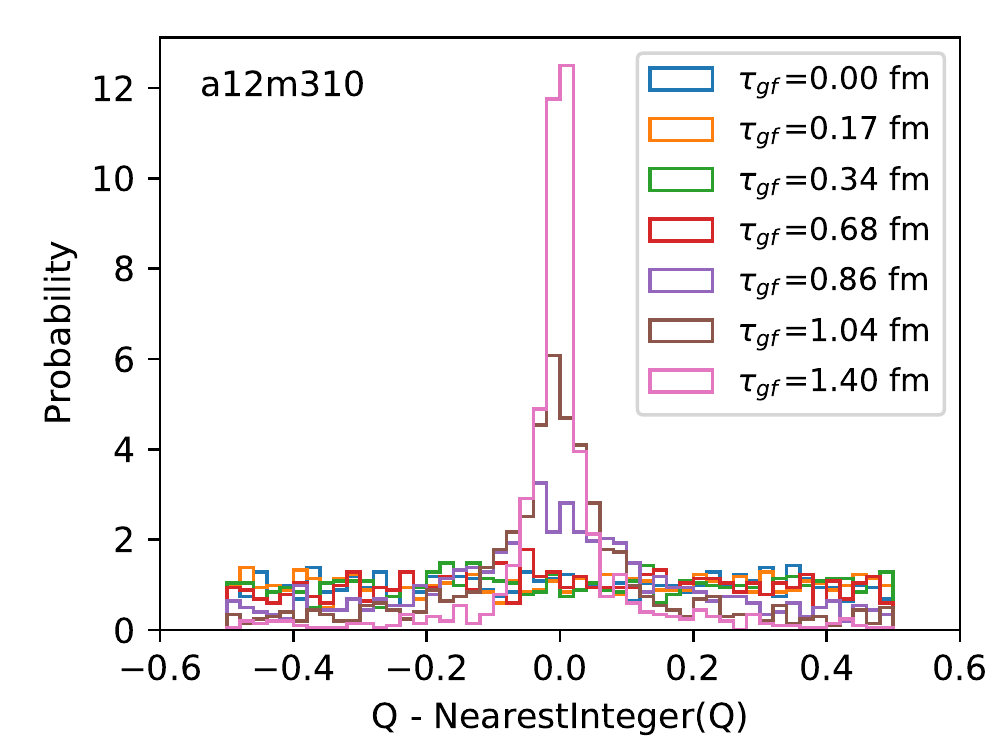}
    \includegraphics[width=0.48\linewidth]{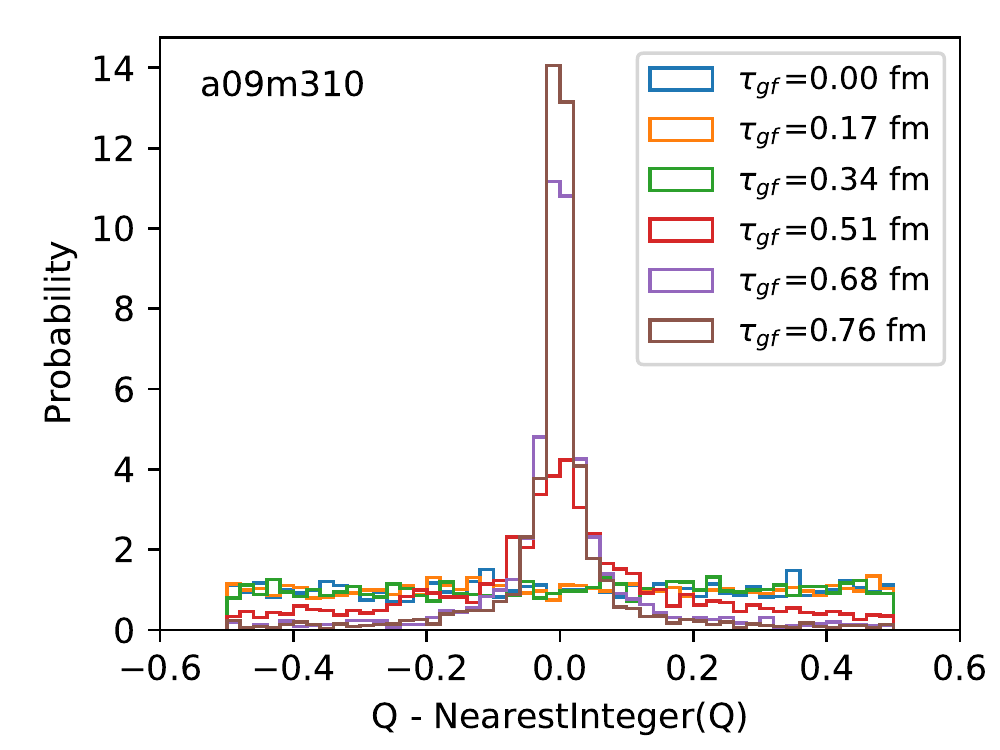}\\
    \includegraphics[width=0.48\linewidth]{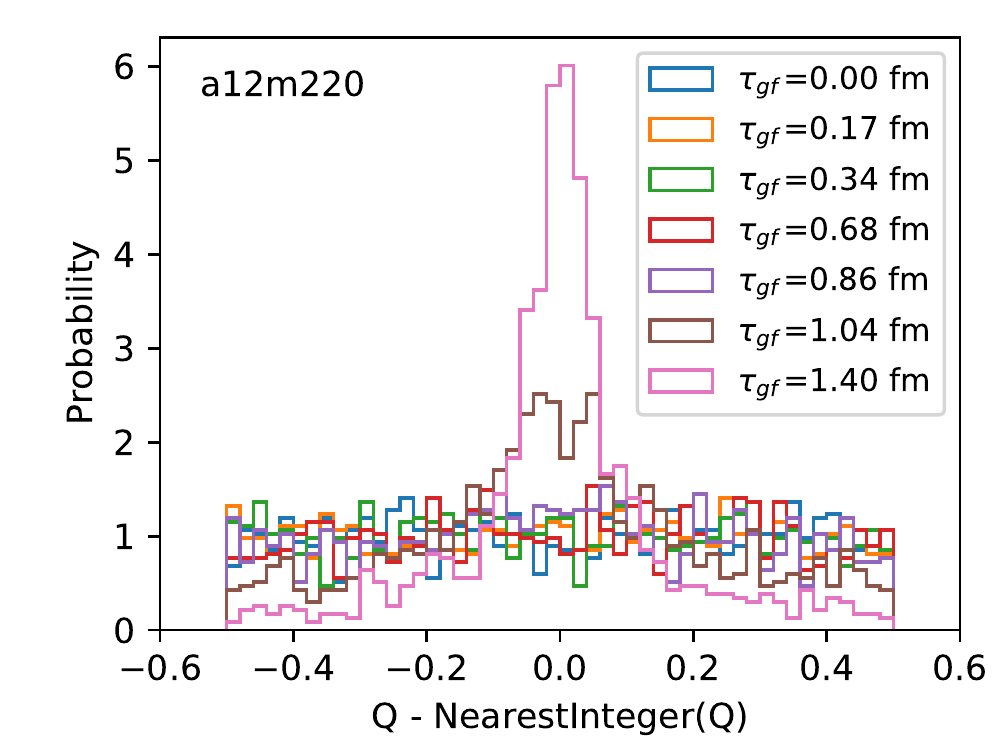}
    \includegraphics[width=0.48\linewidth]{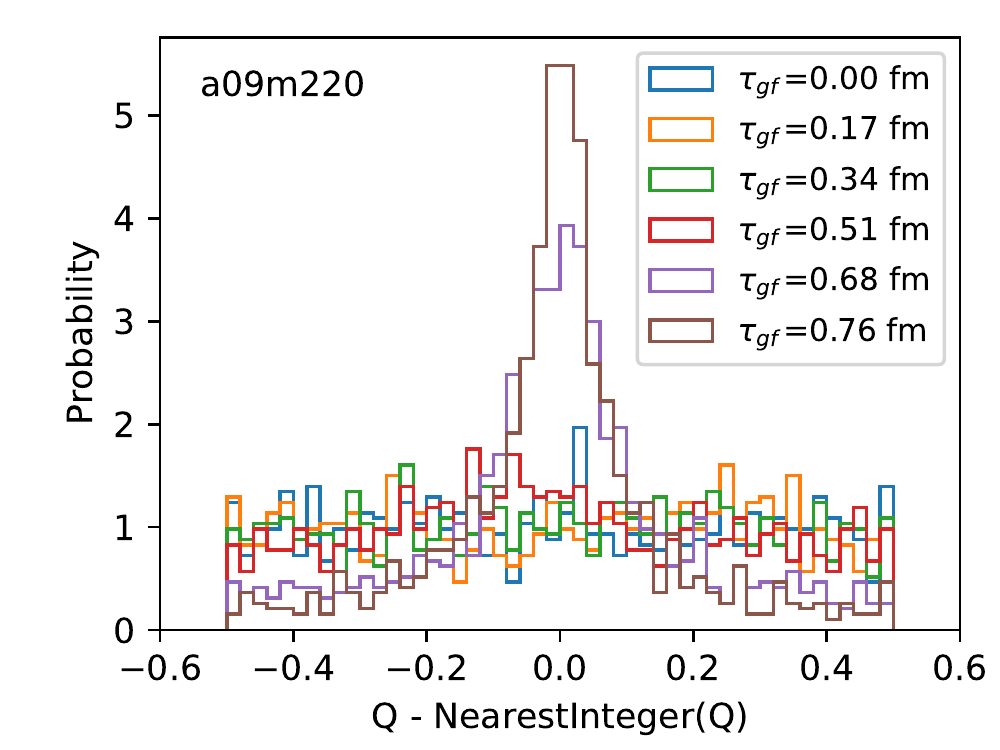}\\
    \includegraphics[width=0.48\linewidth]{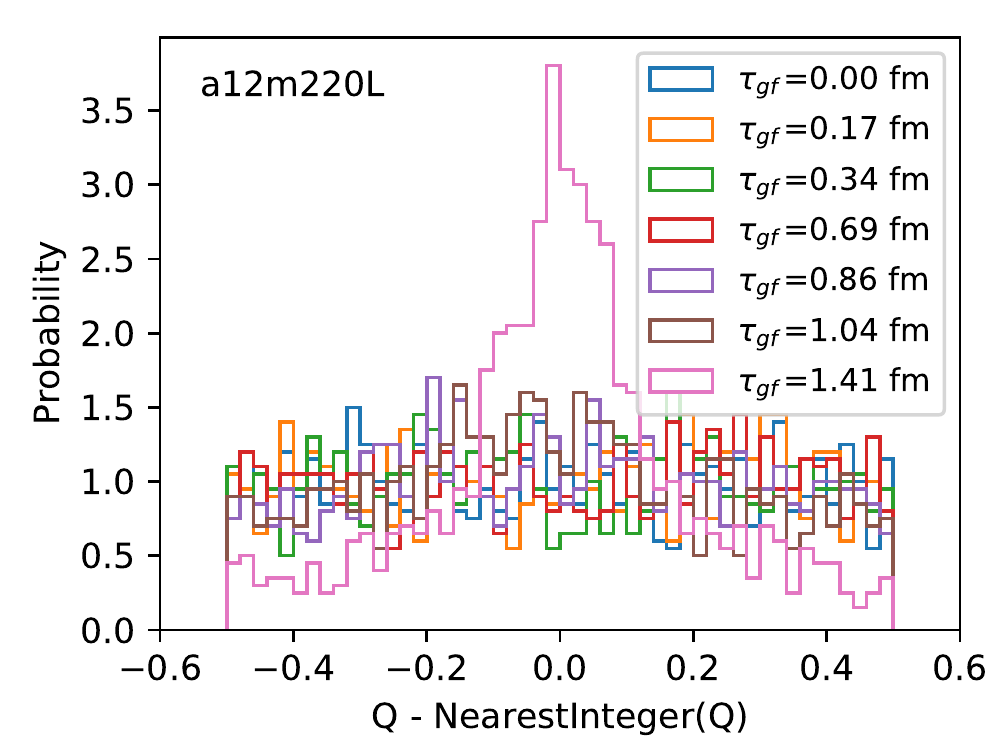}
    \includegraphics[width=0.48\linewidth]{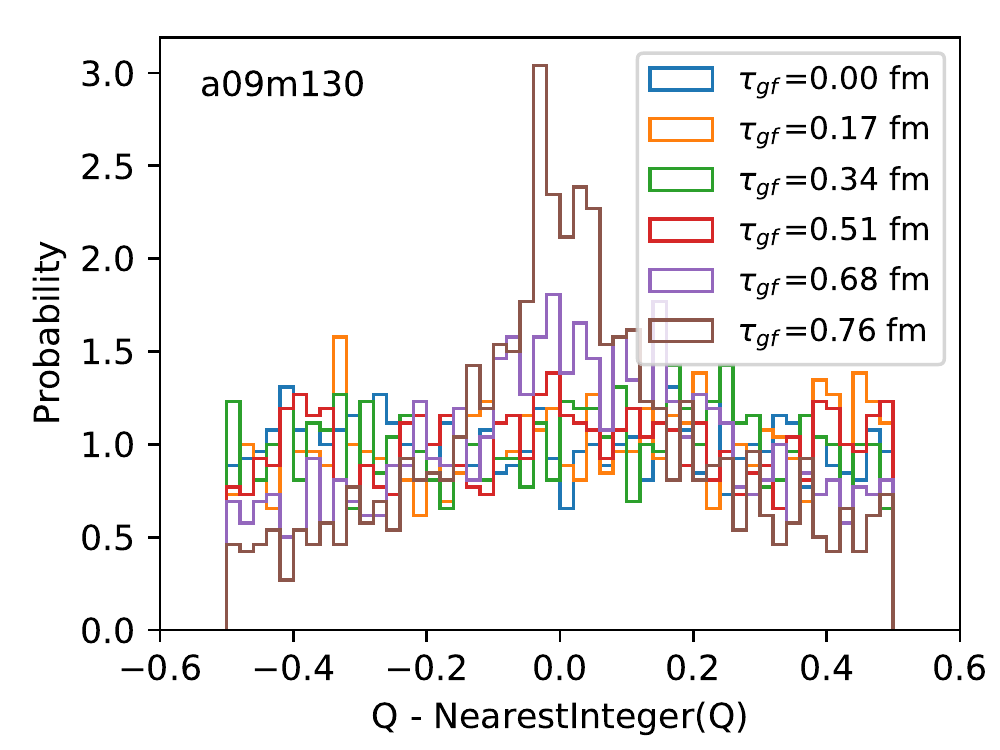}\\
    \includegraphics[width=0.48\linewidth]{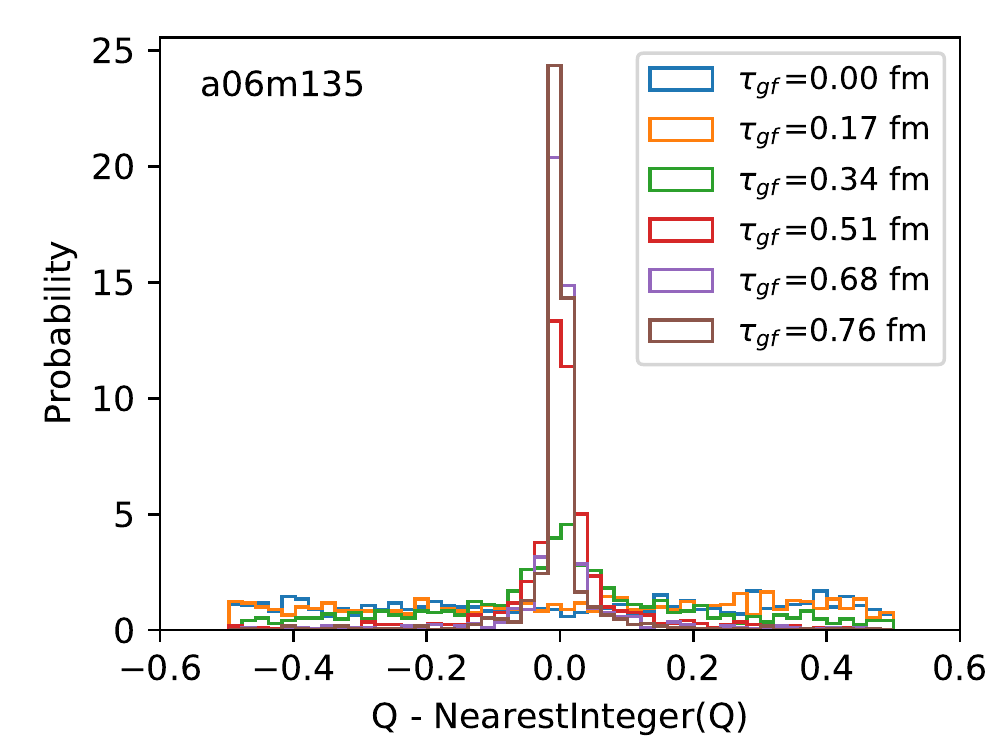}
\caption{The panels show the distribution of the difference, $Q -
  Q_{\rm int}$, of the measured $Q$ from the nearest integer $Q_{\rm
    int}$.
\label{fig:Qinteger}}
  \end{minipage}
\end{figure*}
\vspace{\baselineskip}

\begin{figure*}[tbp]
\begin{minipage}[t]{\linewidth}
  \subfigure{
    \includegraphics[width=0.42\linewidth]{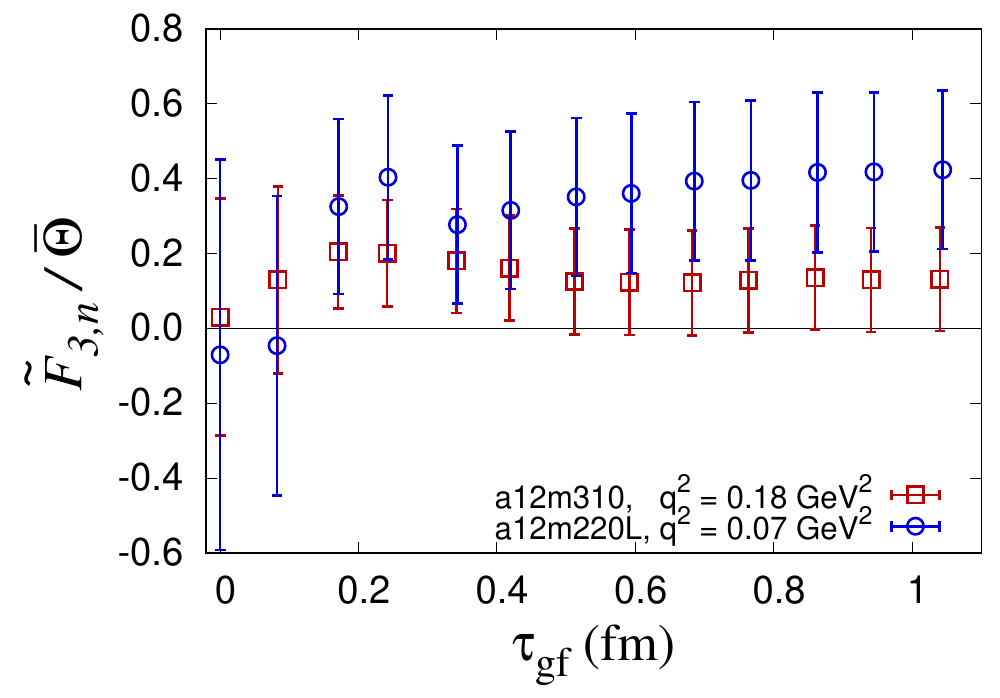}\qquad
    \includegraphics[width=0.42\linewidth]{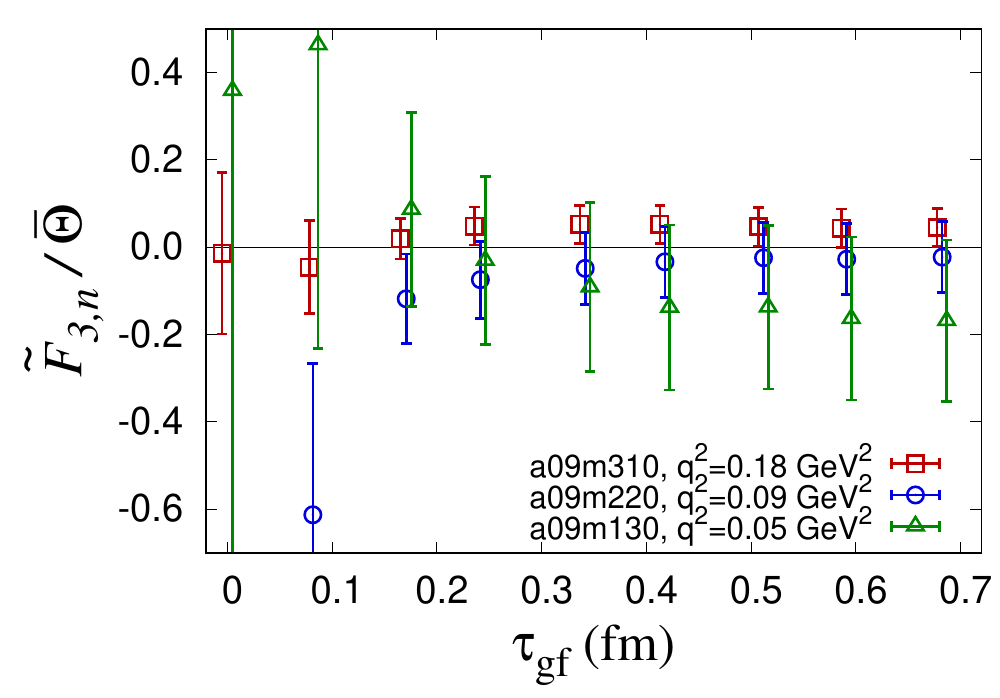}
  }
\caption{Data for ${\widetilde F}_{3,n}/{\overline \Theta}$, 
defined in Eq.~(\ref{eq:F3tilde}), 
at the 
smallest value of $q^2$, respectively, on the $a12$ (left panel) and the $a09$ (right
panel) ensembles. The estimates show no significant change after
$\tau_{\rm gf} \approx 0.4$~fm on the $a09$ ensembles and $\tau_{\rm gf} \approx 0.6$~fm on the $a12$
ensembles.
\label{fig:flowF3}}
\end{minipage}
\end{figure*}

\begin{figure*}[htpb]
  \begin{minipage}[t]{0.45\linewidth}
    \includegraphics[width=0.48\linewidth]{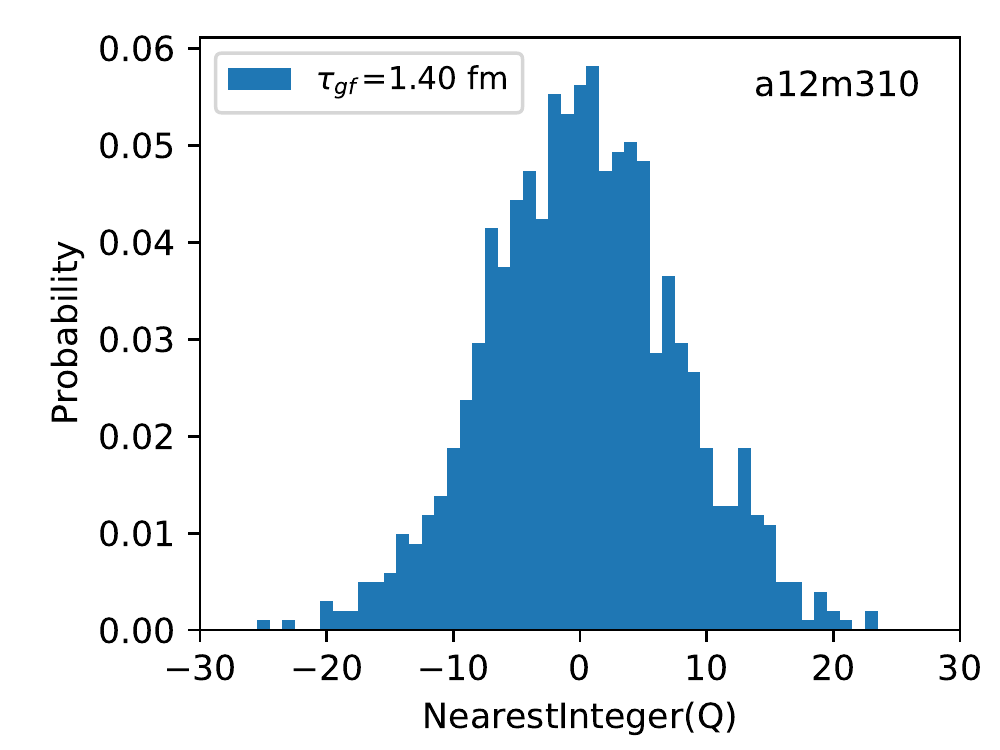}
    \includegraphics[width=0.48\linewidth]{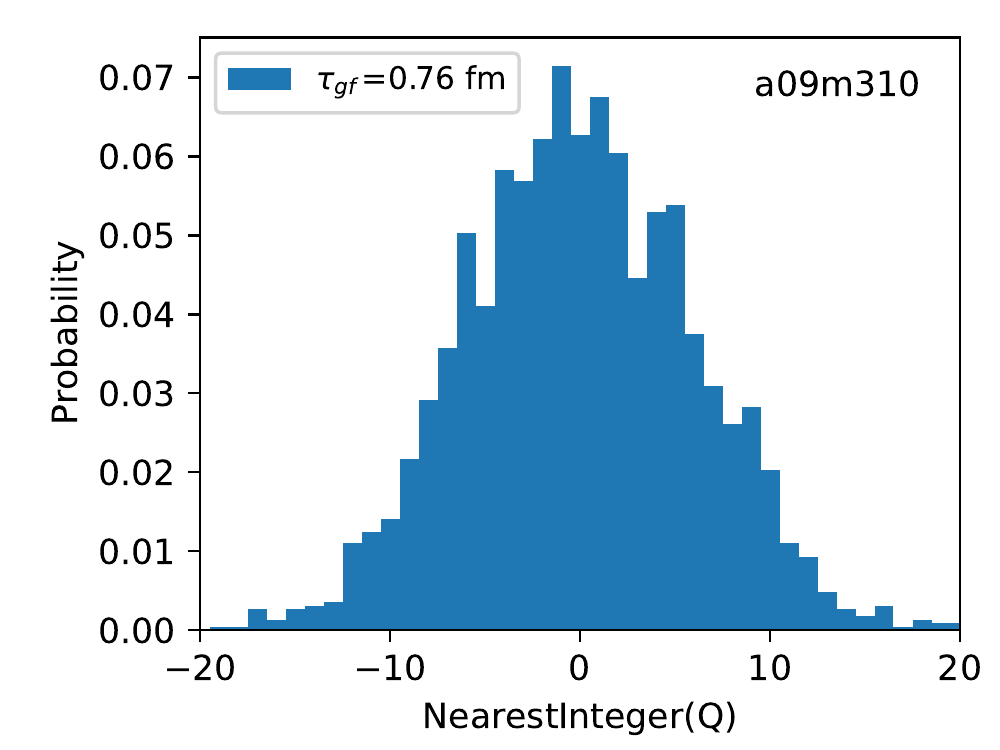}\\
    \includegraphics[width=0.48\linewidth]{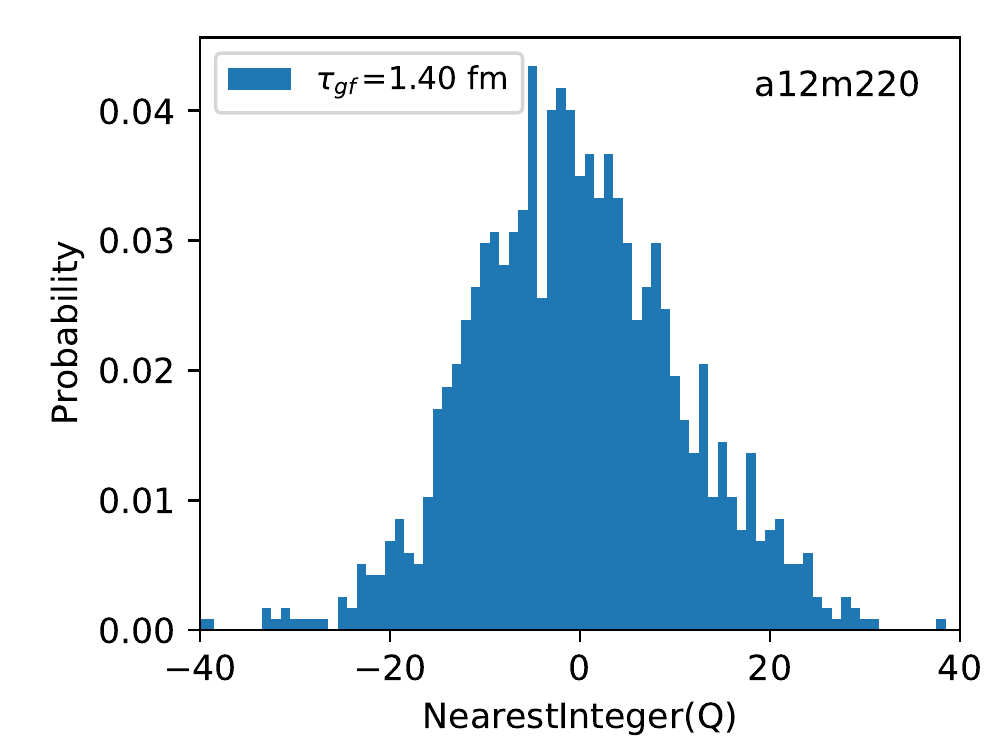}
    \includegraphics[width=0.48\linewidth]{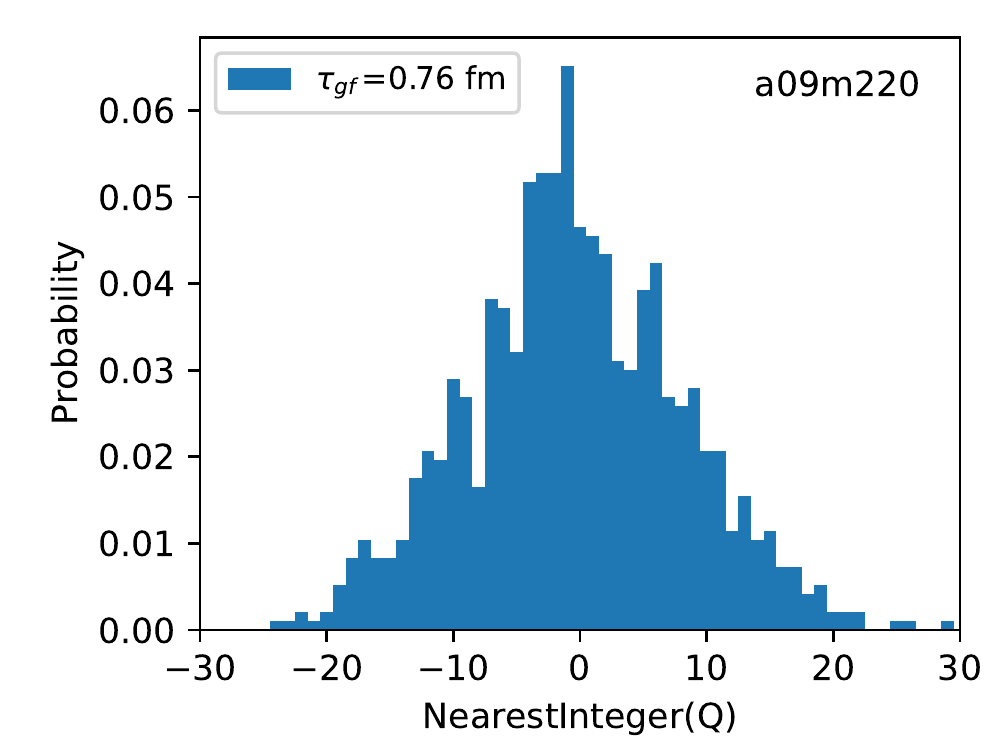}\\
    \includegraphics[width=0.48\linewidth]{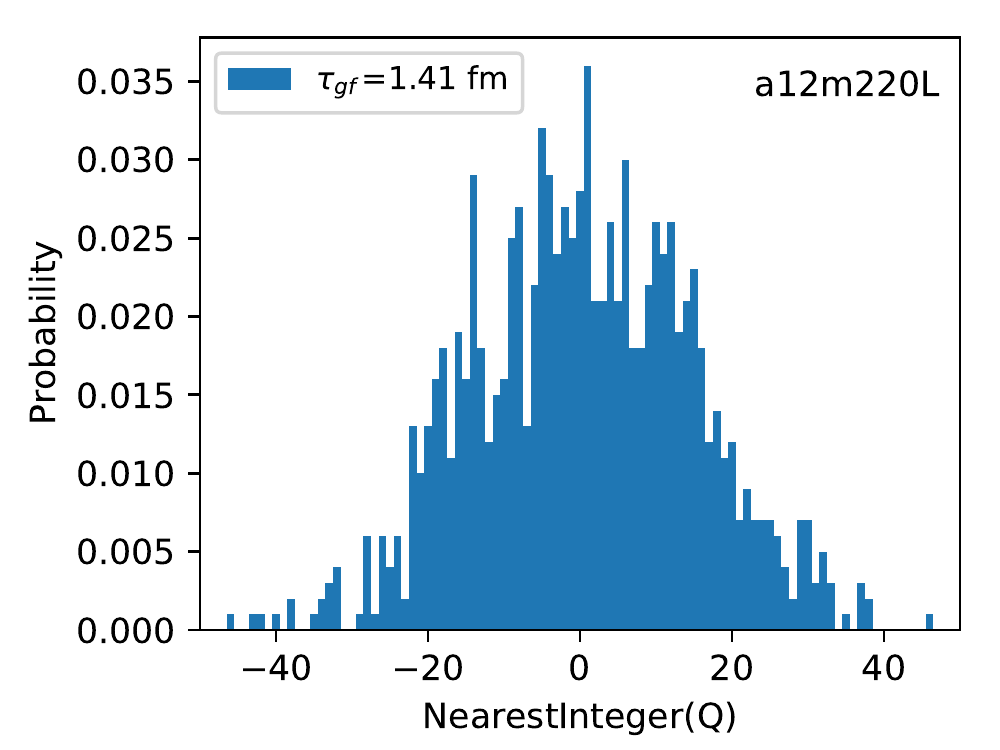}
    \includegraphics[width=0.48\linewidth]{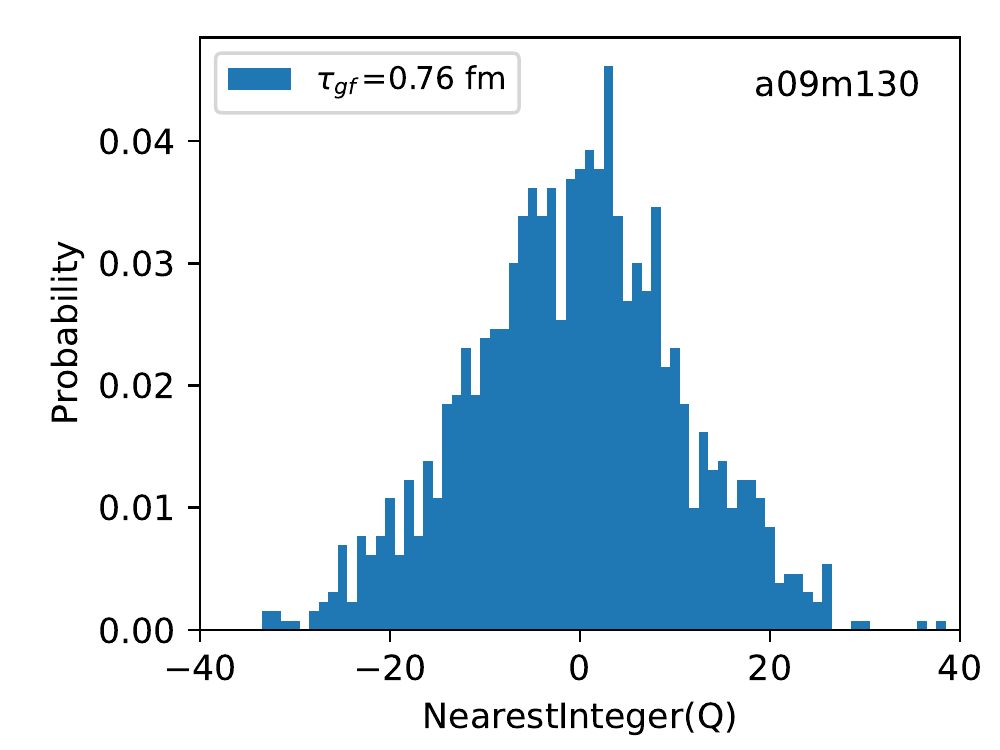}\\
    \includegraphics[width=0.48\linewidth]{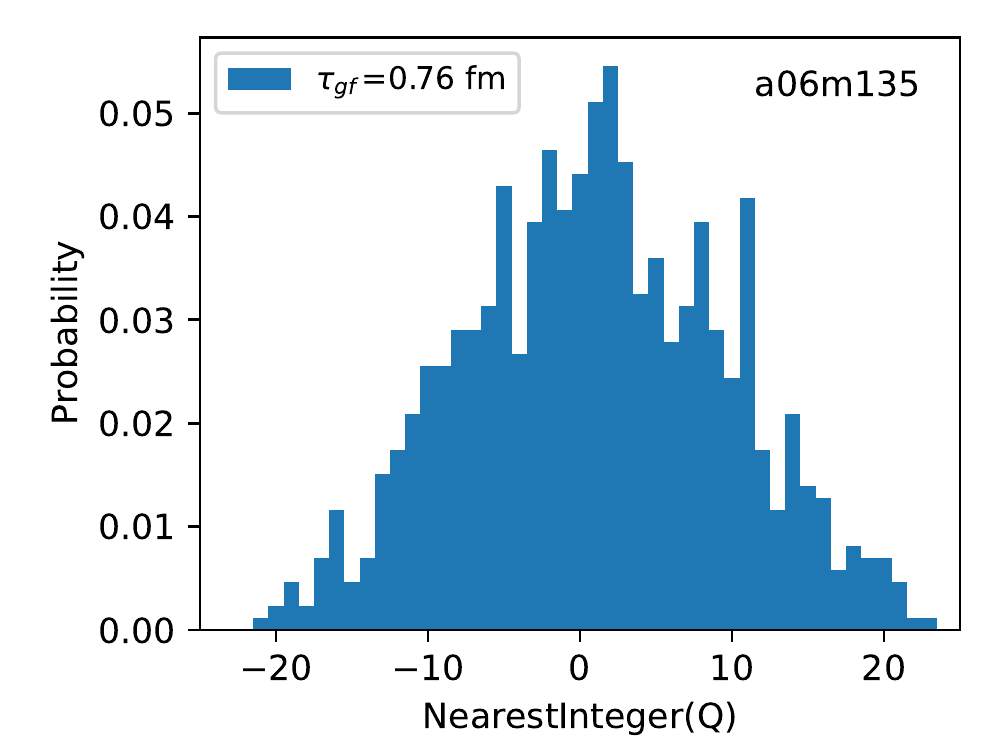}
\caption{The distribution of the nearest integer charge,
  $Q_{\rm int}$, associated with a given configuration at $\tau_{\rm
    gf} \approx 1.4$~fm ($a12$ ensembles) and $0.76$~fm ($a09$ and $a06$ ensembles), by which time the $Q_{\rm int}$ identified with a
  given configuration has stabilized.  
\label{fig:QintDist}}
  \end{minipage}
\hspace{0.05\linewidth}
  \begin{minipage}[t]{0.45\linewidth}
    \includegraphics[width=0.48\linewidth]{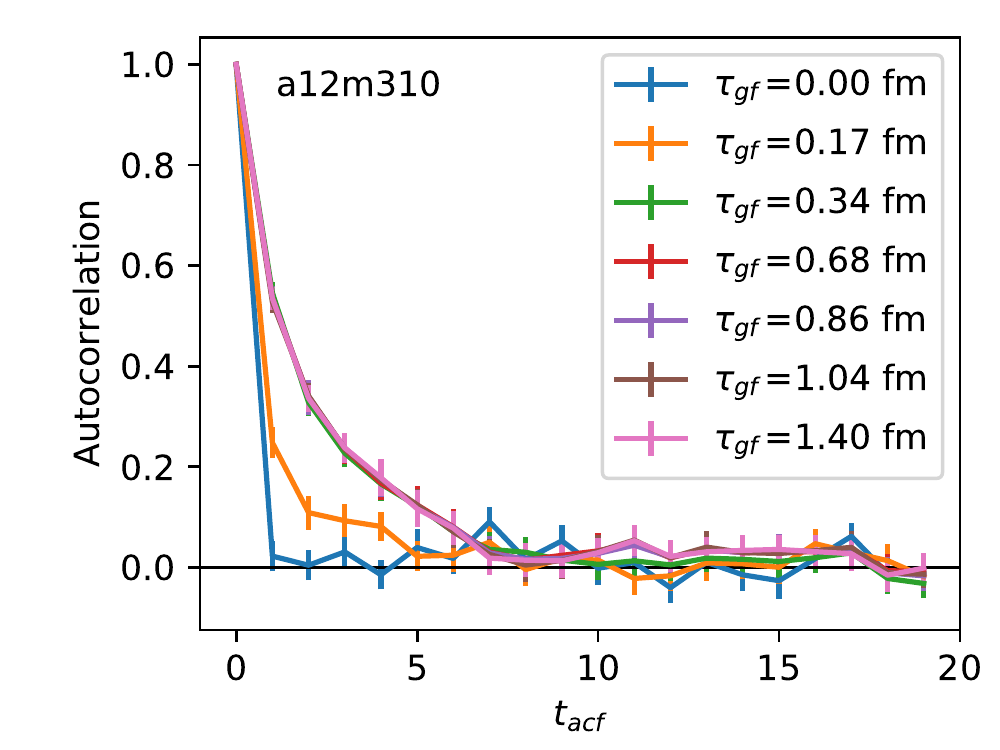}
    \includegraphics[width=0.48\linewidth]{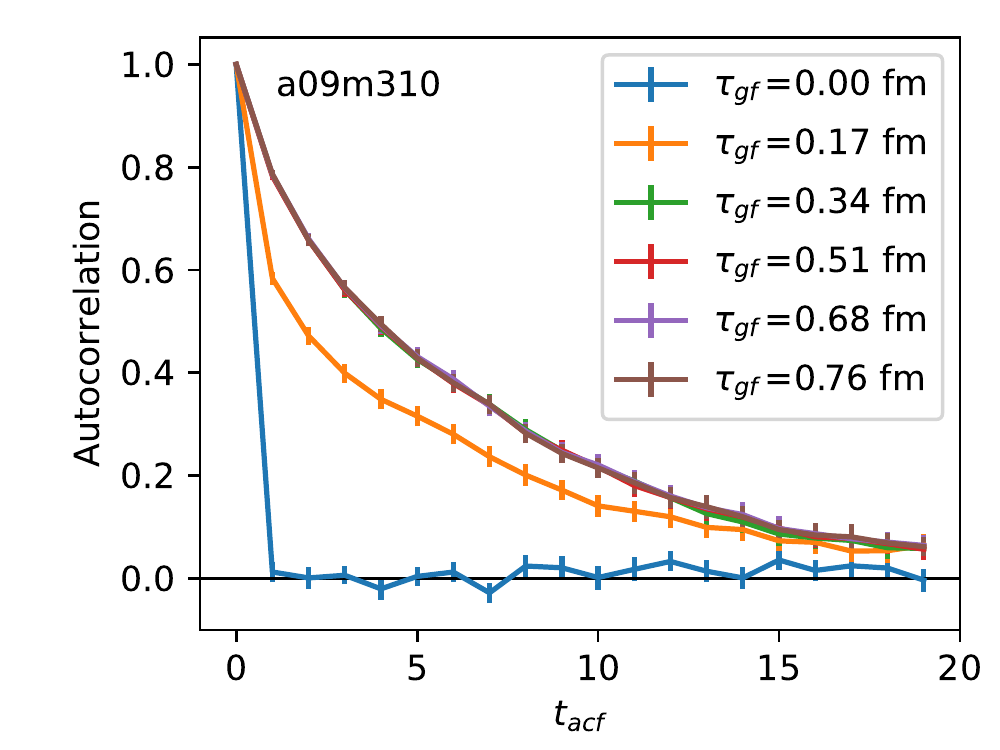}\\
    \includegraphics[width=0.48\linewidth]{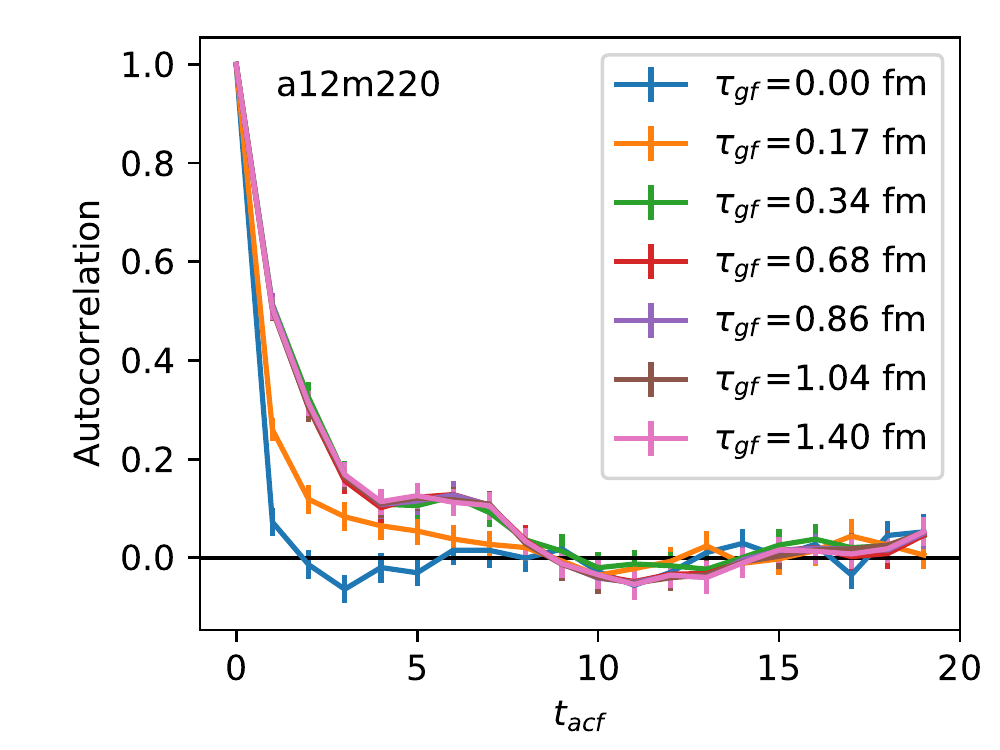}
    \includegraphics[width=0.48\linewidth]{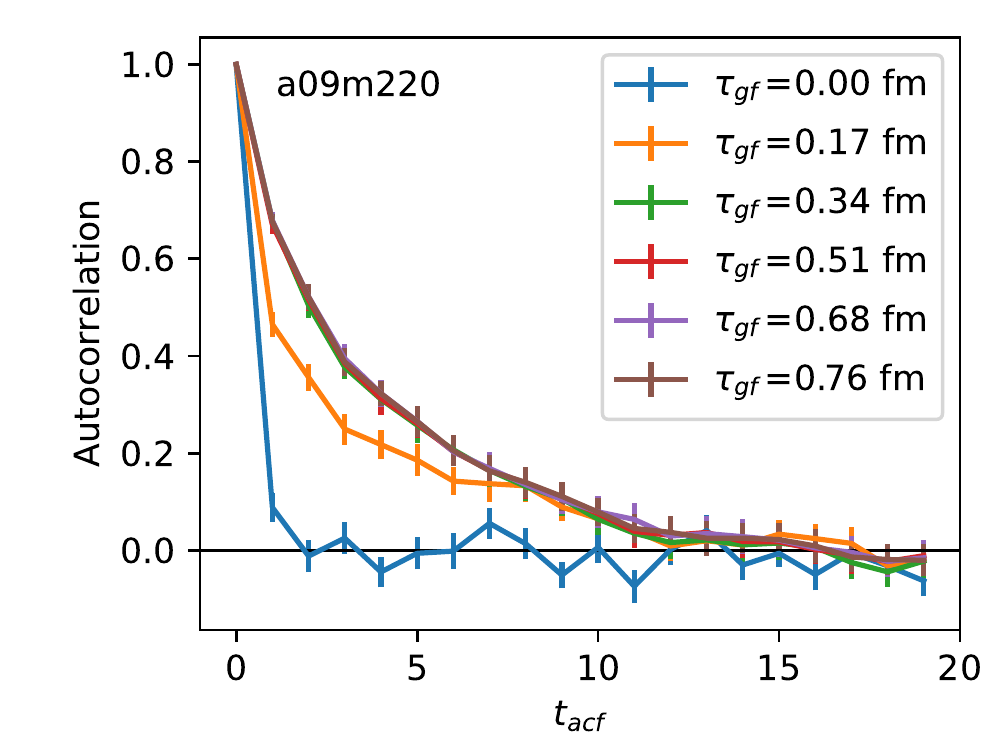}\\
    \includegraphics[width=0.48\linewidth]{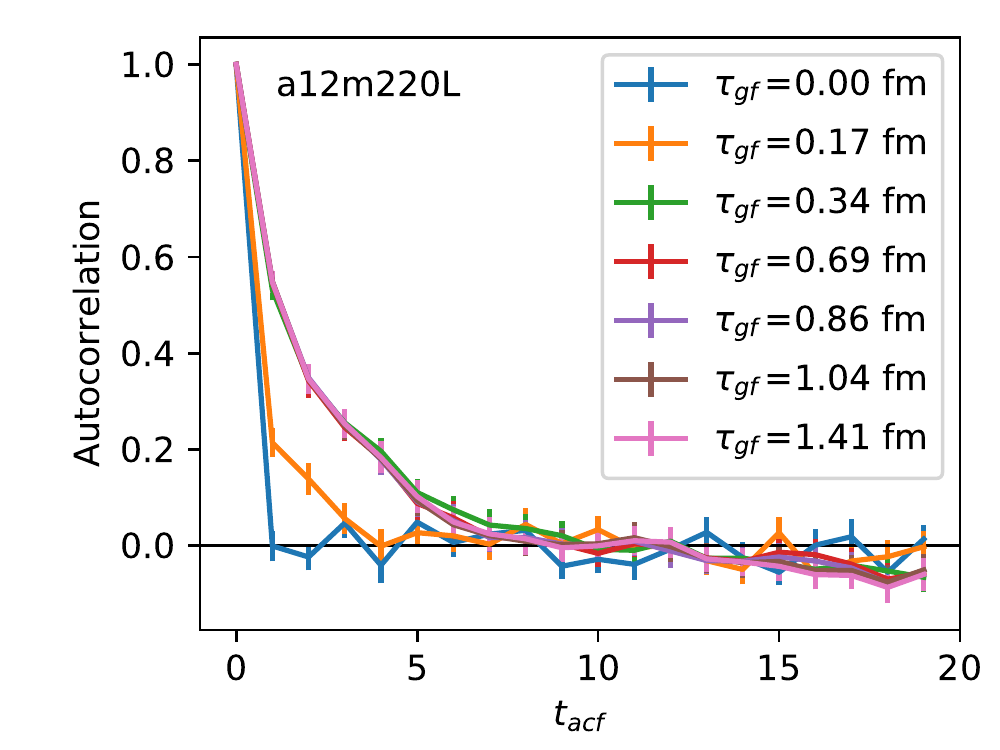}
    \includegraphics[width=0.48\linewidth]{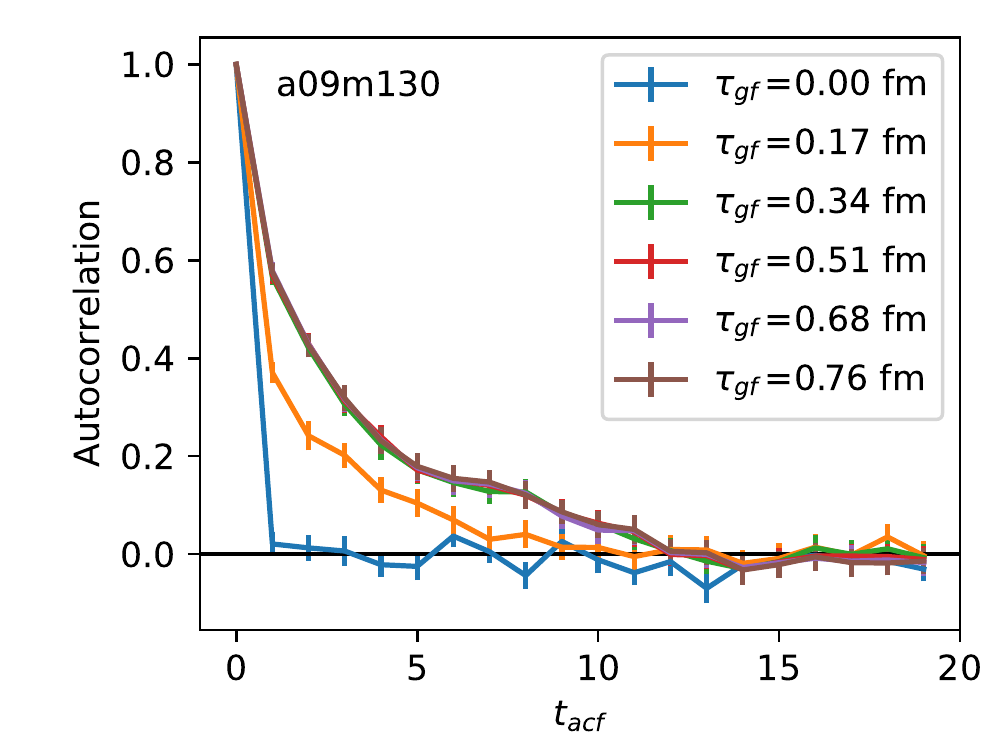}\\
    \includegraphics[width=0.48\linewidth]{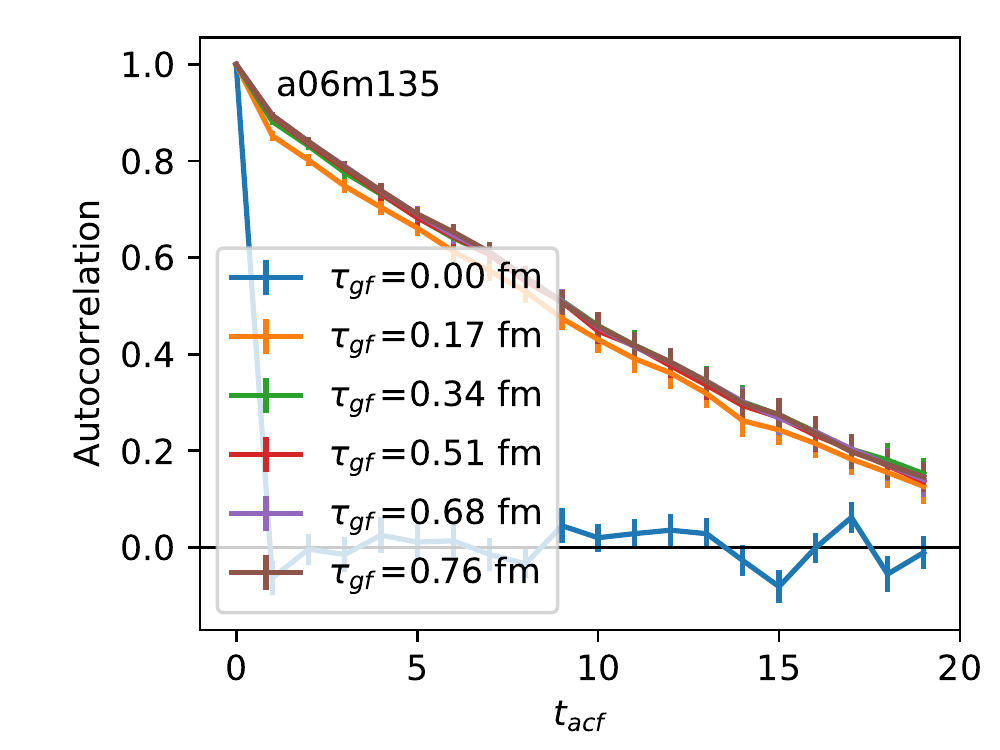}
\caption{The autocorrelation
  function for different values of the flow time.  The data show that the long   $\tau_{\rm gf}$ behavior stabilizes by
  $\tau_{\rm gf} = 0.34$~fm in all cases. 
\label{fig:Qacor}}
  \end{minipage}
\end{figure*}

%

\section{Topological charge under gradient flow}
\label{sec:charge}

We calculate the topological charge using the gradient flow
scheme to implement operator renormalization and to reduce lattice discretization
effects~\cite{Shindler:2015aqa,Luscher:2010iy}. The primary advantage of the scheme is that at finite flow
times\footnote{We use the notation \(\tau_{\rm gf}\equiv \sqrt{8t}\)
for the flow time, where \(t\) is the parameter in the flow equations
in Ref.~\cite{Luscher:2010iy}. We used the Runge-Kutta integrator
given in that reference for integrating the flow
equations, with a step size of $0.01$. Changing the step size to $0.002$ changed the results on topological susceptibility by less than 0.2\%.}, i.e.,
for \(\tau_{\rm gf}>0\), the flow time provides an ultraviolet cutoff, and the continuum limit, \(a\to0\), of all
operators built solely from gauge fields is finite.  Moreover, since
topological sectors arise dynamically as we take the continuum limit,
the gradient flowed topological charge takes on integer values, and no
renormalization is needed to convert it to a scheme that preserves
this property; in particular, correlators of the topological
charge are flow-time independent~\cite{Luscher:2010iy}.

These statements are, however, not true at finite lattice spacing and
volume. At small \(\tau_{\rm gf}\), we get \(O(a^2/\tau^2_{\rm gf})\)
artefacts. In Fig.~\ref{fig:Qdist}, we show the distribution of the
topological
charge $Q$ as a function of the flow time $\tau_{\rm gf}$ in physical
units. Its distribution 
has stabilized by $\tau_{\rm gf} = 0.24$~fm for the $a=0.12$~fm ensembles,  
and by $\tau_{\rm gf} = 0.17$~fm for the $a=0.09$ and $0.06$~fm ensembles.  
The large values of $Q$ that form the long tail of the distribution at
$\tau_{\rm gf} = 0$ are smoothed out, indicating that they are
lattice artifacts. 

In Fig.~\ref{fig:Qinteger}, we show the distribution of the
difference from the nearest integer. This distribution stabilizes more
slowly and it is only by $\tau_{\rm gf} = 1.31$~fm ($\tau_{\rm gf} = 0.76$~fm) on the 
$a\approx 0.12$~fm ($a\approx 0.09$ and $0.06$~fm) ensembles that the charges
are close to integers. The relevant
distribution important for the calculation of the nucleon correlation
functions is, however, likely to be the distribution of $Q$ shown in Fig.~\ref{fig:Qdist}. 
To explore this, we show in 
Fig.~\ref{fig:flowF3} the value of $F_3$ as a function of $\tau_{\rm gf}$ for the 
$a\approx 0.12$ and $0.09$~fm ensembles, and find that 
indeed the correlation functions, and thus $F_3$, do stabilize early but the 
 $\tau_{\rm gf}$ required for the coarser lattices is longer. Thus, to 
be conservative, the results presented below are obtained 
with flow times  $\tau_{\rm gf}(a06)=0.68$~fm,  $\tau_{\rm gf}(a09)=0.68$~fm and 
$\tau_{\rm gf}(a12)=0.86$~fm respectively. 

\begin{figure}[tbp]
    \includegraphics[width=0.96\linewidth]{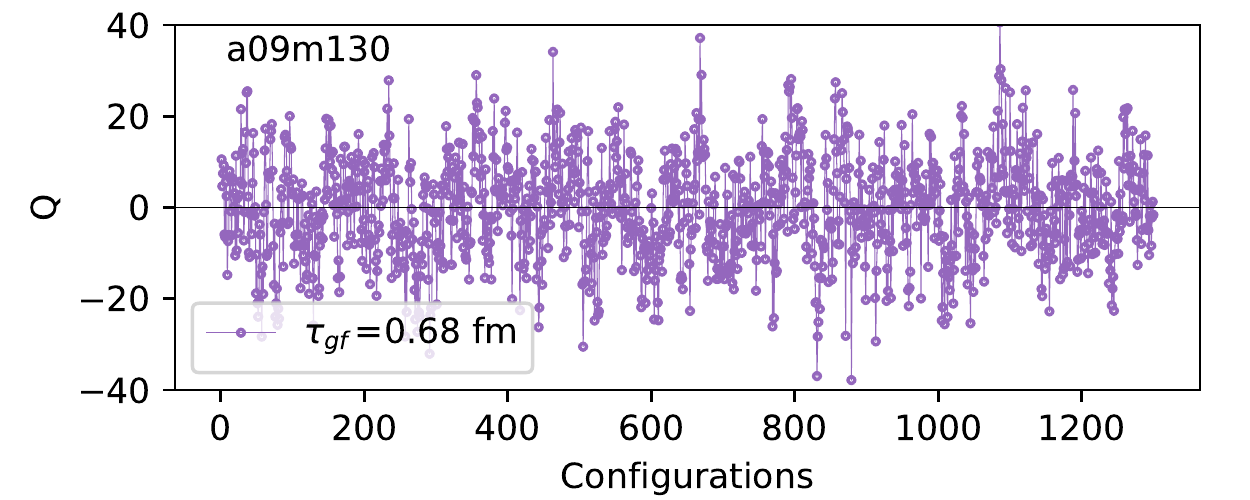}\\
    \includegraphics[width=0.96\linewidth]{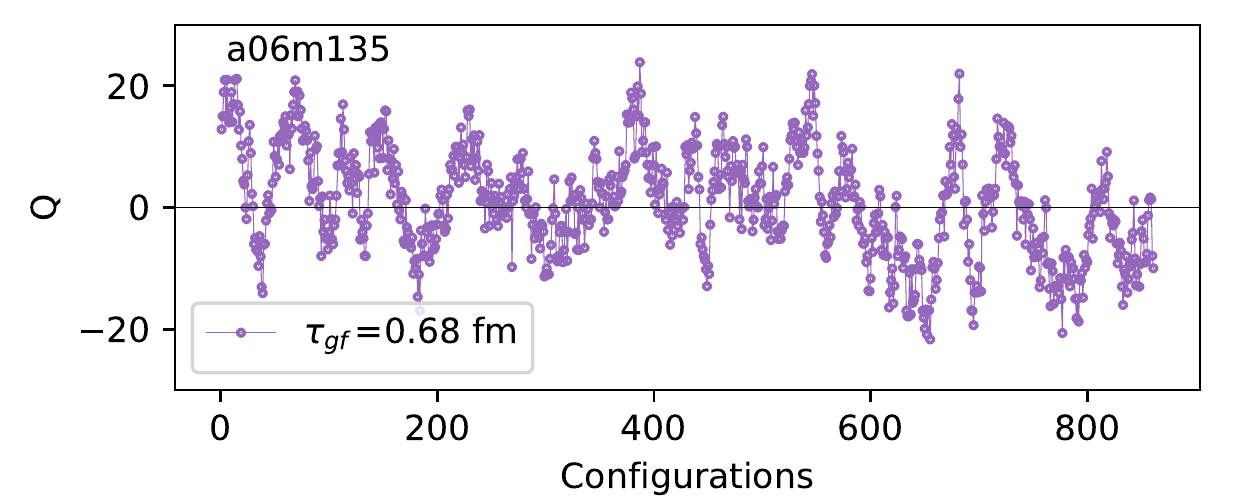}
\caption{The time history of $Q$ on the $a09m130$ (upper) and $a06m135$ (lower) ensembles at
  $\tau_{\rm gf} = 0.68$~fm.  No long time freezing of the topological
  charge is observed.
\label{fig:Q_a09m130}}
\end{figure}

\begin{figure*}[tbp]
\includegraphics[width=0.45\linewidth]{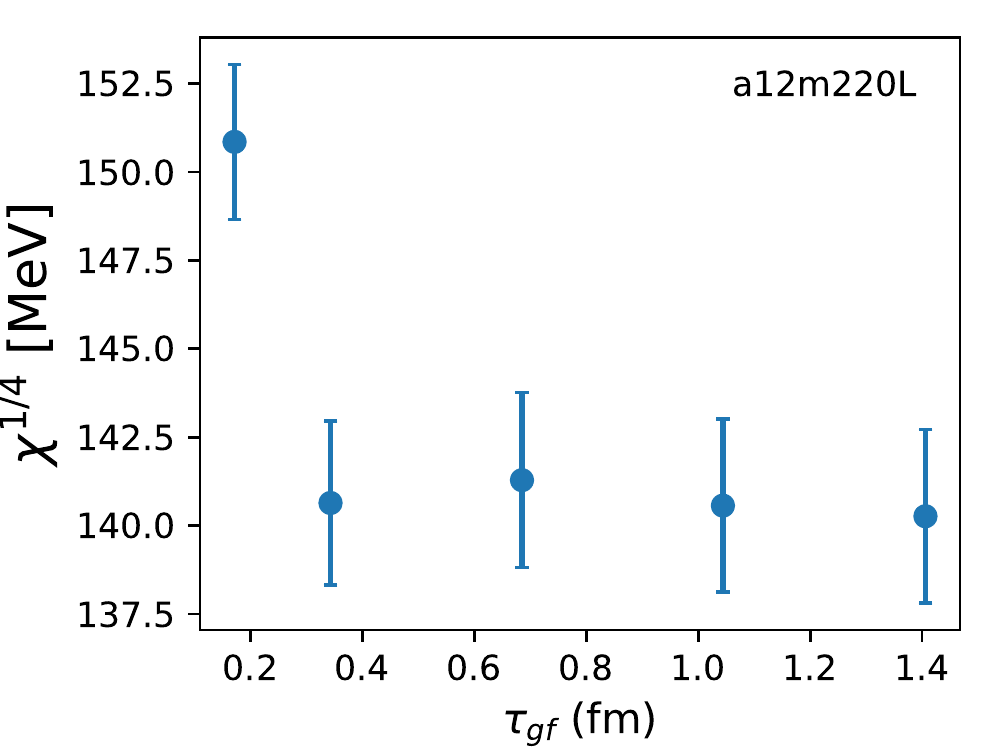}
\includegraphics[width=0.45\linewidth]{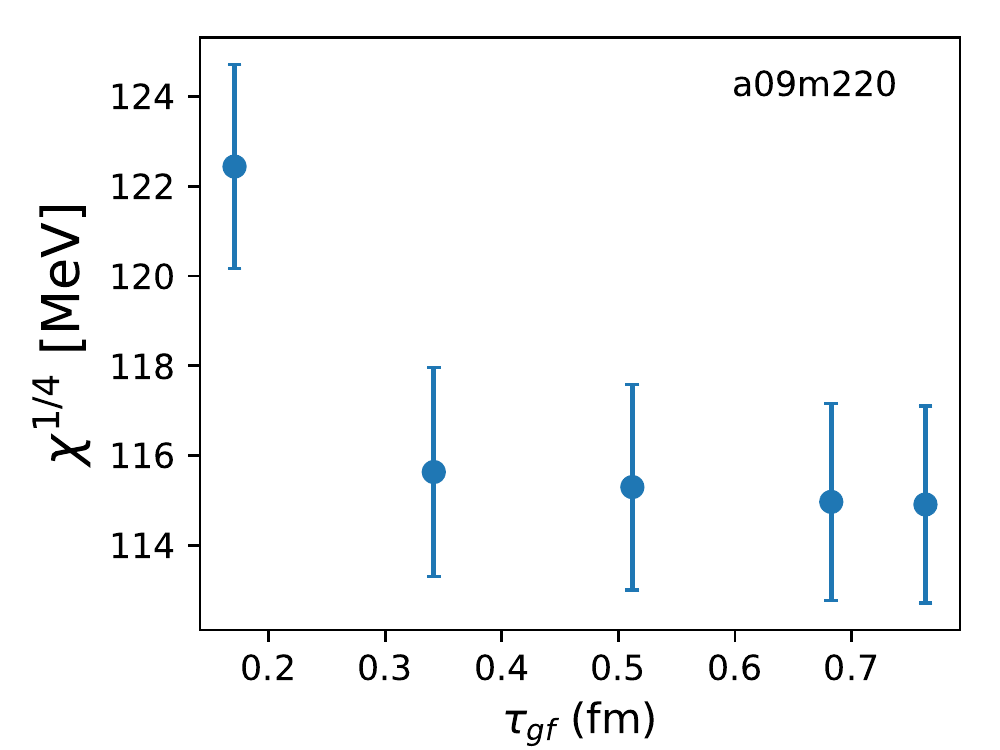}
\caption{Illustration of the flow-time dependence of the topological
susceptibility at small flow times showing that it is almost
independent of the flow time when the flow time is much larger than
the lattice spacing.}
\label{fig:almost_flat}
\end{figure*}
\begin{figure}[tbp]
\includegraphics[width=0.9\linewidth]{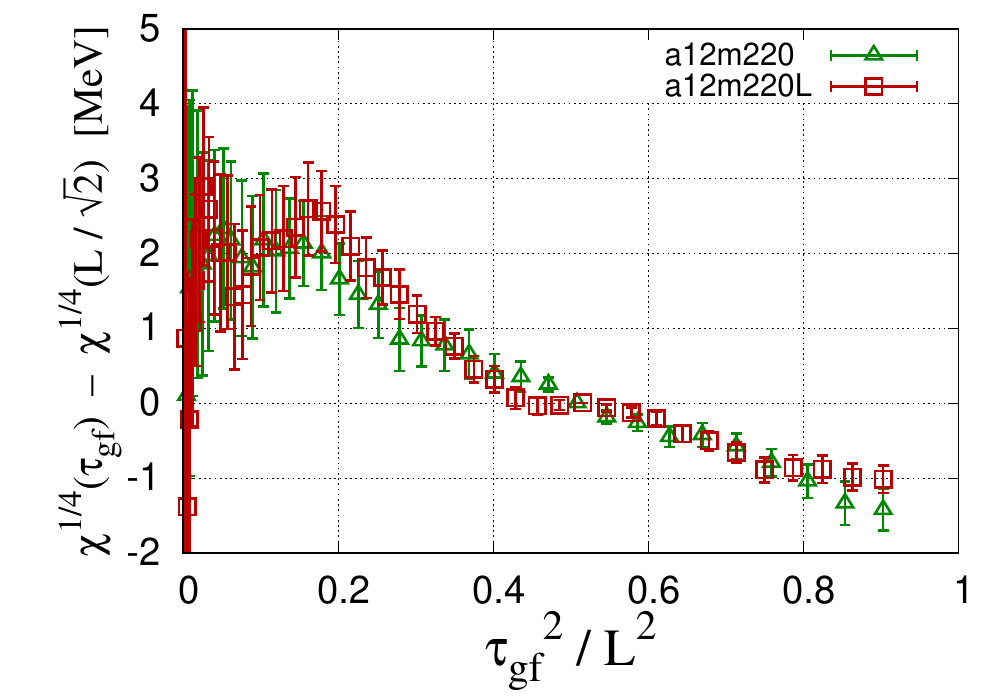}
\caption{Comparison of the flow-time dependence of the topological
susceptibility at large flow times on two ensembles differing only in
lattice volume, showing that the dependence is a finite size effect.}
\label{fig:finitevolume}
\end{figure}

In Fig.~\ref{fig:QintDist}, we show the distribution of the nearest integer, $Q_{\rm int}$, to the topological charge
at $\tau_{\rm gf} \approx 1.4$~fm ($\tau_{\rm gf} = 0.76$~fm) on the
$a\approx 0.12$~fm ($a\approx 0.09$ and $0.06$~fm) ensembles, by which time the
$Q_{\rm int}$ identified with a given configuration has stabilized.
This distribution is approximately symmetric about zero as expected
since $\langle Q \rangle = 0$, and no gaps are visible in the distribution.  In Fig.~\ref{fig:Qacor}, we show the
autocorrelation function of $Q$ versus the flow time. The data show no
significant change after $\tau_{\rm gf} \gtrsim 0.3$~fm, so we can
determine the autocorrelation from these data. We do not observe a
long time freeze in $Q$ in any of the ensembles analyzed as
illustrated using the $a09m130$ and $a06m135$ ensembles at flow time $\tau_{\rm gf} =
0.68$~fm in Fig.~\ref{fig:Q_a09m130}.  The autocorrelation is less than
about 10 configurations for all but the $a06m135$ ensemble. Based on this study, the bin size
used in the single elimination jackknife procedure is given in
Table~\ref{tab:ensembles}. \looseness-1

\section{Topological Susceptibility}
\label{sec:ChiQ}

The topological susceptibility $ \chi_Q$ is defined as 
\begin{equation}
\chi_Q = \int d^4x \langle Q(x)Q(0) \rangle \,.
\label{eq:Tsusceptibility}
\end{equation}
Its value in the pure gauge theory, \(\chi_Q^{\rm quenched}\), is related to the mass of the $\eta^\prime$ meson in a theory with \(N_f\) light flavors in the chiral limit via the axial anomaly, viz., the Witten-Veneziano relation~\cite{Witten:1979vv,Veneziano:1979ec}
\begin{equation}
M_{\eta^\prime}^2 \approx \frac{2N_f}{F_\pi^2} \chi_Q^{\rm quenched} \,,
\end{equation}
{where \(F_\pi\) is the pion decay constant in the convention where its physical value is about 93 MeV. Following Ref.~\cite{Evans:1996kf}, we can include the effects of the quark masses. Including  \(SU(3)\) breaking 
at leading order in $\chi$PT but neglecting the heavier quarks gives 
\begin{eqnarray}
\chi_Q^{\rm quenched} &\approx& \frac{F_\pi^2(M_{\eta^\prime}^2-M_{\eta}^2)}6
     \left(\sqrt{1 - \frac{32\,\delta_{K\pi}^2}{9}}
           + \frac{2\,\delta_{K\pi}}3\right),\nonumber\\
\chi_Q^{\rm quenched}&\approx& \frac{F_\pi^2(M^2_{\eta^\prime} - M^2_\eta)}6\left(1 + 2 \frac{M_\eta^2 - M_K^2}{M^2_{\eta^\prime}-M^2_\eta}\right)           ,
\label{eq:WVmechanism}
\end{eqnarray}
where \(\delta_{K\pi}\equiv(M_K^2-M_\pi^2)/(M_{\eta^\prime}^2-M_\eta^2)\) is an SU(3) breaking ratio.
The two expressions, which can be derived independently, 
give \(\chi_Q^{\rm quenched} \approx (172\ {\rm MeV})^4\) and \((179\ {\rm MeV})^4\) respectively, thus quantifying the accuracy of the expansion.

With dynamical fermions, however, the susceptibility should vanish in the chiral limit. For $SU(N_f)$ flavor group with finite but degenerate quark masses, it 
should behave as~\cite{Crewther:1977ce,DiVecchia:1980yfw,Leutwyler:1992yt}:
\begin{equation}
\frac1{\chi_Q} \approx \frac1{\chi_Q^{\rm quenched}} + \frac{2N_f}{M_\pi^2F_\pi^2}\,.
\end{equation}
For \(N_f=2\) light flavors and the strange quark, but 
neglecting the heavier quarks that give negligible corrections,
leading order chiral perturbation theory ($\chi$PT) modifies this to 
\begin{equation}
\frac1{\chi_Q} \approx \frac1{\chi_Q^{\rm quenched}}
+ \frac{4}{M_\pi^2F_\pi^2}\left (1 - \frac{M_\pi^2}{3 M_\eta^2}\right)^{-1}\,.
\label{eq:WVchiral}
\end{equation}
}

We calculate $\chi_Q$ on the 2+1+1 flavor HISQ ensembles, which are
$O(a)$ improved.  The results are given in
Table~\ref{tab:ensembles}. In addition to the seven ensembles used to
calculate $F_3$, we include data from the $a06m310$ and $a06m220$
ensembles. We remind the reader that the MILC collaboration has
previously highlighted the issue of frozen topology on these
ensembles~\cite{Bernard:2017npd}, which is why we do not use them in
the calculation of $F_3$.

As discussed in Section~\ref{sec:charge}, the topological
susceptibility at finite flow time needs no renormalization, and
should be independent of flow time up to \(O(a^2/\tau_{\rm
gf}^2)\) effects. As shown in Fig.~\ref{fig:almost_flat}, this is true
up to a small, almost linear, downward drift with increasing flow
time. In Fig.~\ref{fig:finitevolume}, we compare the results
on \(a12m220\) and \(a12m220L\) ensembles, and show that this is
a \(\tau^2_{\rm gf}/L^2\) effect, where \(L\) is the lattice
size.\footnote{For asymmetric lattices like ours, we expect the
smaller spatial extent to dominate the finite volume effect.} At the
flow times and volumes we use in the calculation, this is a small
effect and therefore neglected.

\begin{figure*}[tbp]
  \subfigure{
    \includegraphics[width=0.47\linewidth]{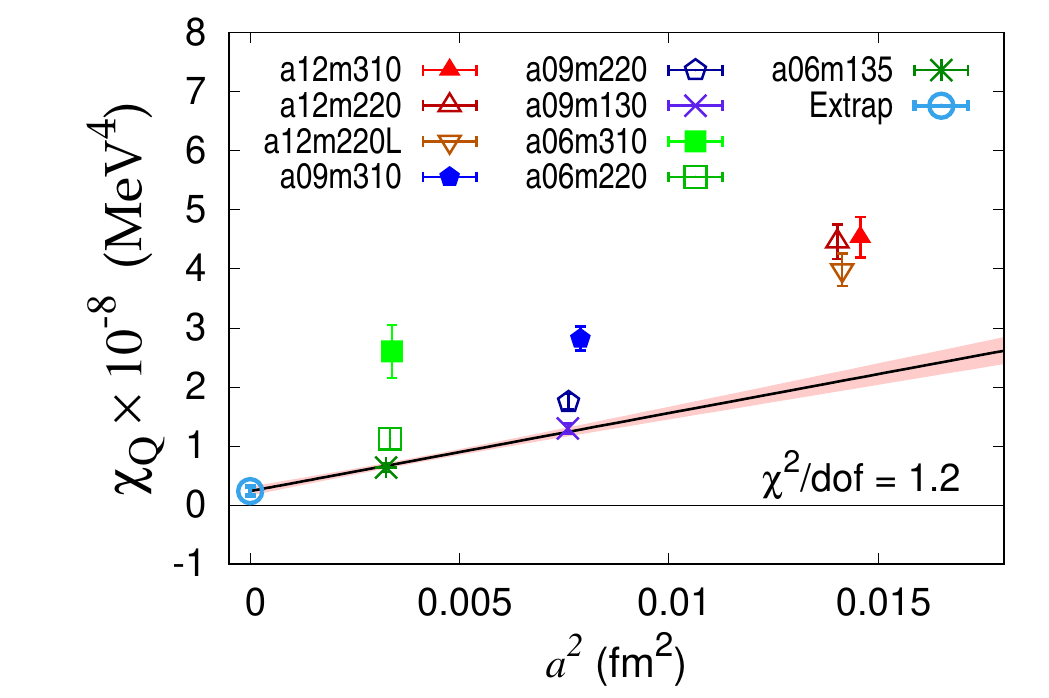}\qquad 
    \includegraphics[width=0.47\linewidth]{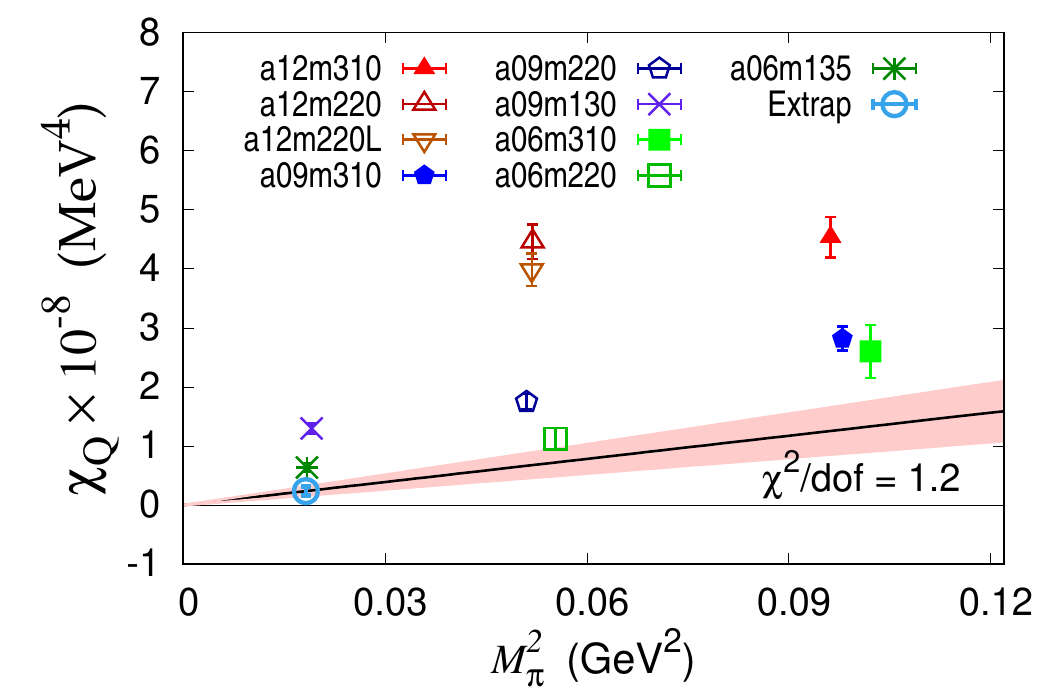}
  }
\caption{Fits to the data for the topological susceptibility, $\chi_Q$, using the ansatz 
given in Eq.~\protect\eqref{eq:ChiCC}. }
\label{fig:ChiCC}
\end{figure*}

To obtain $\chi_Q$ at $M_\pi = 135$~MeV and $a = 0$, we 
use the fit ansatz
\begin{equation}
\chi_Q (a, M_\pi) = c_1 a^2 + c_2 M_\pi^2 + c_3 a^2 M_\pi^2  \, ,
\label{eq:ChiCC}
\end{equation}
which assumes $\chi_Q$ is zero in the chiral-continuum limit.  
We do not find a viable $\chi^2$/dof on including
all nine data points.  Reasonable fits are found on neglecting (i) all
three $a \approx 0.12$~fm points and (ii) all three $a \approx
0.12$~fm and the $a06m310$ point.  These two fits give $\chi_Q =
[70(6)~{\rm MeV}]^4$ and $\chi_Q = [63(9)~{\rm MeV}]^4$, respectively,
at $M_\pi = 135$~MeV. We take the average $\chi_Q = [66(9)(4)~{\rm
MeV}]^4$ as our best estimate, the larger of the two errors and an
additional systematic uncertainty, which is half the difference. These
results are in good agreement with the expected value,
\((79~{\rm MeV})^4\), {obtained using the physical meson masses and decay constants in Eqs.~\eqref{eq:WVmechanism} and~\eqref{eq:WVchiral}}. The data and the fit case (i) are shown in Fig.~\ref{fig:ChiCC}.

\section{Calculation of the \texorpdfstring{\CPV}{CPV} phase
  \texorpdfstring{$\alpha$}{\textalpha}}
\label{sec:alpha}

In a field theory in which parity is not conserved, the definition of parity of
a composite state, {\it e.g.,} the neutron state, needs care~\cite{Pospelov:1999ha,Pospelov:2000bw,Shintani:2005xg}.  To explain
this, we start with the most general spectral decomposition of the time-ordered
2-point nucleon correlator 
\begin{equation}
\langle \Omega | {\cal T} {N({\bm p},\tau) \overline N({\bm p},0)} | \Omega
\rangle = \sum_{i,{\bm s}} e^{-E_i \tau }\; {\cal A^\ast}_i {\cal A}_i \; {\cal M}^{\bm s}_i \,, 
\label{eq:N2pt}
\end{equation}%
where ${\cal A}_i$ is the amplitude for creating state $i$, \(E_i\) is its energy,
the Euclidean time $\tau$ is the separation between the source and the
sink, and, for notational convenience, we are assuming a discrete spectrum. A common choice on the lattice of the neutron interpolating operator $N$ is
\begin{equation}
    N \equiv \epsilon^{abc} [{d}^{a{\rm T}} C \gamma_5 \frac{1+\gamma_4}2 u^b\;] d^c\,,
\end{equation}
where $C= \gamma_2 \gamma_4$ (the sign is conventional and does not affect the nucleon correlators we study;
see Appendix~\ref{sec:appendix0} for details of our convention) is the charge conjugation matrix, $a$, $b$,
$c$ are the color indices and $u$, $d$ are the quark flavors. 
The $4 \times 4$ spinor matrix ${\cal M}^{\bm s}_i$ in Eq.~\eqref{eq:N2pt} depends on the state
    and the momentum ${\bm p}$. Its most general form consistent with Lorentz covariance is\footnote{Up to a possible extra factor of \(\gamma_5\), which, however, is prohibited by PT symmetry in our calculations.}
\begin{eqnarray}
\sum_{\bm s}{\cal M}^{\bm s}_i &=& e^{i \alpha_i \gamma_5 } \frac{({
{ - i} 
\slashed p_i} +
  M_i)}{2E_i^p} e^{i \alpha_i^\ast \gamma_5 }\\&\equiv& e^{i \alpha_i \gamma_5 } \sum_s
{u^i_N}({\bm p},{\bm s}) {\overline u^i_N}({\bm p},{\bm s})e^{i \alpha^\ast_i \gamma_5 } 
\,,
\end{eqnarray}%
where \(p^4_i\equiv iE_i\). It is clear that because of the presence of the phases $\alpha_i$, the parity operator that transforms the spinor associated with the \(i^{\rm th}\) asymptotic state is \({\cal P}_{\alpha_i} \equiv e^{i\alpha_i\gamma_5}{\cal P}e^{-i\alpha_i\gamma_5}\), where \({\cal P}\equiv \eta \gamma_4\) is the usual parity operator for a particle with intrinsic parity \(\eta\). The phases
$\alpha_i$ depend on the realization of discrete symmetries: If the interpolating
field is chosen such that \({\cal P}\) implements parity in the free theory, $\mathop{\rm
  Im}\alpha_i = 0$ for a PT symmetric theory, $\mathop{\rm Re}\alpha_i = 0$ 
for the CP symmetric theory, $\alpha_i = 0$ for a P symmetric theory. For
our case of only \CPV, all $\alpha_i$ are, therefore, real,
which will be implicit except in Appendix~\ref{sec:appendix1}. It is important to note that the value of $\alpha_i$ depends on
the interpolating operator $N$, the state, and the source of 
\CPV. Its value for the ground state can be extracted from
the large $\tau$ behavior of the imaginary part of the nucleon 2-point
function. Consider\looseness-1
 \begin{eqnarray}
r_\alpha (\tau) &\equiv&
\frac{\Im C_{\textrm{2pt}}^{{P}}(\tau)}
         {\Re C_{\textrm{2pt}}(\tau)}\\
    & \equiv &
      \frac{\mathop{\rm Im}\Tr \left[ {\gamma_5}  \frac12 (1+\gamma_4) 
            \langle N(\tau) \overline N(0) \rangle \right]}
          {\mathop{\rm Re}\Tr \left[ \frac12 (1+\gamma_4) 
           \langle N(\tau) \overline{N}(0) \rangle \right]}\\
&=&
\frac{\sum_i M_i \sin (2 \alpha_i) \ |{\cal A}_i|^2 / (2 E_i) \ e^{- E_i \tau} }{ 
\sum_i (E_i + M_i \cos (2 \alpha_i) ) \ |{\cal A}_i|^2 / (2 E_i) \ e^{- E_i \tau}}\,.\nonumber\\
\end{eqnarray}
Keeping only the first two states one gets
\begin{eqnarray}
r_\alpha (\tau) &\approx& 
\frac{M_0 \sin (2 \alpha_0)}{E_0 + M_0 \cos (2 \alpha_0)}
\\
&\times &
\frac{1 + \frac{M_1 E_0}{M_0 E_1} \frac{\sin (2\alpha_1)}{\sin (2 \alpha_0)} | \tilde{\cal A}_1|^2 \
e^{-(E_1 - E_0)\tau} }{1 + \frac{(E_1 + M_1 \cos (2 \alpha_1)) E_0}{(E_0 + M_0 \cos (2 \alpha_0)) E_1} \ | \tilde{\cal A}_1|^2 
\ e^{-(E_1 - E_0)\tau}}\,,
\nonumber
\end{eqnarray}
where $\tilde{\cal A}_i = {\cal A}_i/{\cal A}_0$. 
At zero three-momentum ($E_i = M_i$) the above expression simplifies to 
\begin{equation}
    r_\alpha (\tau) \approx
    \tan \alpha_0 \times \frac{1 + \frac{\sin (2 \alpha_1)}{\sin (2 \alpha_0)} |\tilde{\cal A}_1|^2 \ e^{- (M_1 - M_0)\tau}}{1 + \frac{\cos^2 (\alpha_1)}{\cos^2 ( \alpha_0)} |\tilde{\cal A}_1|^2 \ e^{- (M_1 - M_0)\tau}}\,.
\label{eq:ralpha}
\end{equation}

\begin{figure*}[tbp]
  \subfigure{
    \includegraphics[width=0.32\linewidth]{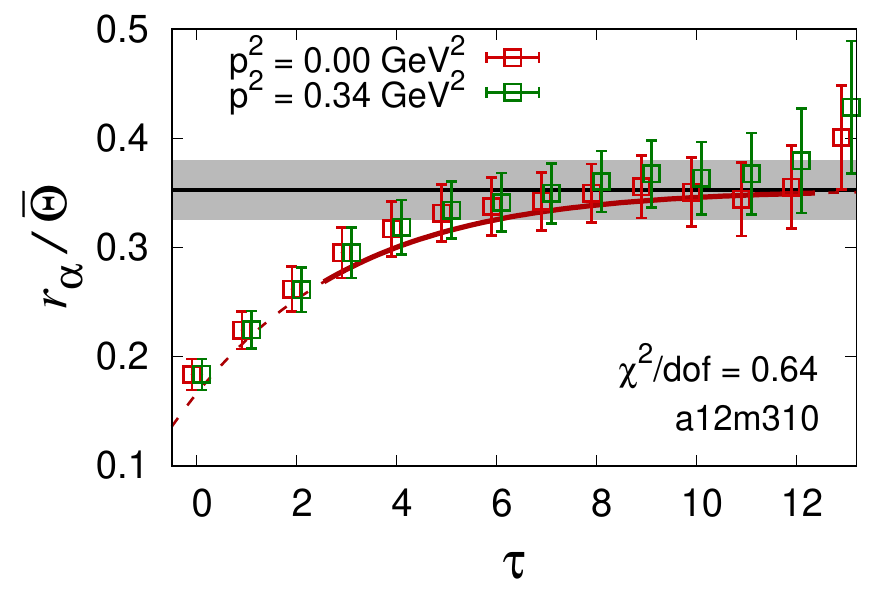}
    \includegraphics[width=0.32\linewidth]{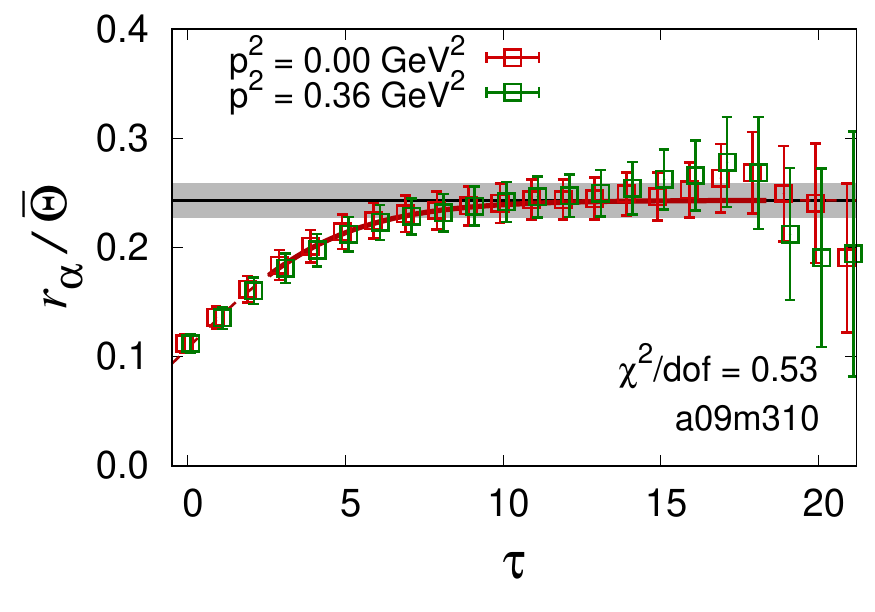}
  }
  \subfigure{
    \includegraphics[width=0.32\linewidth]{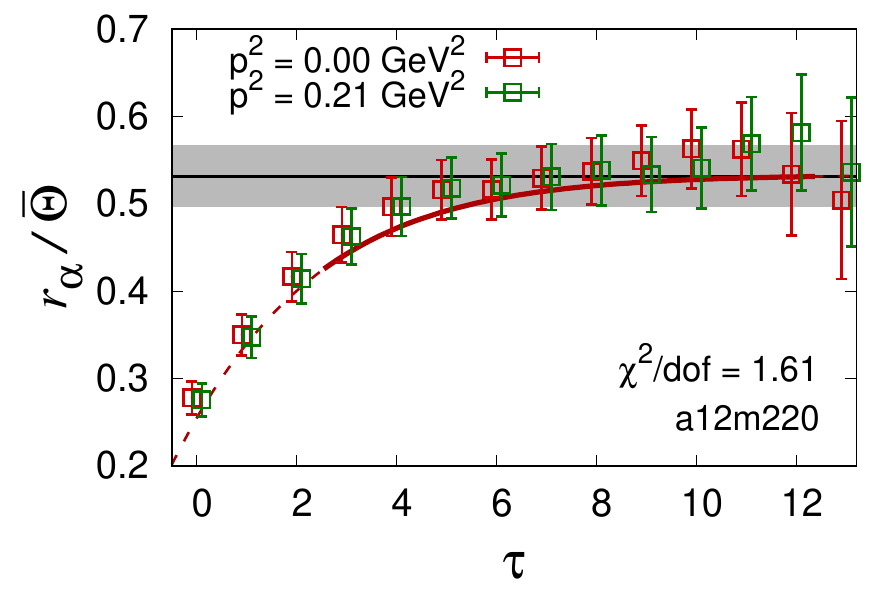}
    \includegraphics[width=0.32\linewidth]{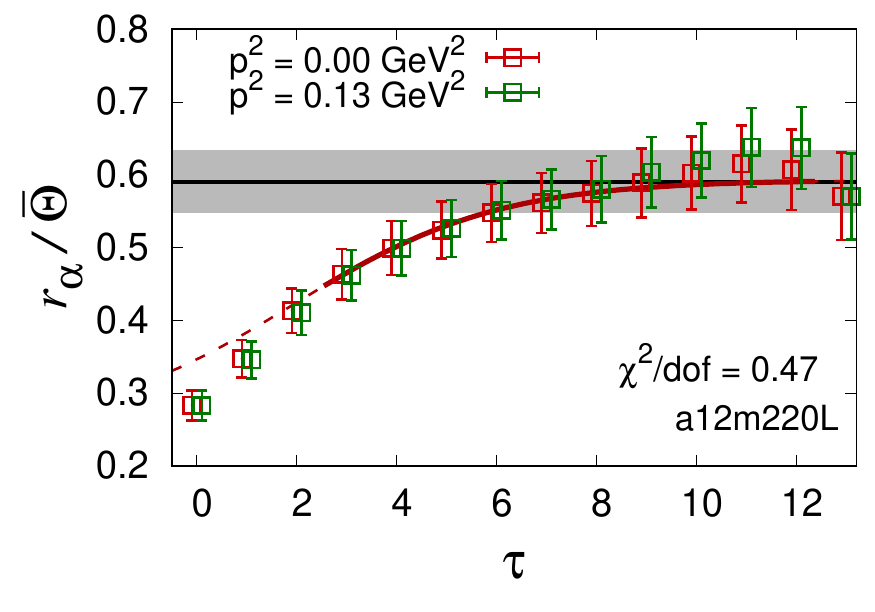}
    \includegraphics[width=0.32\linewidth]{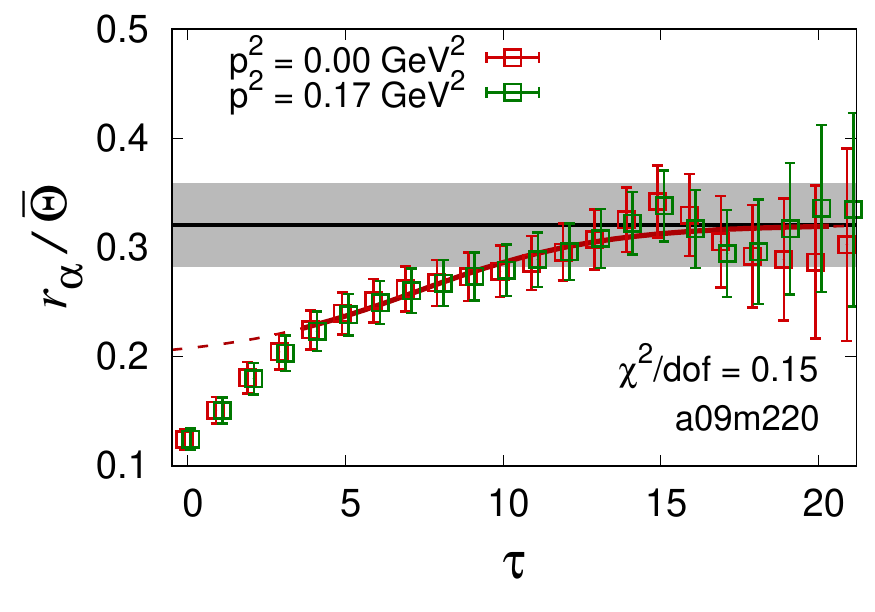}
  }
  \subfigure{
    \includegraphics[width=0.32\linewidth]{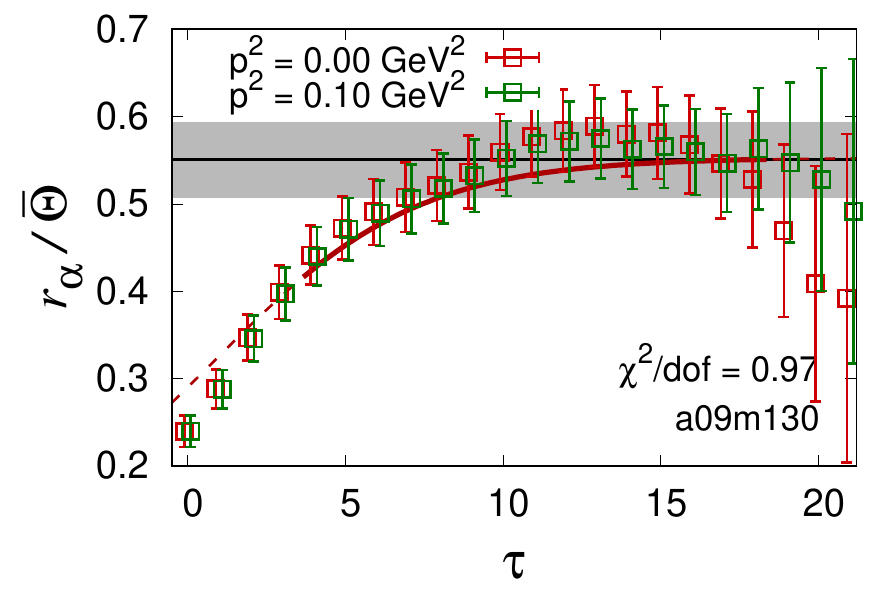}
    \includegraphics[width=0.32\linewidth]{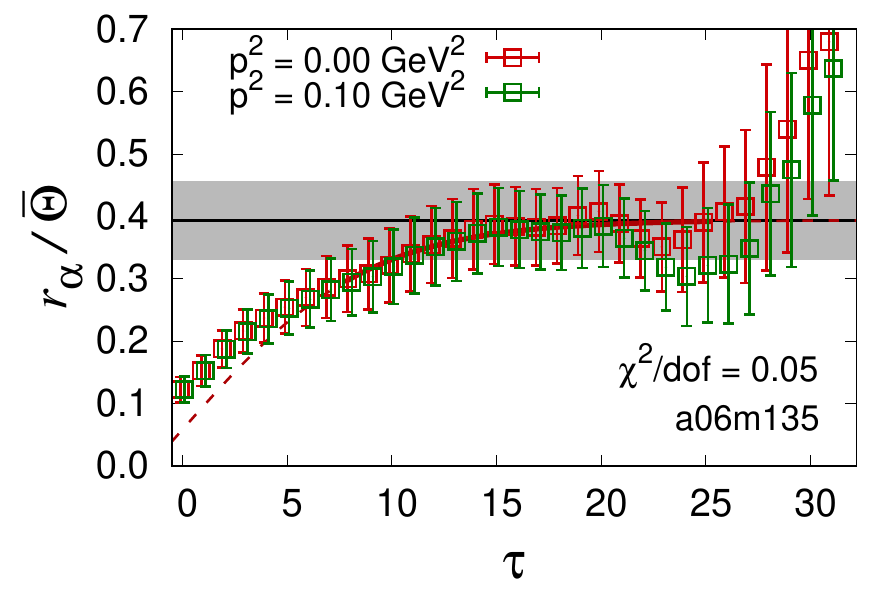}
  }
\caption{The extraction of the phase $\alpha_N/{\overline \Theta}$ with ${\overline \Theta}=0.2$
for the ground state nucleon on the seven ensembles {from the asymptotic value of \(r_\alpha\) defined
in Eq.~\eqref{eq:ralpha}}. It is a Lorentz scalar and  
independent of the momentum as confirmed by the lattice data. The $\chi^2$/dof values presented are from fully correlated fits, except for the case of the $a09m220$ and $a06m135$ ensembles on which we use uncorrelated fits to avoid instabilities.
\label{fig:alpha}}
\end{figure*}

The data for $r_\alpha$ versus $\tau$ are shown in Fig.~\ref{fig:alpha} for all seven ensembles. 
The $\alpha_0$ for the ground state obtained from the two-state fit agrees
with the plateau at large $\tau$, where the lowest state dominates, and
is independent of momentum.

\section{Three-point Functions in the presence of the phase \texorpdfstring{$\alpha$}{\textalpha}}
\label{sec:3pt}


In the presence of the phase $\alpha_N\equiv\alpha_0$ corresponding to the ground-state nucleon~\cite{Abramczyk:2017oxr}, 
the most straightforward way to extract the matrix element of the
electromagnetic current $J^{\rm EM}_\mu$ within the neutron ground state in the
presence of \CPV\ is to calculate the correlation function
\begin{eqnarray}
e^{-i \alpha_N \gamma_5 } \left.\langle \Omega | {N({\bm p'},\tau) J^{\rm EM}_\mu({\bm q},t) \overline N({\bm p},0)} | \Omega \rangle\right|_{\not{\rm CP}}  e^{-i \alpha_N \gamma_5 } \span\omit\span\nonumber\\
 \vrule width0pt\qquad\qquad\qquad           &\propto&( - i  \slashed{p}' + M_N)  O^\mu    ( - i \slashed{p} + M_N)\,,
\label{eq:V1tt}
\end{eqnarray}
where \(p'\equiv p+q\) and 
\begin{eqnarray}
O^\mu &\equiv&  \gamma^\mu   F_1  + \frac{1}{2 M_N} \sigma^{\mu \nu} q^\nu \left( F_2  - i F_3 \gamma_5 \right)  
\nonumber \\
&&\qquad\qquad{}+ 
\frac{F_A}{M_N^2}  (\slashed q   q^\mu  - q^2 \gamma^\mu) \gamma_5\,.
\label{eq:3ptA}
\end{eqnarray}
Here, the current $J^{\rm EM}_\mu$ is inserted at times $t$ between the neutron
source and sink operators located at time $0$ and $\tau$, and a sum over the spin labels is implicit. We also assume that \(t\) and \(\tau\) are large enough that only the ground state dominates the correlation function. This form results from the 
realization that 
$\gamma_4$ remains the parity operator for the ground state nucleon when working with the interpolating field defined to be 
\(e^{-i \alpha_N \gamma_5 } N\) instead of \(N\) in  all correlation functions.

This approach, however,
requires, evaluating the full $ 4 \times 4$ matrix of 3-point
correlation functions.  In our calculation, we have implemented the
spin projection using 
\begin{equation}
{\cal P}_{3pt} \equiv \frac12 (1+\gamma_4) (1+i\gamma_5\gamma_3) \,,
\label{eq:projection}
\end{equation}
so the contribution of a nonzero $\alpha_N$ has to be incorporated at
the time of the decomposition of the matrix element into the form
factors. 
As discussed in Appendix~\ref{sec:appendix1}, 
by taking a suitable ratio of 3- and 2-point functions, 
one can isolate the four-vector ${\cal V}_\mu$ encoding the nucleon ground state contribution 
to the matrix element of the electromagnetic current, 
\begin{eqnarray}
\label{eq:V1t}
{\cal V}^\mu &\equiv& 
            \frac{1}{4} \mathop{\rm Tr} \left[ e^{i \alpha_N \gamma_5}  {\cal P}_{3pt}   e^{i \alpha_N \gamma_5}\right.\nonumber\\&&\qquad\left.    ( - i  \slashed{p}' + M_N)  O^\mu    ( - i \slashed{p} + M_N)  \right]~, 
\end{eqnarray}
%
%
where $O^\mu$ is given in Eq.~(\ref{eq:3ptA}). 
The full expressions 
for ${\cal V}_{1,2,3,4}$, along with a general strategy for
extracting $F_3$, from the four coupled complex equations is given in Appendix~\ref{sec:appendix1}. 
 
%
%
%

To extract $F_3$, the \CPV\ part of the three-point functions, a
very significant simplification of the analysis and improvement in the signal 
is achieved by subtracting the ${\overline \Theta}=0$ contribution from
each component of the  current in Eq.~\eqref{eq:V1t} 
before making the excited state fits and decomposing
the resulting ground state matrix element in terms of form factors. 
This is implemented by analyzing the ground state contribution in terms of the  
combination $\bar {\cal V}_\mu = {\cal V}_\mu ({\overline \Theta}) - {\cal V}_\mu (0)$.
Working to first order in ${\overline \Theta}$, and recalling that
$s_{\alpha_N} \equiv \sin\alpha_N \cos\alpha_N \sim \alpha_N \sim O({\overline \Theta})$, and $F_3 \sim O({\overline \Theta})$,
the expressions for the ground state contributions of the three-point functions $\bar {\cal V}_{1,2,3,4}$ in
terms of form factors simplify to
\begin{subequations}
\label{eq:Vinear}
\begin{eqnarray}
\bar {\cal V}_1  &=& - \frac{1}{2} q_1  q_3 G_3  \,, \\
\label{eq:Vbar1}
\bar {\cal V}_2  &=& - \frac{1}{2} q_2  q_3 G_3  \,, \\
\label{eq:Vbar2}
\bar {\cal V}_3  &=& \frac{1}{2} \Big( 2 M_N (E_N - M_N)  \, s_{\alpha_N}  G_1 \ - \ q_3^2 G_3 \Big) \, , \\
\label{eq:Vbar3}
\bar {\cal V}_4  &=& \frac{i}{2} \Big( q_3 (E_N + M_N) G_3 - 2 q_3 M_N s_{\alpha_N} G_1 \Big) \, \nonumber\\
                 &=& {i q_3 M_N} \Big( \frac{(E_N + M_N)}{2M_N} F_3 - s_{\alpha_N} G_E \Big) \,, 
\label{eq:Vbar4}
\end{eqnarray}
\end{subequations}
where $G_1 = F_1 + F_2 $ and $G_3 = F_3 + s_{\alpha_N} F_2$.  
We solve the above system for $G_1$ and $G_3$.
At $q^2=0$ there is a further simplification because 
$G_1 (0) = Q_N + F_2(0)$ where $Q_N$ is the nucleon charge. With this, we get
\begin{equation}
F_3(0) = G_3(0) - s_{\alpha_N} \left( G_1 (0) - Q_N \right) \,.
\label{eq:F3at0}
\end{equation}
Though the nucleon anomalous magnetic moment \(G_1(0) - Q_N = F_2(0) \equiv \kappa_N\) has been measured very precisely, the largest contribution to
\(G_3\) comes from \(s_{\alpha_N}F_2\), and the statistical
error is much smaller when extrapolating \(G_3(q^2)-s_{\alpha_N} (G_1(q^2) - Q_N)\), rather than 
extrapolating only \(G_3(q^2)\) and then combining it with 
\(s_{\alpha_N} \mu_N\) to get the right hand side of Eq.~\eqref{eq:F3at0}.  
Also, note
that $G_3(q^2)$ can be obtained uniquely from $\bar {\cal V}_1$ and
$\bar {\cal V}_2$ for a number of values of $q^2$, which provides a useful check.  
One can extend Eq.~\eqref{eq:F3at0} to define 
\begin{equation}
{\tilde F}_3(q^2) \equiv G_3(q^2) - s_{\alpha_N} \left( G_1 (q^2) - Q_N \right) \,.
\label{eq:F3tilde}
\end{equation}
To get $F_3(0) = {\tilde F}_3(0)$, we find better control by 
extrapolating ${\tilde F}_3(q^2)$ to $q^2 \to 0$. 

\begin{figure*}[tbp]
  \subfigure{
    \includegraphics[width=0.325\linewidth]{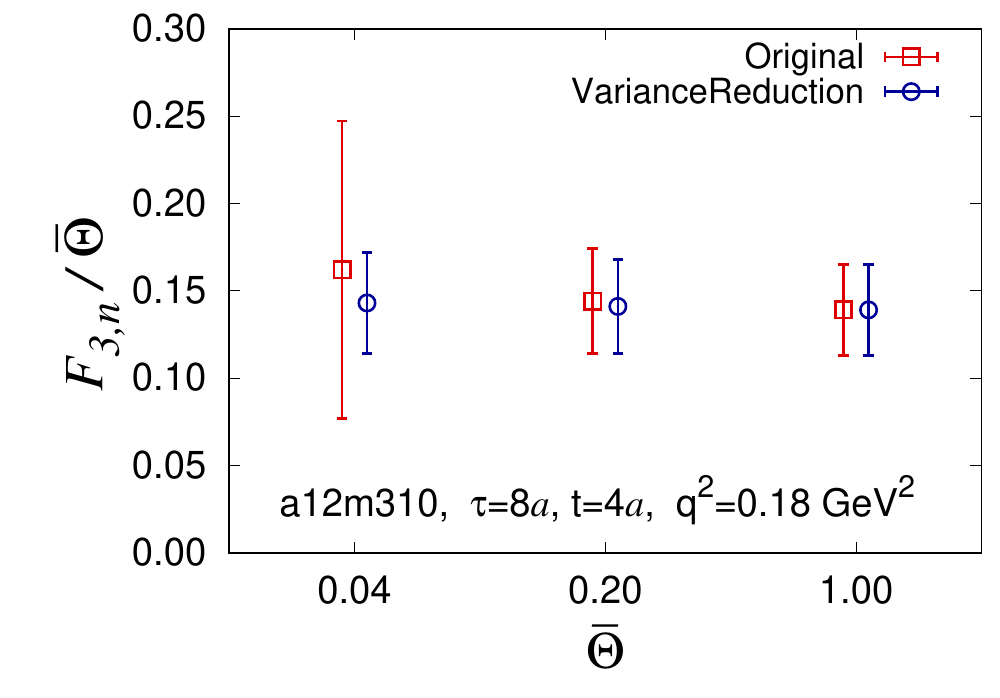}
    \includegraphics[width=0.325\linewidth]{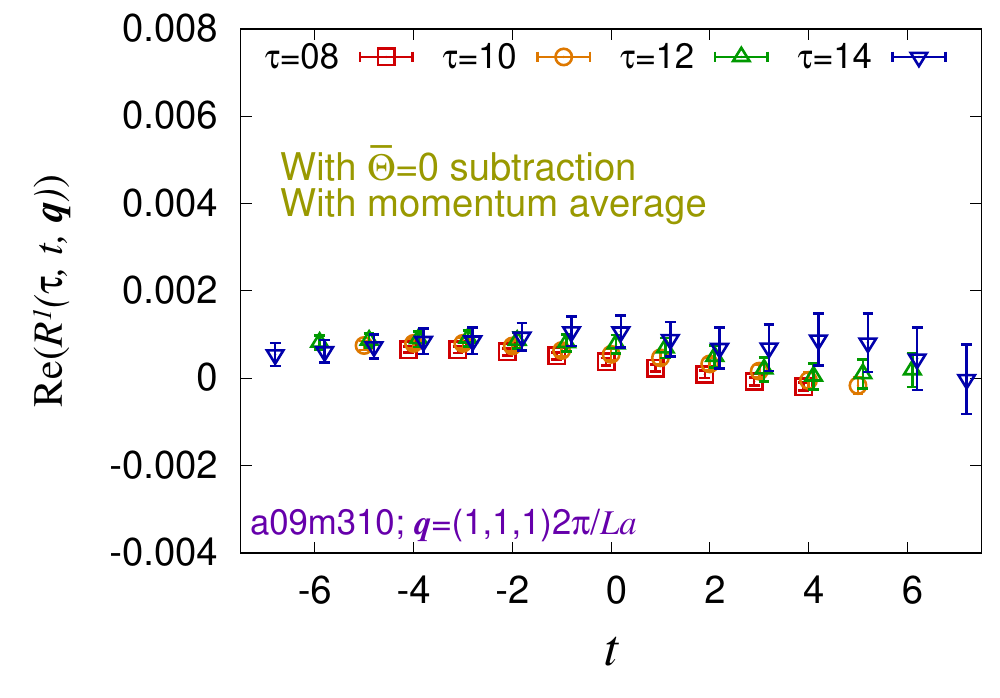}
    \includegraphics[width=0.325\linewidth]{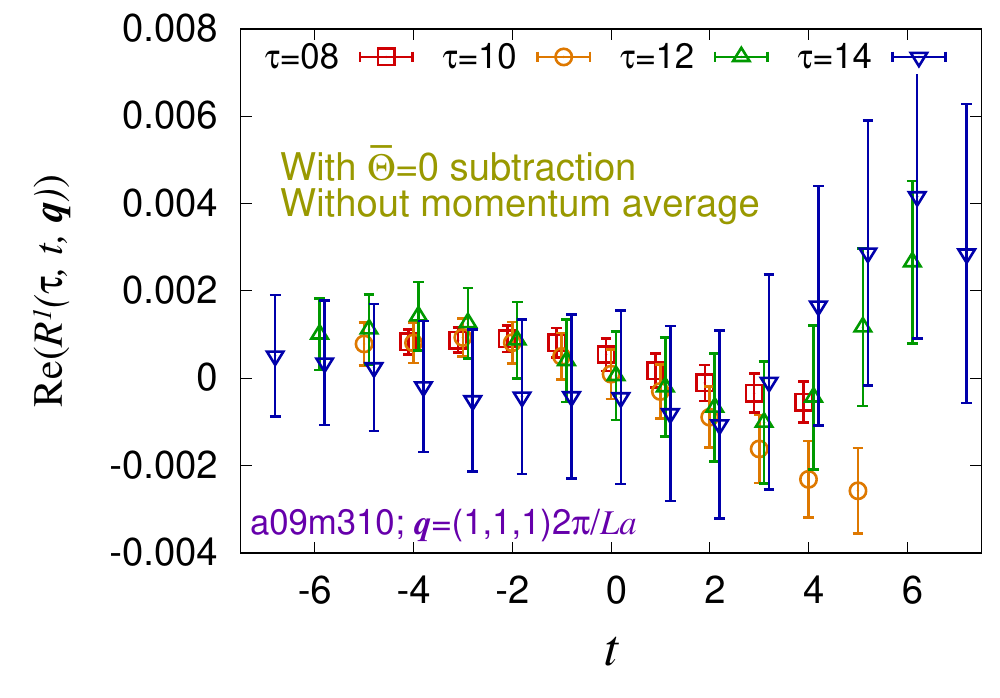}
  }
\caption{The improvement in signal under subtraction of the
${\overline \Theta}=0$ contribution and averaging over equivalent momenta.  The
panel on the left shows, using data from the $a12m310$ ensemble, (i) the improvement in $F_{3,n}$ as
${\overline \Theta} \to 0$ and (ii) even $\overline \Theta = 1$ is in the linear regime.  The panel in
the middle shows the signal in 
$R^1(\tau,t,\bm{q})$ 
with both the
${\overline \Theta}=0$ and momentum averaging on the $a09m310$ ensemble with
${\overline \Theta}=0.2$ and $\bm q = (1,1,1)2\pi/La$, while that on the right
is without averaging over equivalent momenta. 
\label{fig:eps_sub}}
\end{figure*}

The subtraction of the ${\overline \Theta}=0$ contribution also allows
averaging of the three point functions over momenta related by cubic invariance, as seen by   comparing the
simpler Eqs.~\eqref{eq:Vinear} with Eqs.~\eqref{eq:components}. We illustrate
the improvement in the signal in Fig.~\ref{fig:eps_sub}.
The averaging over equivalent cases (over momenta related by cubic symmetry and
over $\overline{ \cal V}_1$ and $\overline{ \cal V}_2$) significantly
reduces the statistical errors and improves the analysis of excited state contamination (ESC) discussed next.

\section{Removing ESC in  \texorpdfstring{$F_3$}{F\unichar{"2083}}}
\label{sec:ESC}




In order to extract the ground state contribution 
$\bar {\cal V}_\mu$ from lattice data on the 
ratio  $R^\mu (\tau, t, \bm{q})$ of three- and two-point functions 
defined in Eq.~(\ref{Rmu}), 
we need to remove  all excited states that make a
significant contribution. 

\begin{figure*}[tbp]
  \subfigure{
    \includegraphics[width=0.45\linewidth]{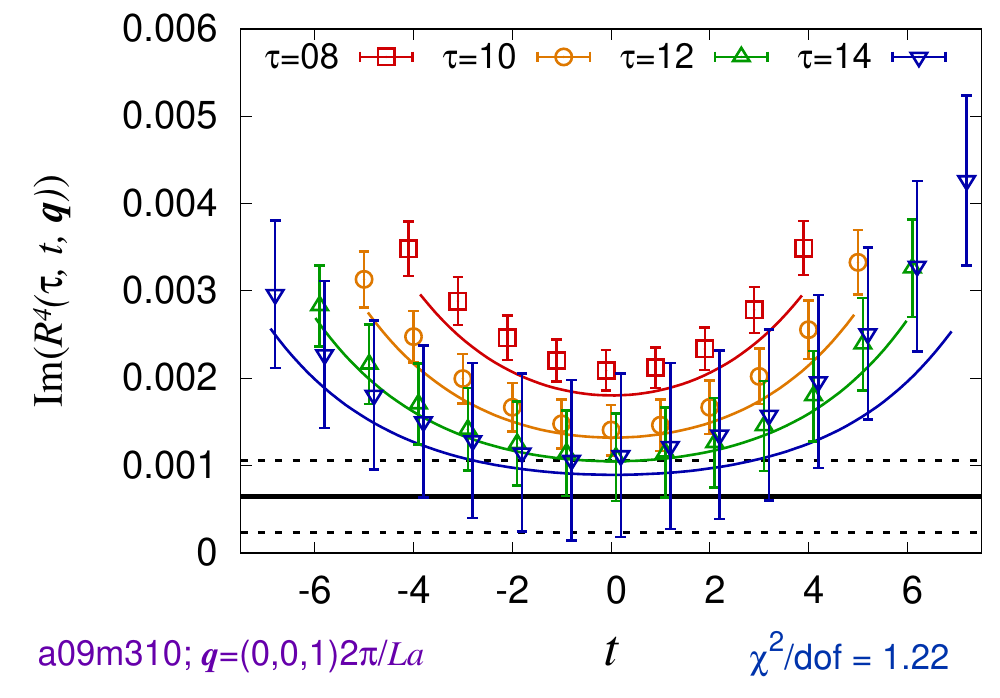}
    \includegraphics[width=0.45\linewidth]{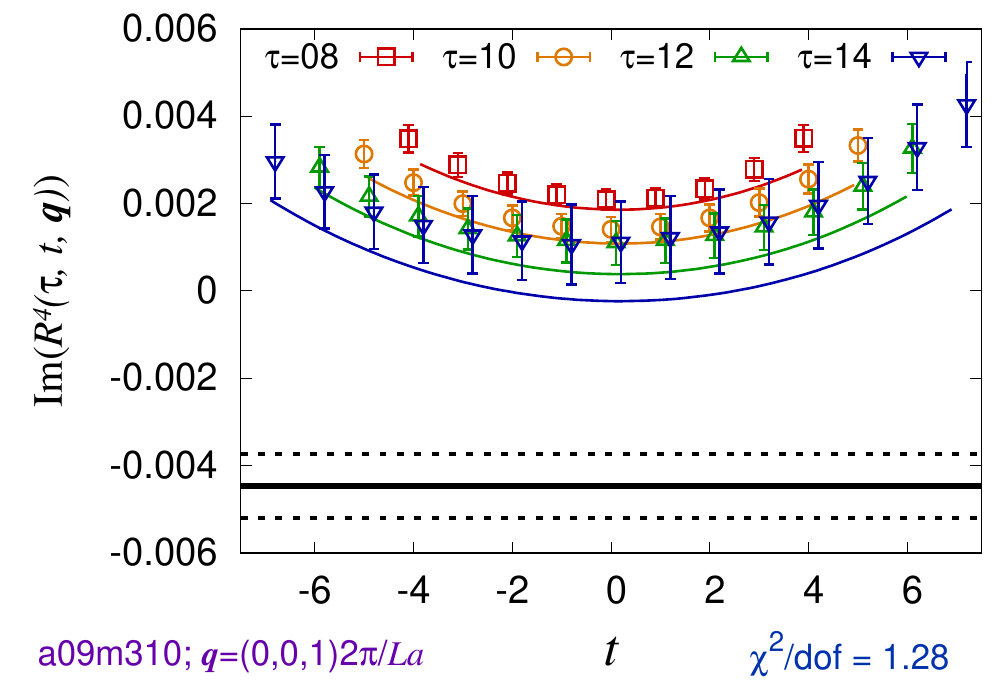}
  }
  \subfigure{
    \includegraphics[width=0.45\linewidth]{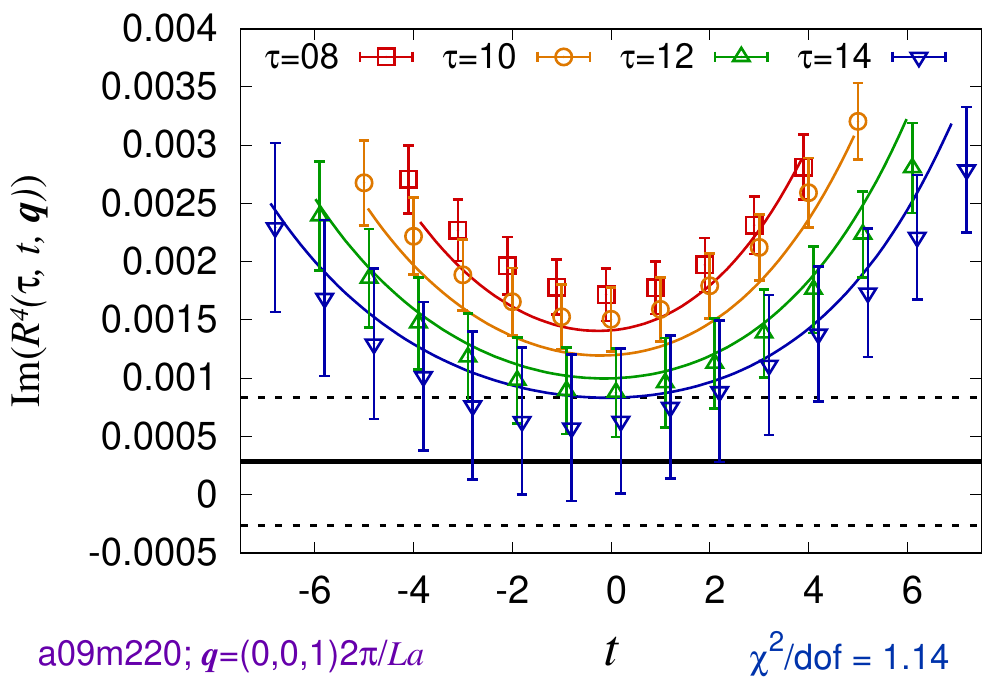}
    \includegraphics[width=0.45\linewidth]{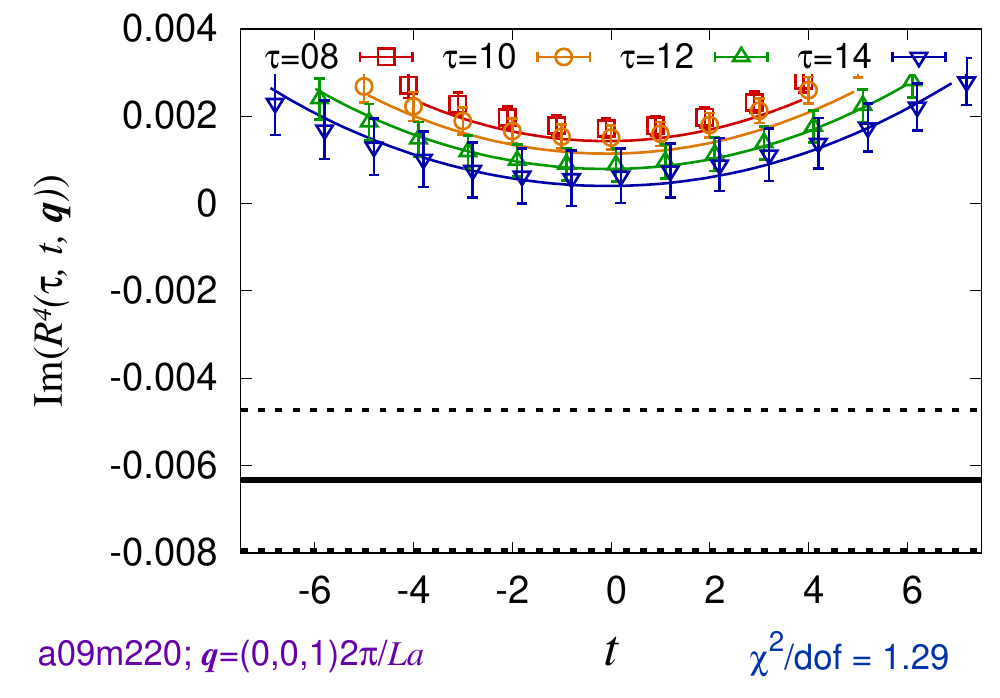}
  }
  \subfigure{
    \includegraphics[width=0.45\linewidth]{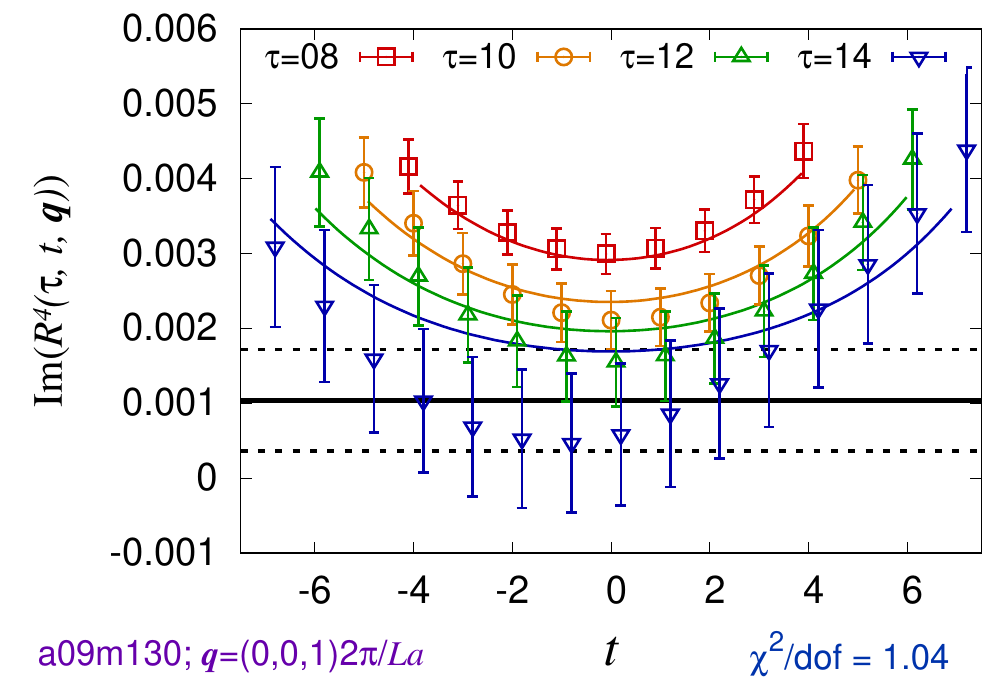}
    \includegraphics[width=0.45\linewidth]{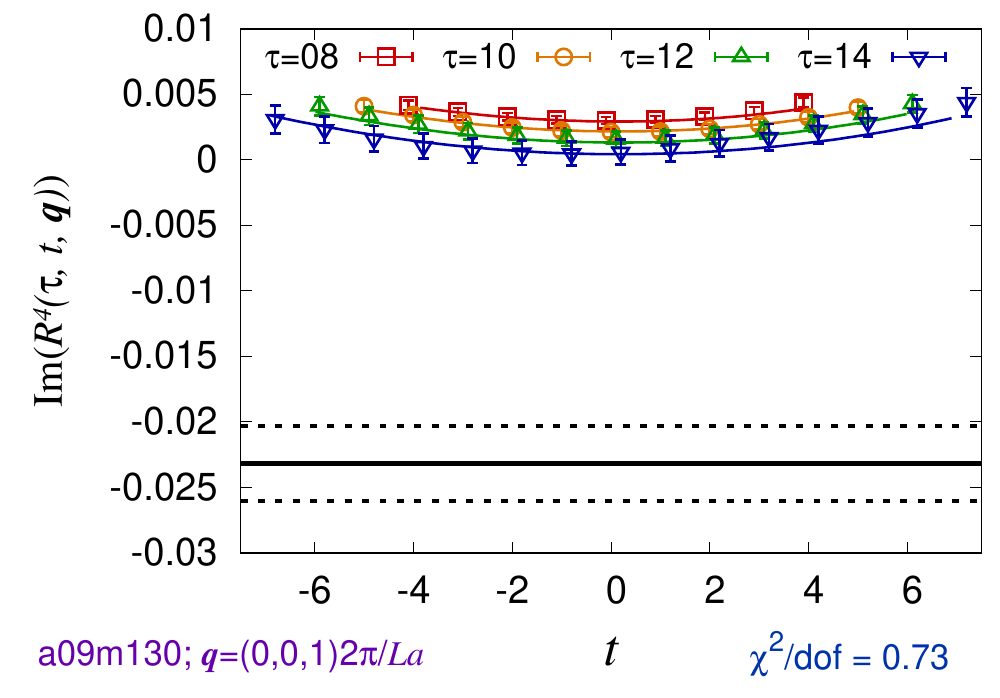}
  }
\caption{Comparison of the two-state fit to the ratio 
$\Im (R^4(\tau,t,\bm{q}))$ 
defined in Eqs.~\protect\eqref{Rmu} with the first excited-state
energies taken from a three-state fit to the two-point function (left panels)
and set equal the non-interacting energy of the $N \pi$ state (right
panels).  The data for the three ensembles with $a \approx 0.09$~fm
are shown in the three rows. The $\chi^2$/dof of the two sets of
fits are comparable, but the extrapolated 
ground sate value (solid black line) is vastly different. 
The data are shown for $\bm q =
(0,0,1)2\pi/La$ and the four largest values of $\tau$. 
All data are with ${\overline \Theta}=0.2$.
\label{fig:ESCV4}}
\end{figure*}

We have analyzed data on $R^\mu (\tau, t, \bm{q})$ in terms of a two-state 
fit, following two strategies. 
In the first, we have taken the first excited-state energies  from a three-state fit to the two-point function. 
In the second strategy, we have set the first excited-state energy 
to the non-interacting energy of the $N \pi$ state, 
motivated by the $\chi$PT expectation 
that the leading excited state is the $N\pi$ state, with amplitude  of the
same size as the ground state contribution (see  Appendix~\ref{sec:ESCappendix} 
for more details). 
In Fig.~\ref{fig:ESCV4} we
compare the two strategies for  
$\Im (R^4 (\tau, t, \bm{q}))$. 
The $\chi^2$/dof of the fits are similar for the
two cases on all three ensembles, but the ground state estimate is
vastly different and thus the contribution to the nEDM. 
With the current data, picking between them is the key unresolved 
challenge for this calculation. The very large extrapolation for $\tau \to \infty$ in the $N \pi$
case, however, leads us to question whether a two-state fit is sufficient if the
$N \pi$ state is included and whether a similar effect might contaminate 
our extraction of \(\alpha_N\). We therefore first perform the analysis taking the excited state 
energy, $E_1$, from a three-state fit to the two-point function and return to an analysis including a $N \pi$ 
state in Sec.~\ref{sec:Npi}. 

\begin{figure*}[tbp]
  \subfigure{
    \includegraphics[width=0.325\linewidth]{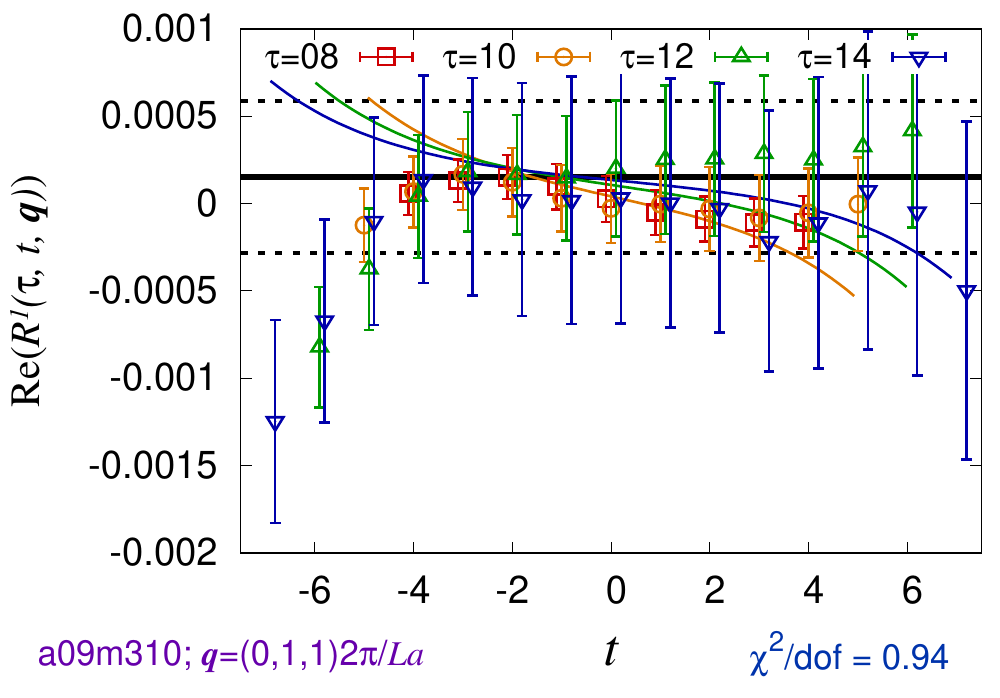}
    \includegraphics[width=0.325\linewidth]{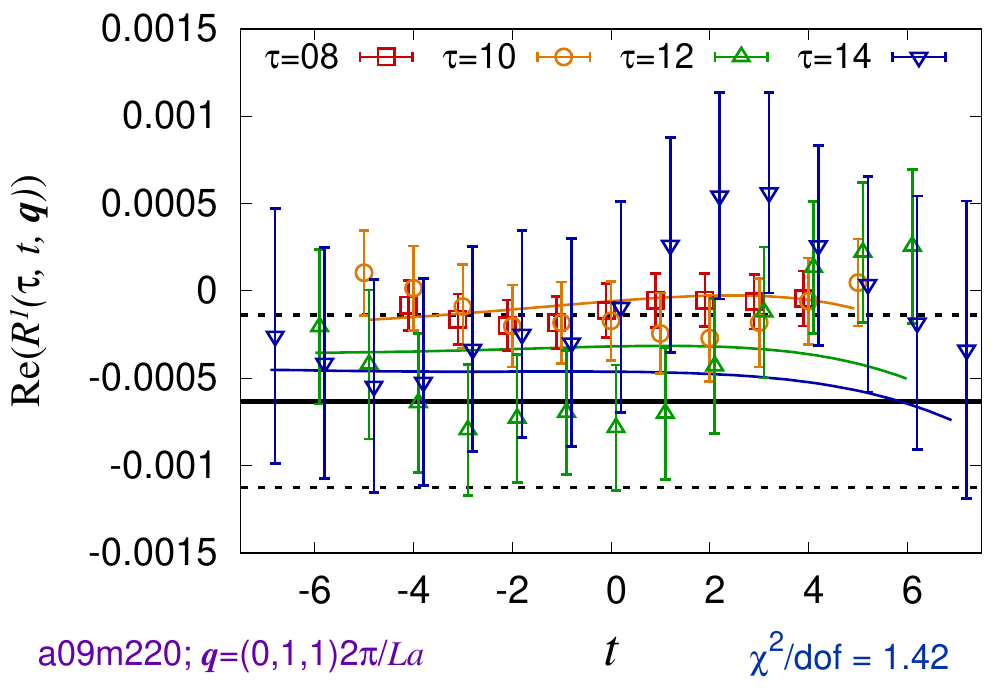}
    \includegraphics[width=0.325\linewidth]{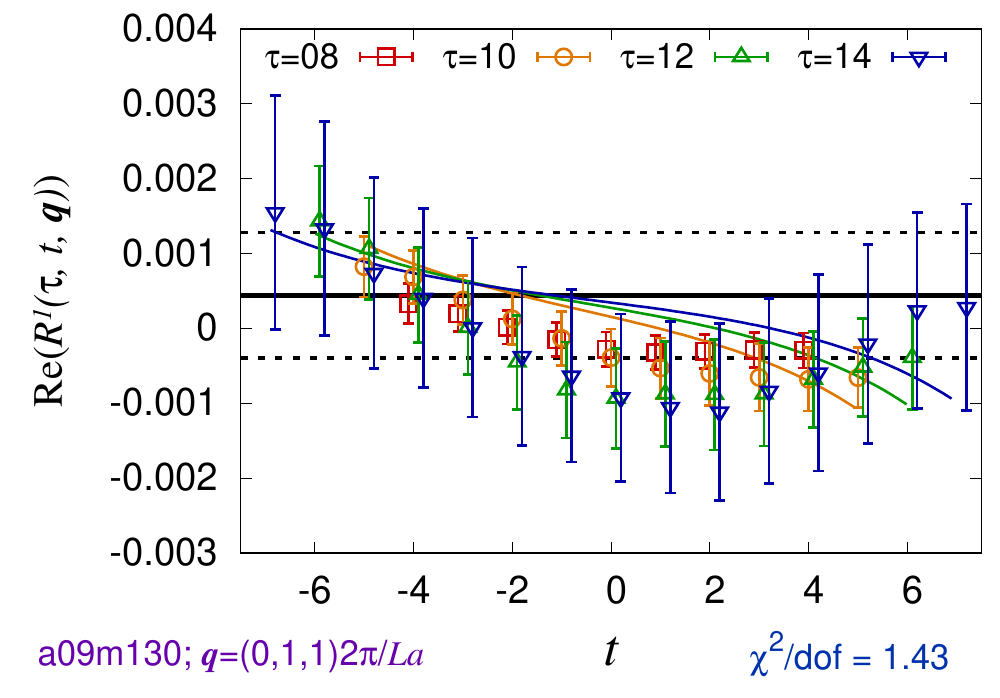}
  }
\caption{Examples of unstable two-state fits to the ratio 
$\Re (R^1(\tau,t,\bm{q}))$ 
defined in Eqs.~\protect\eqref{Rmu} with the first excited-state
energies taken from a three-state  fit to the two-point function. The data are 
for the three ensembles with $a \approx 0.09$~fm, for $\bm q =
(0,1,1)2\pi/La$ and the values of $\tau$ are specified in the labels. 
All data are with ${\overline \Theta}=0.2$.
\label{fig:ESCV1}}
\end{figure*}

A second issue arising from the small signal in $F_3$ is that two-state fits
to many of the correlation functions with the full covariance matrix
are unstable with respect to variations in the values of $\tau$ and
$t_{\rm skip}$, the number of points skipped in the fits adjacent to the source and sink for each $\tau$. Examples of this are shown in Fig.~\ref{fig:ESCV1} for 
$\Re (R^1 (\tau, t, \bm{q}))$. 
This has two consequences for the analysis. First, 
we have carried out the final analysis using only the
diagonal elements of the covariance matrix. We have, however, checked
that in cases where fully covariant fits are possible, the two results are
consistent. Since we use uncorrelated fits for removing excited-state 
contamination, we do not quote a $\chi^2$/dof for these fits. Second, the
system of four equations, Eqs.~\eqref{eq:Vinear}, over determines $G_3$ and $G_1$. While 
we solve the full set of equations as explained in appendix~\ref{sec:appendix1}, the  data from 
$\Re (R^{1,2} (\tau, t, \bm{q}))$,  
which have poor signal, do not make a significant contribution. 
We have checked this by removing them from the analysis and the results are essentially unchanged, i.e., 
the results are dominated by 
$\Re (R^3 (\tau, t, \bm{q}))$ 
and 
$\Im (R^4 (\tau, t, \bm{q}))$. 

\section{Extrapolation of \texorpdfstring{$F_3(q^2)$}{F\unichar{"2083}(q\unichar{"B2})} to \texorpdfstring{$q^2 \to 0$}{q\unichar{"B2}\unichar{"2192}0}}
\label{sec:qsq}

The ansatz used to extrapolate $F_3(q^2)$ to $q^2 \to 0$ is given in
Eq.~\eqref{Hdef} with one caveat.  We use $\tilde F_3(q^2)$, defined
in Eq.~\eqref{eq:F3tilde}, instead of $F_3(q^2)$ as they are consistent
to leading order and the extraction of $\tilde F_3(q^2)$ is better controlled.
We examine three fits based on Eq.~\eqref{Hdef}:
\begin{itemize}
\item
Linear: the quantities $d_i$ and $S_i^\prime$ are free parameters and $H_i$ is set to zero.
\item
$\chi$PT: Only $d_i$ is a free parameter, $S_i^\prime$ are given in
Eq.~\eqref{radius}, $\overline g_0$ in Eq.~\eqref{eq:g0def}, and the
$H_i$ in Eq.~\eqref{FD}.
\item
$\chi$PTg0: Same as $\chi$PT except $\overline g_0$ is left as a free parameter. 
\end{itemize}

\begin{figure*}[tbp]
  \subfigure{
    \includegraphics[width=0.32\linewidth]{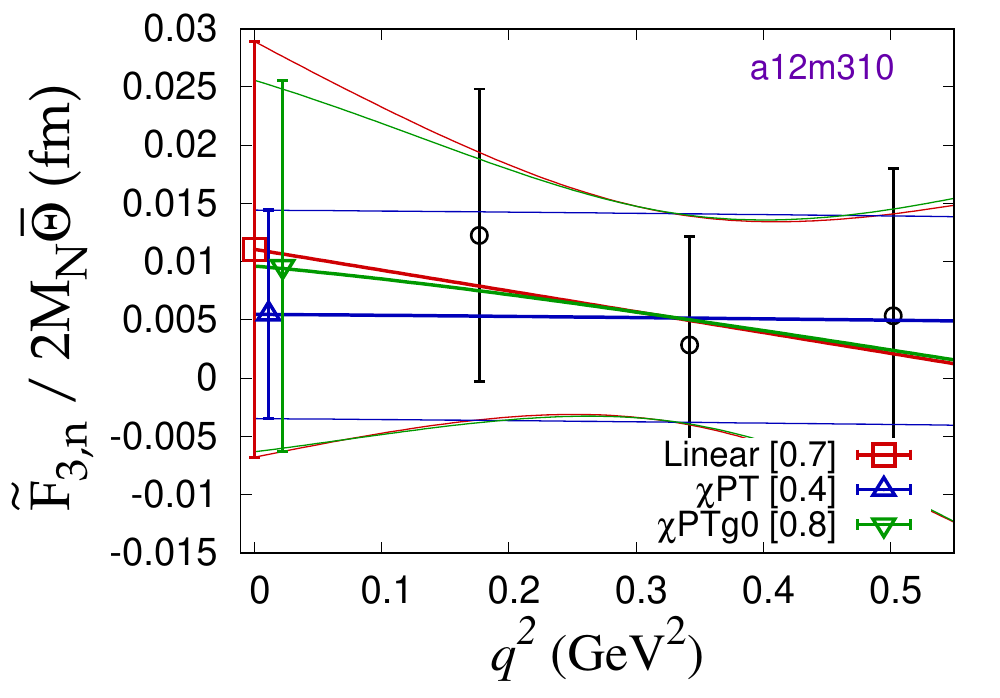}
    \includegraphics[width=0.32\linewidth]{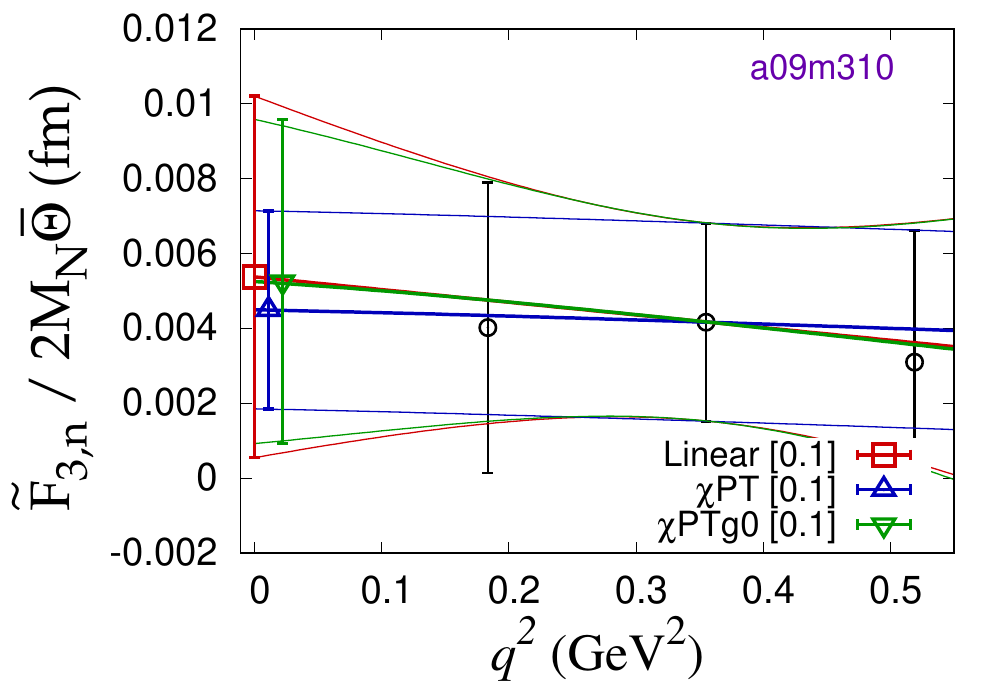}
  }
  \subfigure{
    \includegraphics[width=0.32\linewidth]{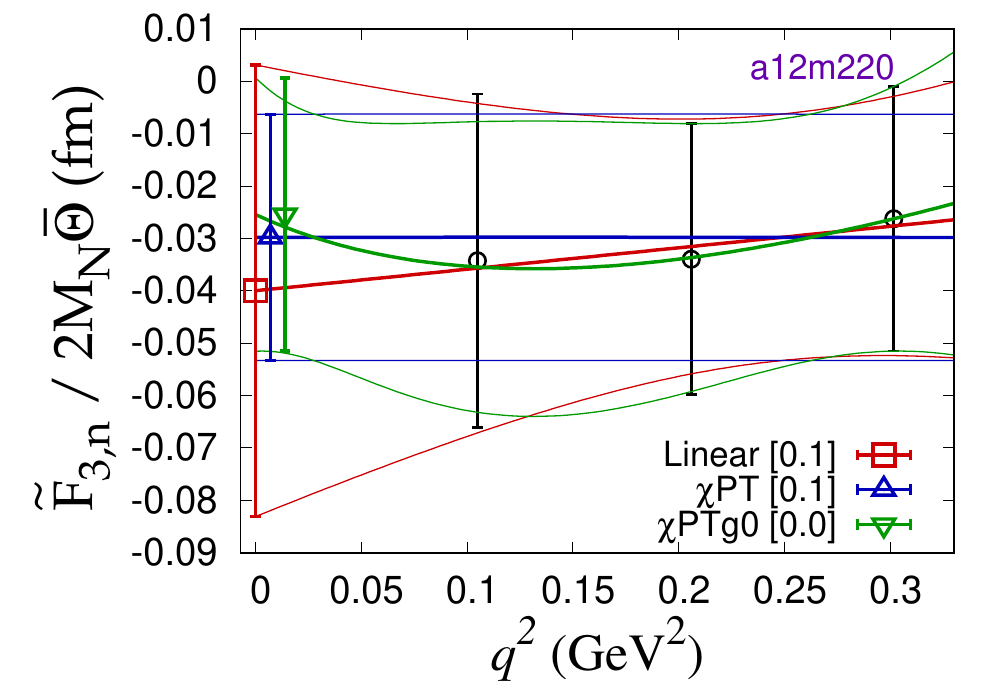}
    \includegraphics[width=0.32\linewidth]{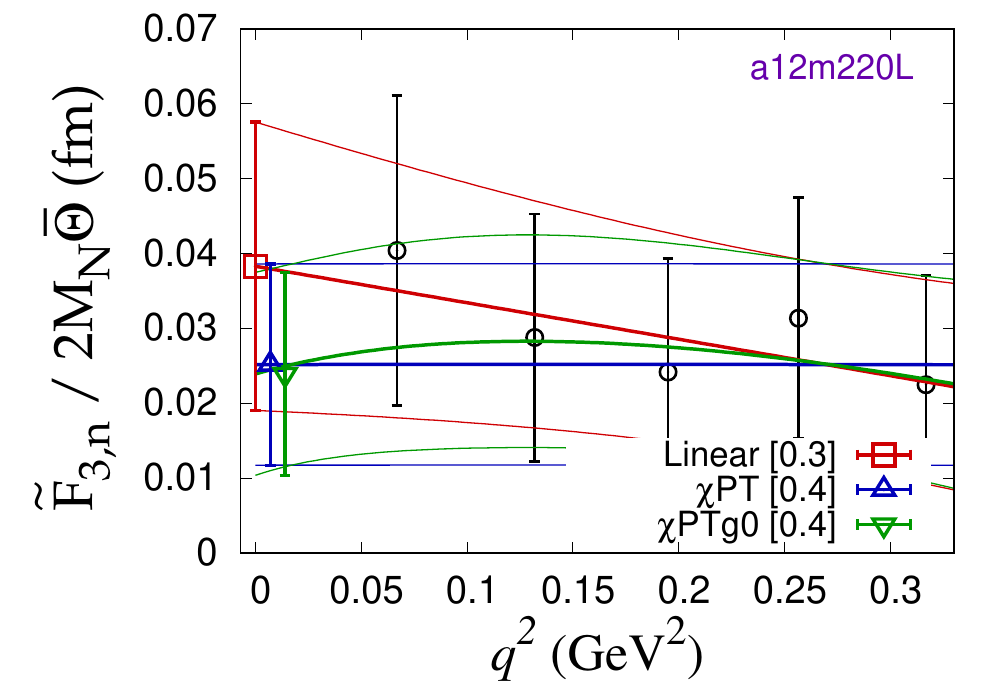}
    \includegraphics[width=0.32\linewidth]{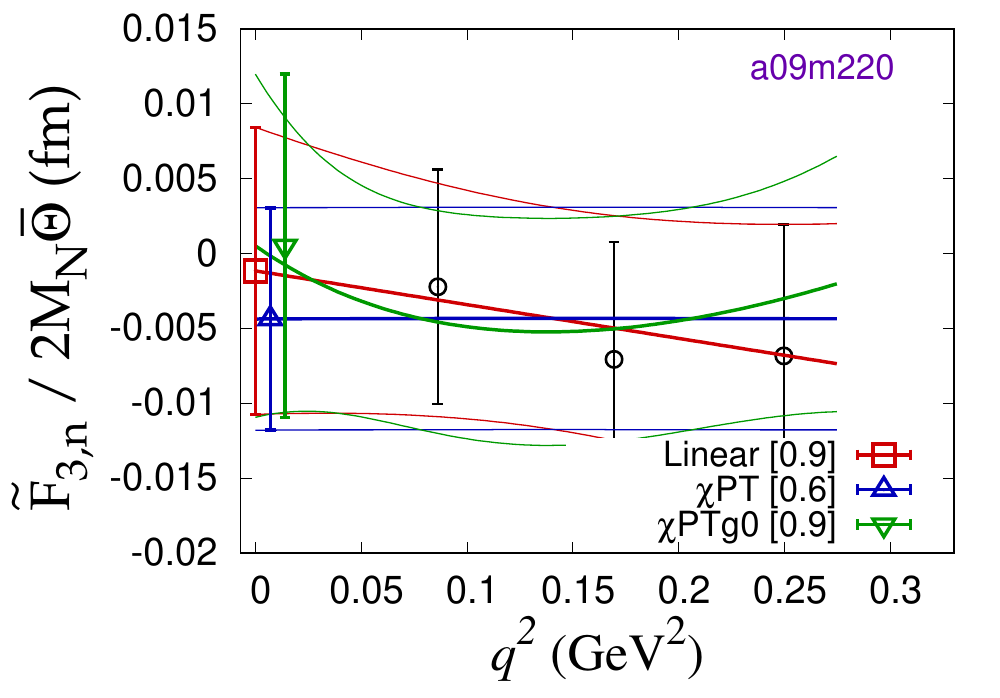}
  }
  \subfigure{
    \includegraphics[width=0.32\linewidth]{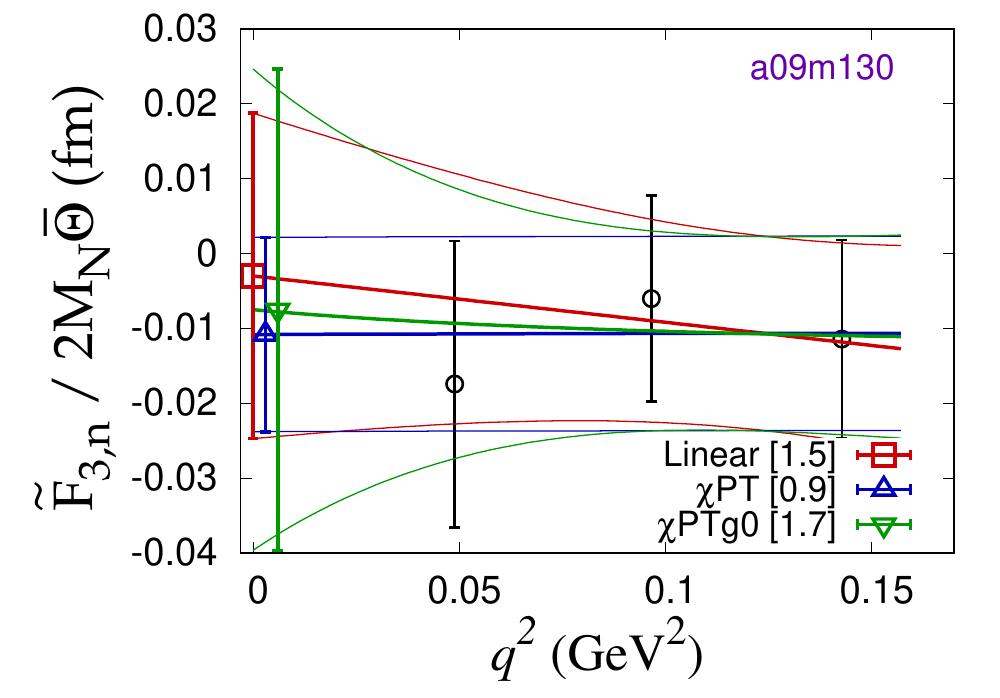}
    \includegraphics[width=0.32\linewidth]{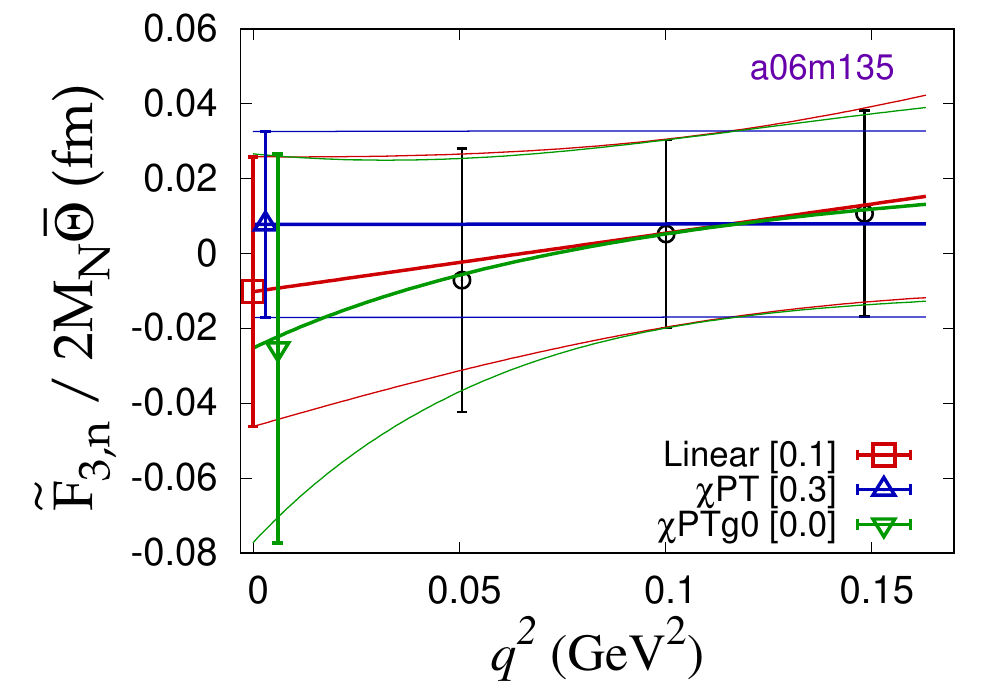}
  }
\caption{The extrapolation of 
$\tilde F_3(q^2)$ to $q^2 \to 0$ using Eq.~\protect\eqref{Hdef} for the neutron. 
The three fit ansatz, Linear, $\chi$PT and $\chi$PTg0, are defined in the text. The $\chi^2$/dof of the 
fits are given within square parentheses. All data are with ${\overline \Theta}=0.2$.
\label{fig:qsqN}}
\end{figure*}

\begin{figure*}[tbp]
  \subfigure{
    \includegraphics[width=0.32\linewidth]{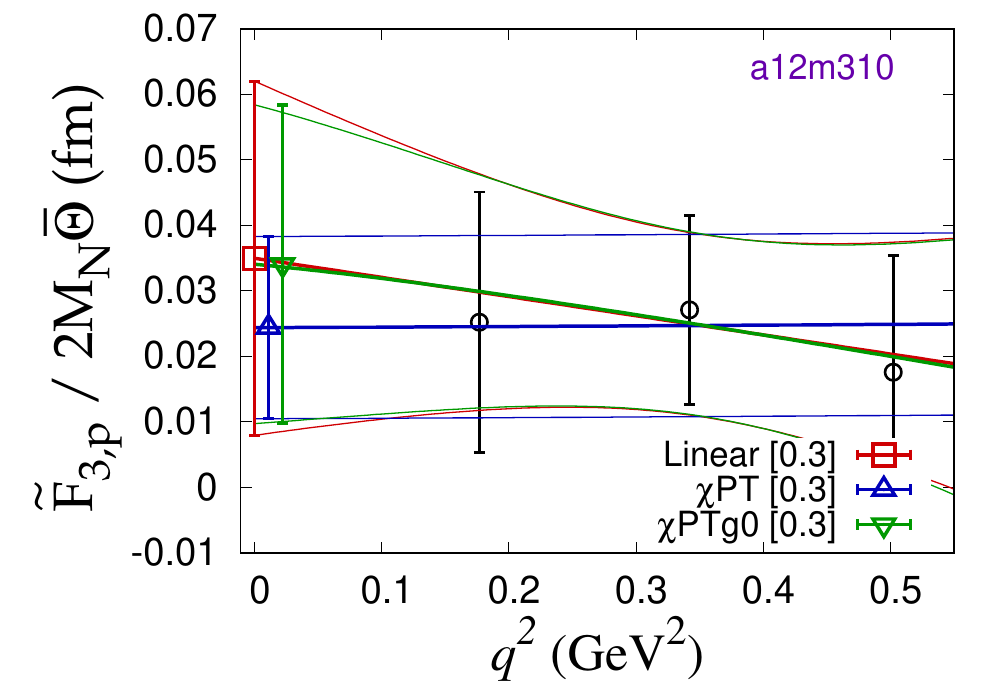}
    \includegraphics[width=0.32\linewidth]{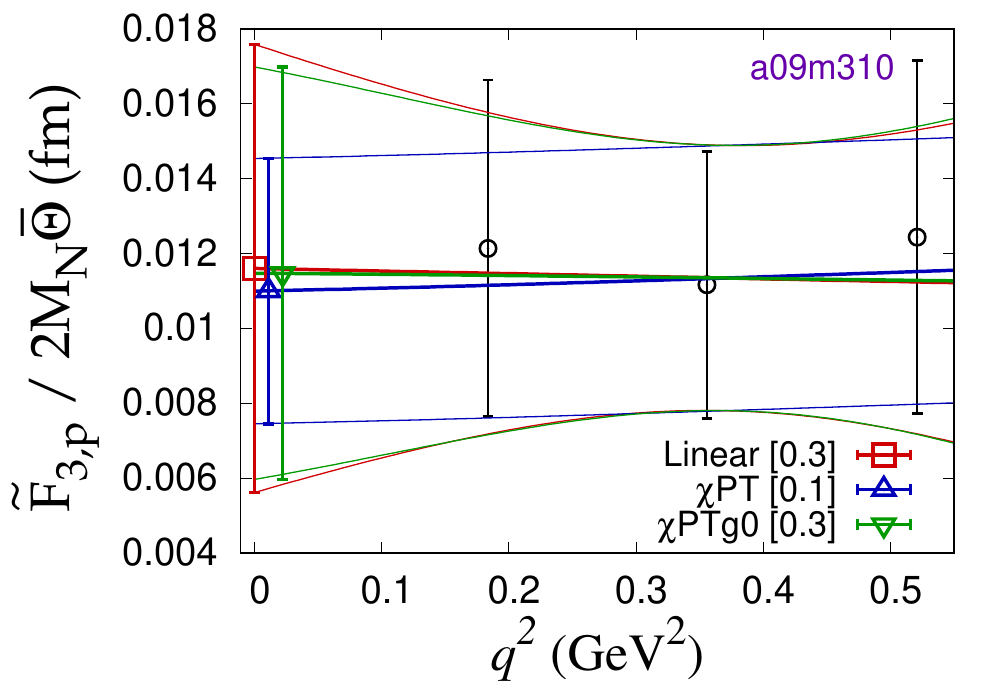}
  }
  \subfigure{
    \includegraphics[width=0.32\linewidth]{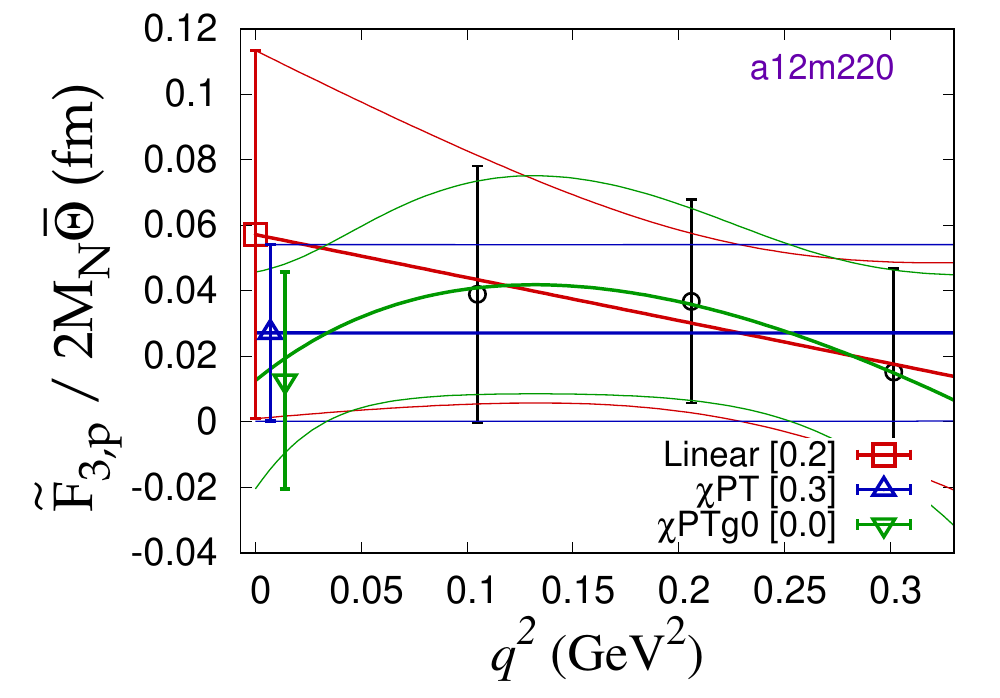}
    \includegraphics[width=0.32\linewidth]{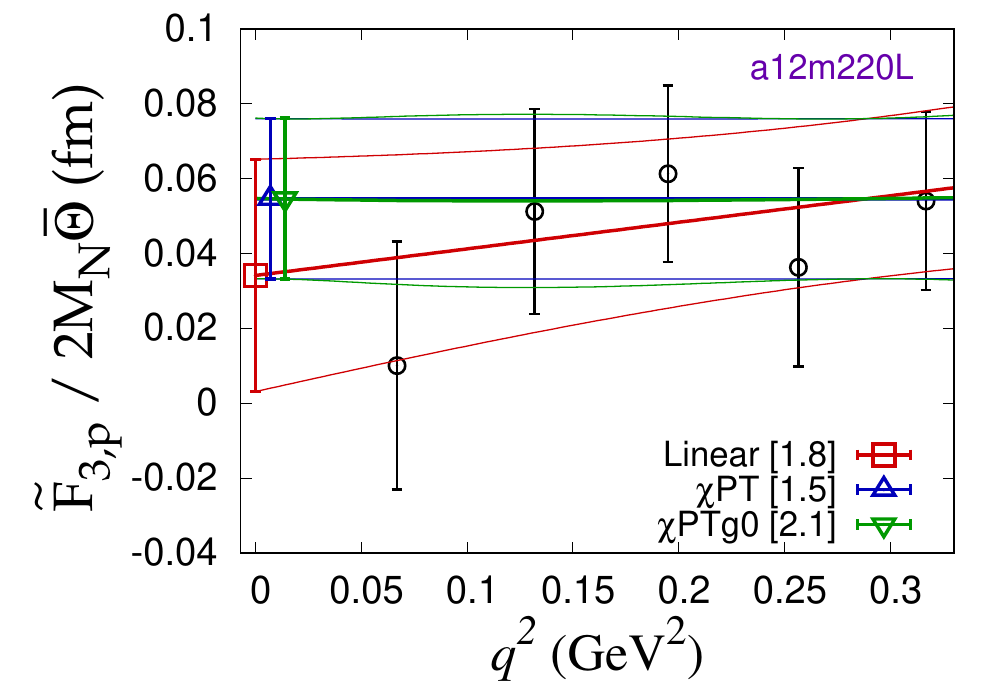}
    \includegraphics[width=0.32\linewidth]{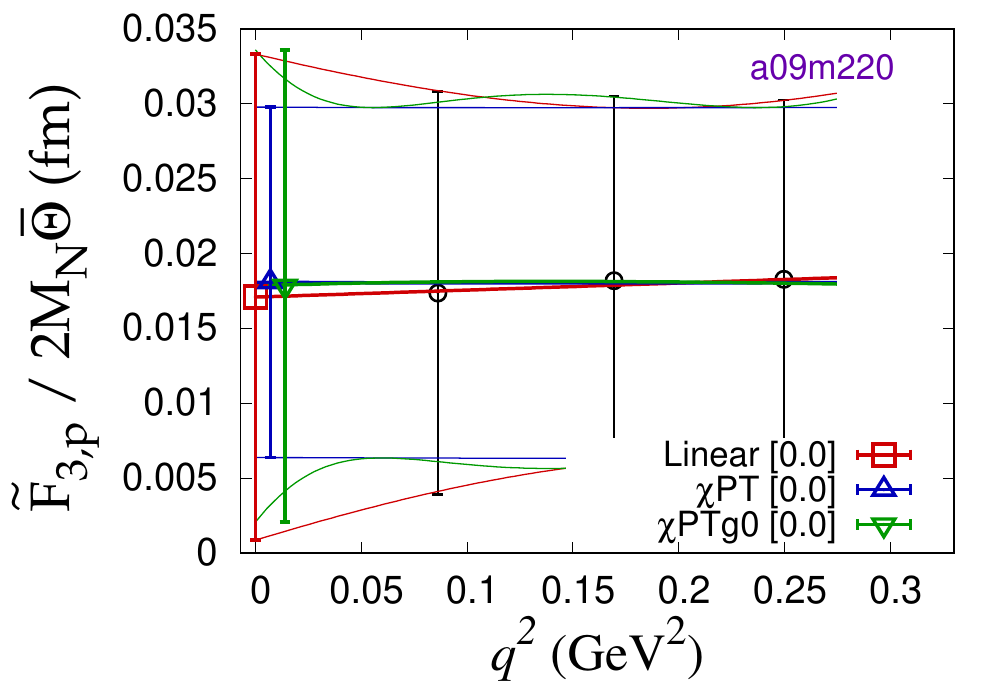}
  }
  \subfigure{
    \includegraphics[width=0.32\linewidth]{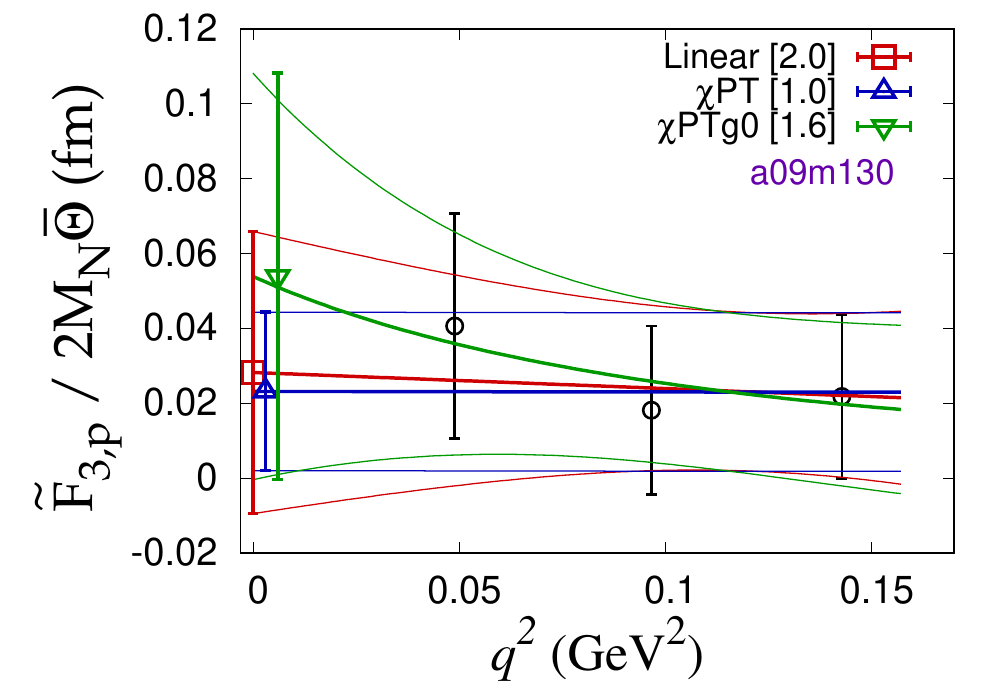}
    \includegraphics[width=0.32\linewidth]{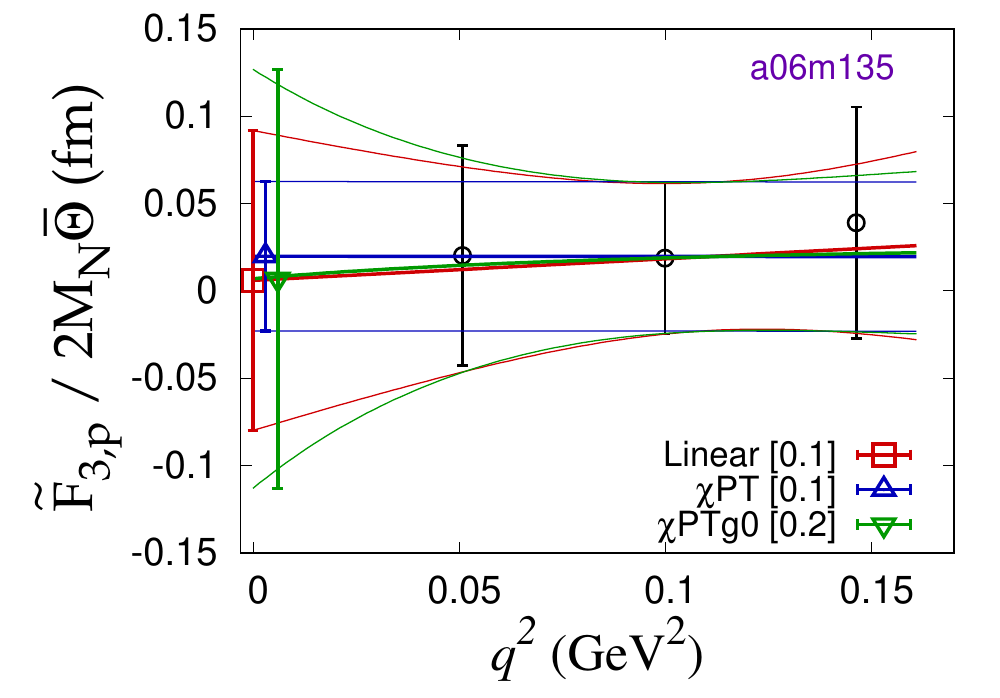}
  }
\caption{The extrapolation of $\tilde F_3(q^2)$ to $q^2 \to 0$ using Eq.~\protect\eqref{Hdef} for the proton. 
The rest is the same as in Fig.~\protect\ref{fig:qsqN}. 
\label{fig:qsqP}}
\end{figure*}

The data and fits for the neutron and proton are presented in
Figs.~\ref{fig:qsqN} and~\ref{fig:qsqP}.  The data are, within errors,
flat in all cases and the extrapolated values from the three types of
fits are consistent. Since in most cases, we have reliable data at only three
values of $q^2$, we take the final result from the $\chi$PT fit. At the end, we will take
the difference between the Linear and $\chi$PT fits to estimate the associated 
systematic uncertainty. 

\section{Additional \texorpdfstring{$O(a)$}{O(a)} artifacts}
\label{sec:Oa}

Before performing a chiral-continuum extrapolation of the results, in this section we 
justify our continuum extrapolation formula for $d_n ({\overline \Theta})$ that includes an $M_\pi$-independent term that does not vanish in the chiral limit, i.e., a term proportional to $am_q^0$.

There are multiple sources of \(O(a)\) corrections that we need to consider. First, since our clover coefficient \(c_{SW}\) is set to its tadpole-improved tree-level value, the action, and hence all matrix elements, have residual \(O(\alpha_s a)\) corrections. Because of the use of smeared gauge fields, however, the tadpole-improved tree-level approximation is extremely good, and these are expected to be tiny effects.  Second, the vector current we insert is not improved~\cite{Bhattacharya:2005rb}, and, hence, we expect its renormalization coefficient to have \(O(am_q)\) corrections.  Such multiplicative terms, however, are unimportant near the chiral-continuum limit, where the \CPV\ form factors vanish.  A third source of \(O(a)\) effects is the required improvement of the vector current by an \(O(am_q^0)\) mixing with the derivative of the tensor current, which can give rise to a nonzero \(F_3\), but only in the presence of CP violation in the theory. Since the topological charge does not introduce CP violation in the chiral limit, we would expect the behavior of \(d_n\) to be dominantly \(O(a^2)\) in the chiral limit, if these were the only \(O(a)\) effects.

In Appendix~\ref{sec:appendix2}, we analyze the Wilson-clover theory based on the framework of a continuum EFT  for the lattice action and the 
axial Ward Identities. Following Refs.~\cite{Bochicchio:1985xa, Testa:1998ez,Guadagnoli:2003zq}, 
we show that the topological charge gives \(O(a)\) \CPV\ corrections, and identify this as effectively due to the insertion of the isoscalar quark chromo-EDM operator, which
the topological term can mix with.  Since this term is expected to survive in the chiral limit, we include an \(O(a m_q^0)\) term in our chiral continuum fits.

\section{Chiral-continuum extrapolation and Results}
\label{sec:Results}

\begin{figure*}[tbp]
  \subfigure{
    \includegraphics[width=0.425\linewidth]{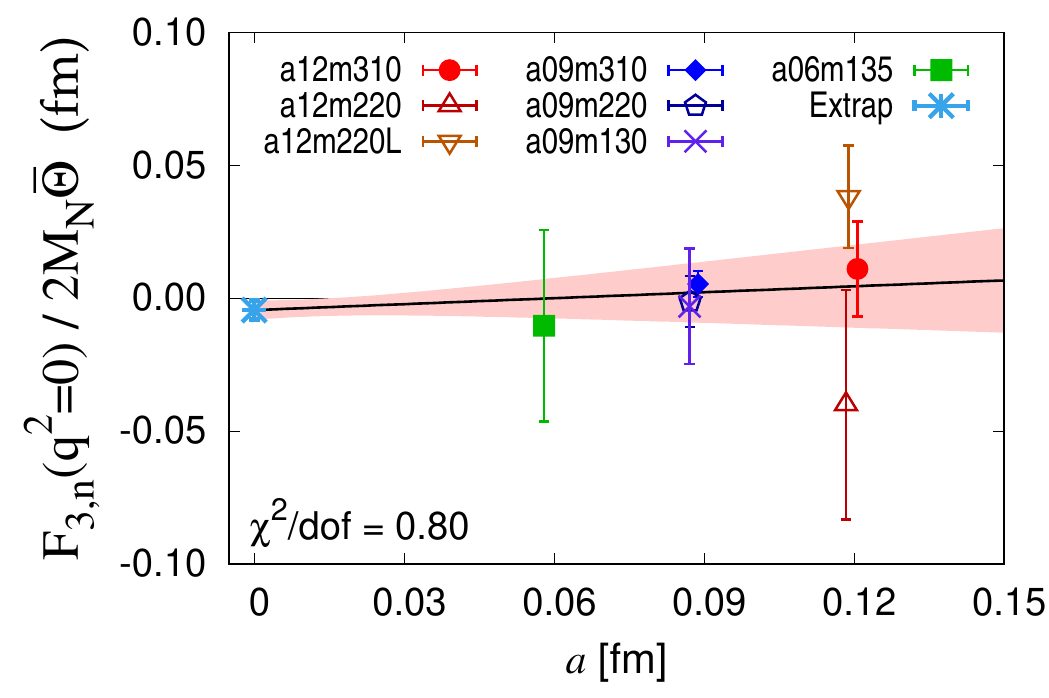}
    \includegraphics[width=0.425\linewidth]{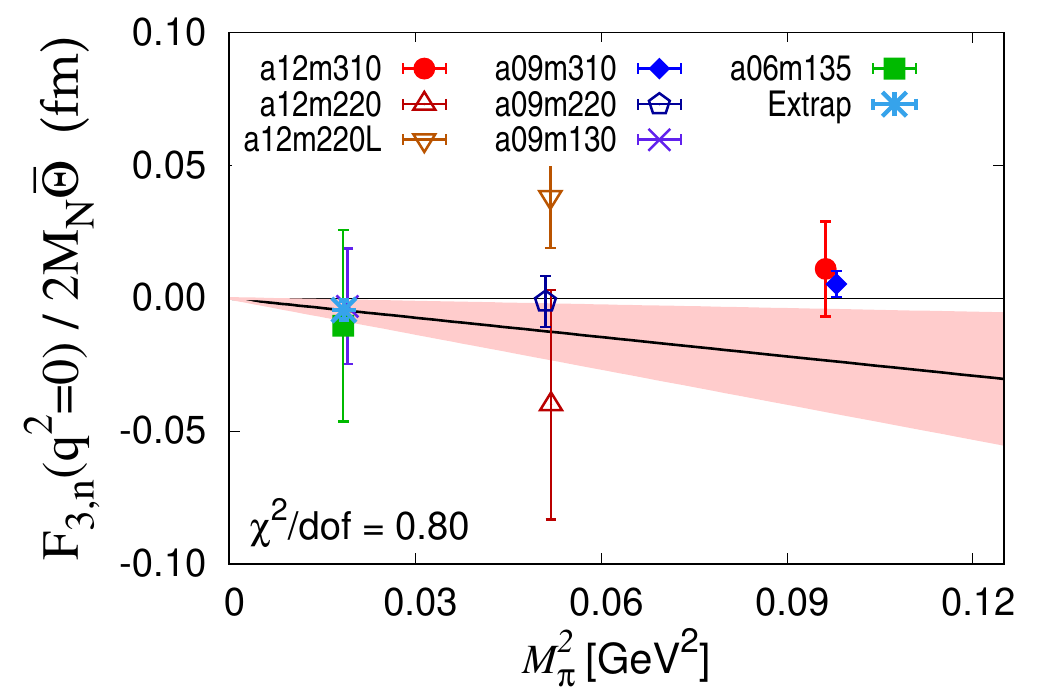}
  }
  \subfigure{
    \includegraphics[width=0.425\linewidth]{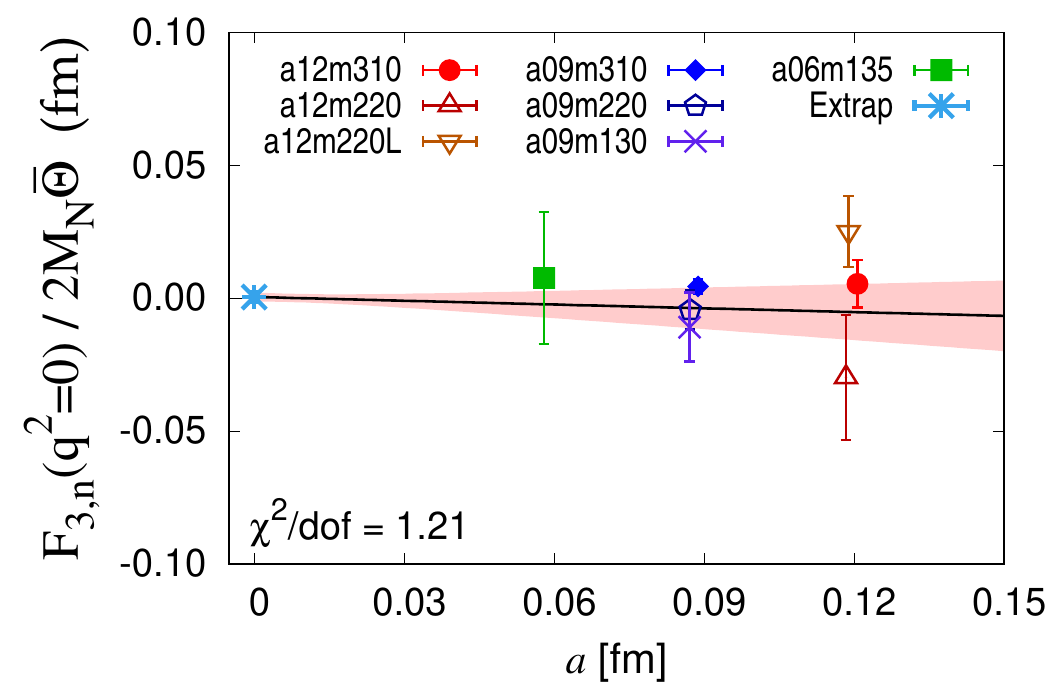}
    \includegraphics[width=0.425\linewidth]{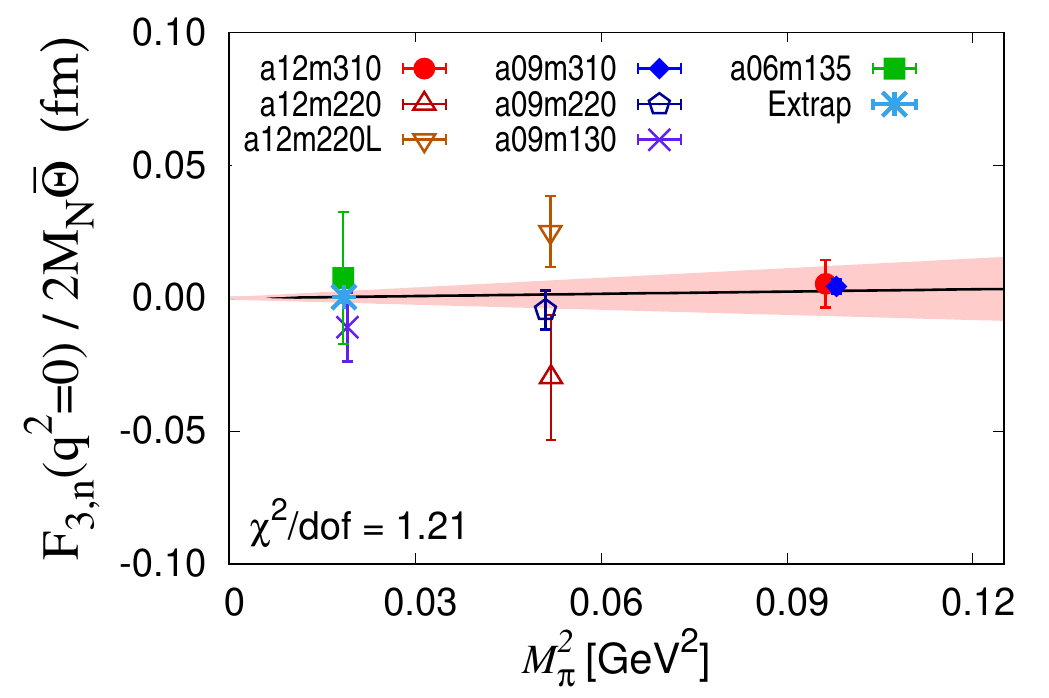}
  }
  \subfigure{
    \includegraphics[width=0.425\linewidth]{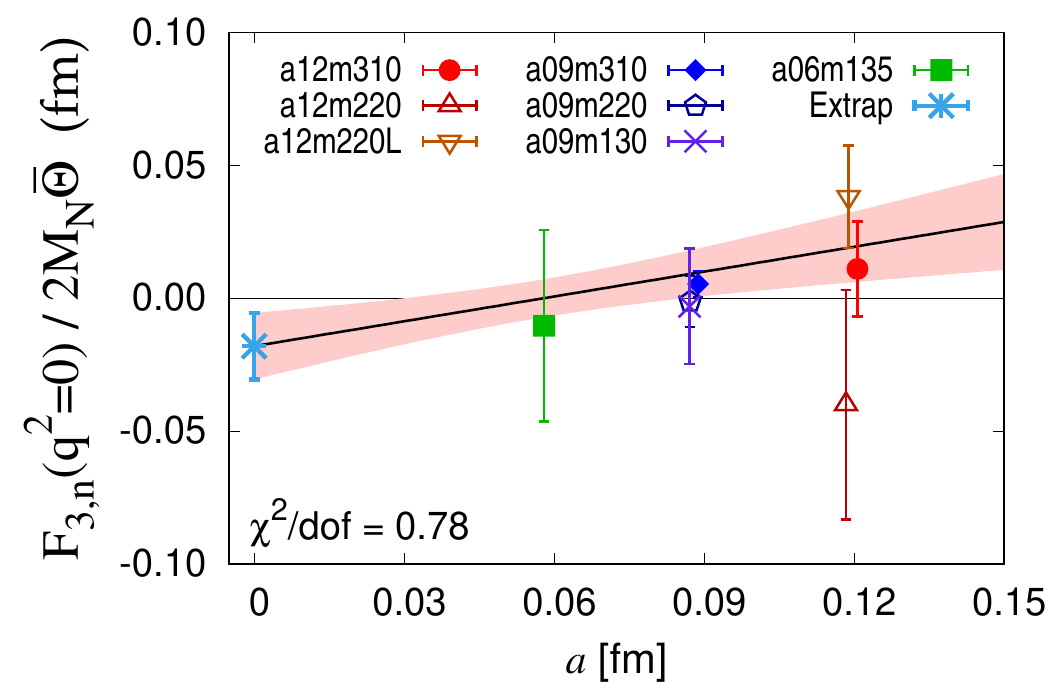}
    \includegraphics[width=0.425\linewidth]{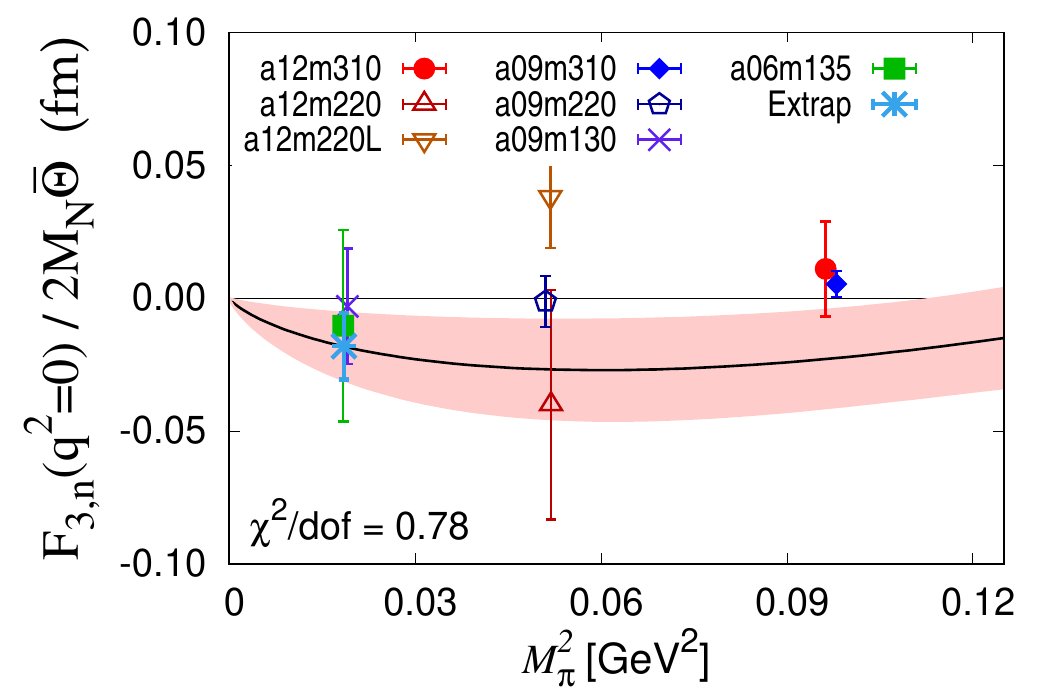}
  }
  \subfigure{
    \includegraphics[width=0.425\linewidth]{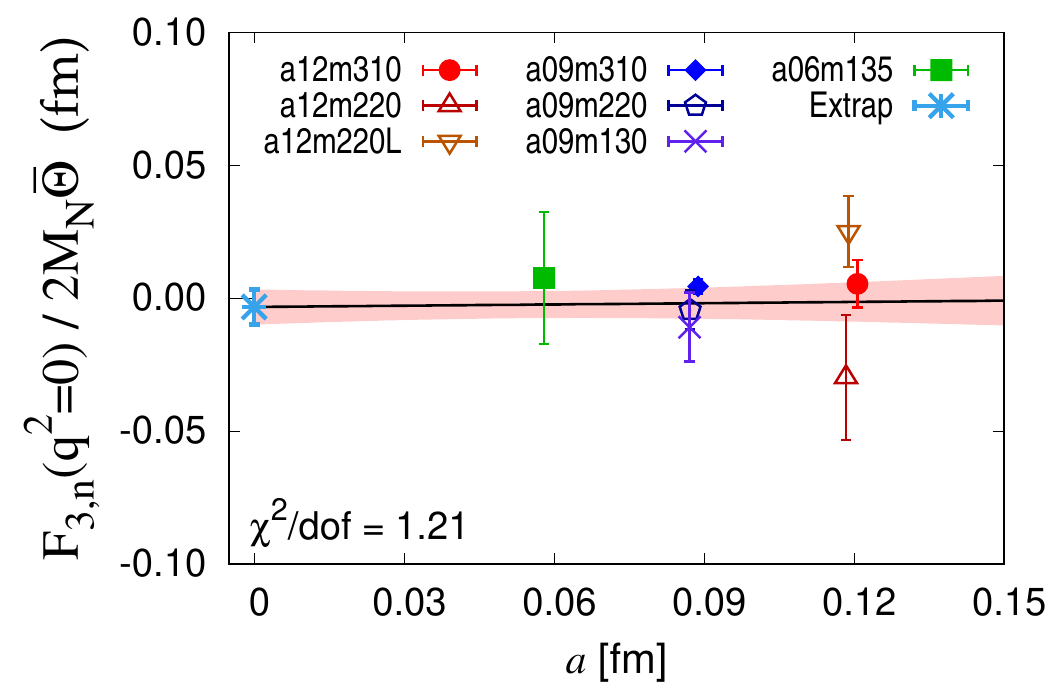}
    \includegraphics[width=0.425\linewidth]{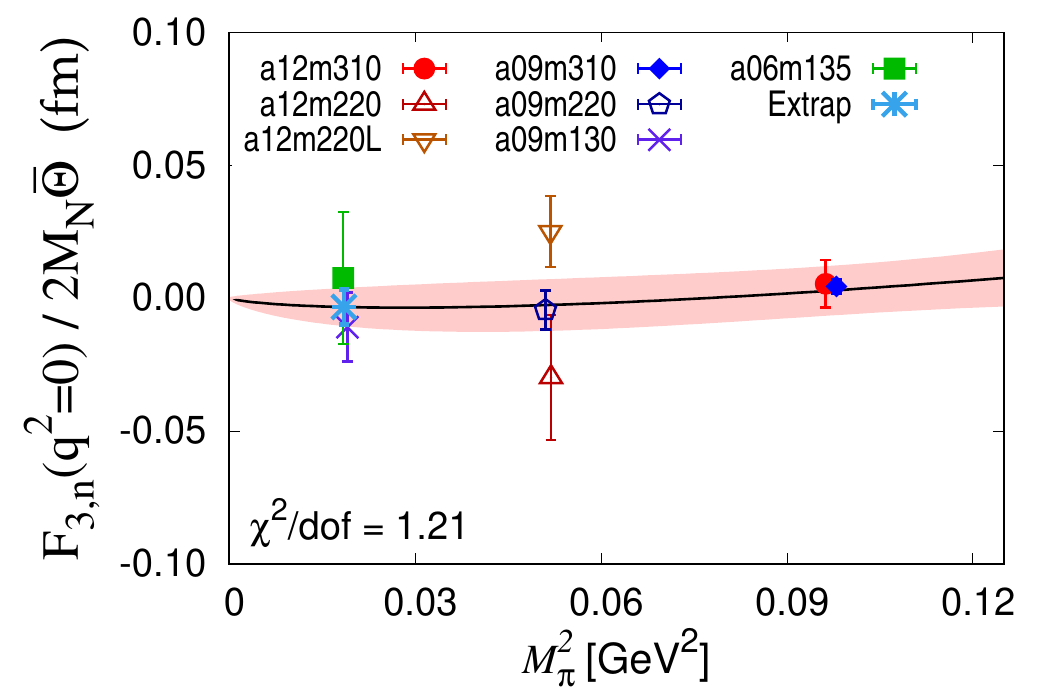}
  }
\caption{The chiral-continuum extrapolation of $d_n$ using the ansatz given in Eq.~\eqref{eq:CCfit}. 
The four rows show (i) a linear CC fit to the data obtained using a
linear extrapolation in $q^2$ discussed in Sec.~\protect\ref{sec:qsq};
(ii) a linear CC fit to the data obtained using the $\chi$PT
extrapolation in $q^2$; (iii) a $\chi$PT CC fit to the data obtained
using a linear extrapolation in $q^2$; and (iv) a $\chi$PT CC fit to
the data obtained using the $\chi$PT extrapolation in $q^2$. All data 
are with ${\overline \Theta}=0.2$.
\label{fig:CCn}}
\end{figure*}

\begin{figure*}[tbp]
  \subfigure{
    \includegraphics[width=0.425\linewidth]{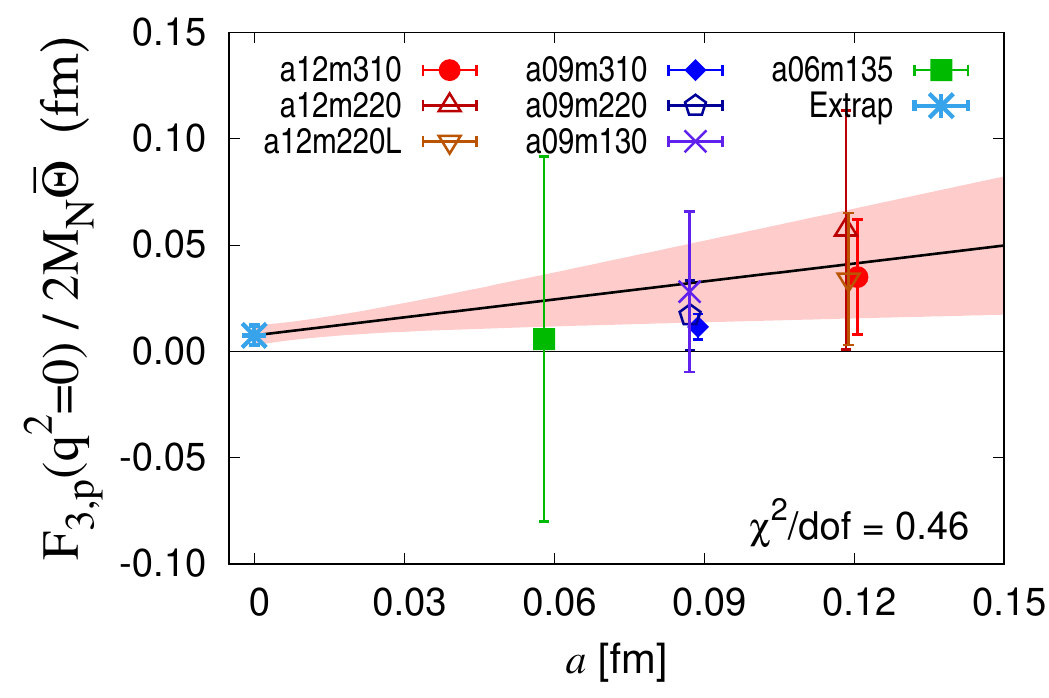}
    \includegraphics[width=0.425\linewidth]{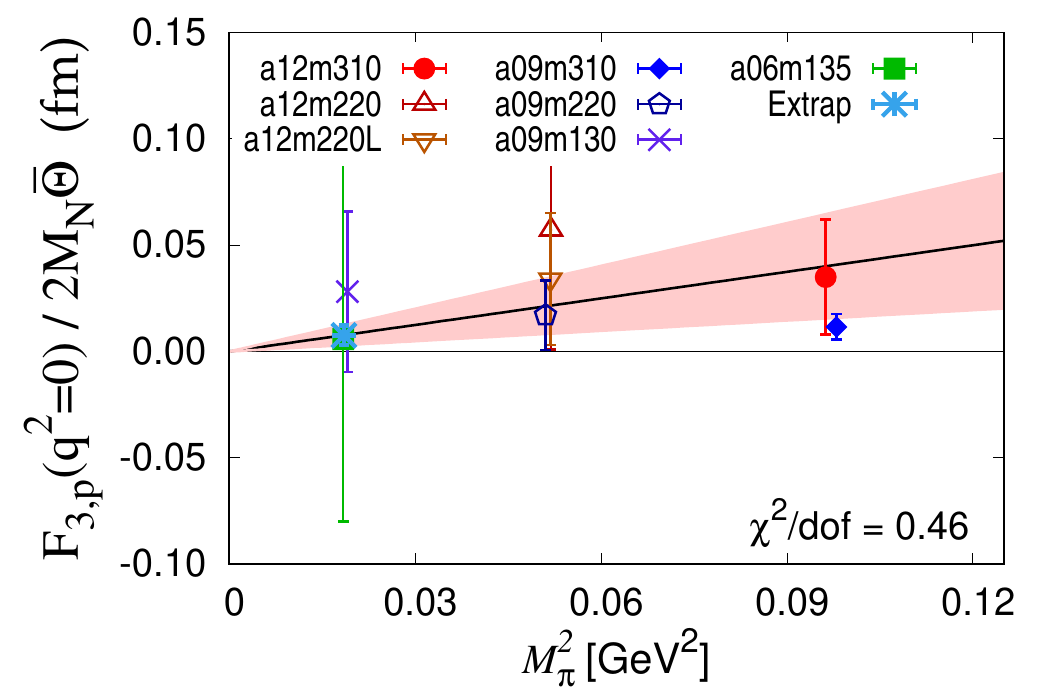}
  }
  \subfigure{
    \includegraphics[width=0.425\linewidth]{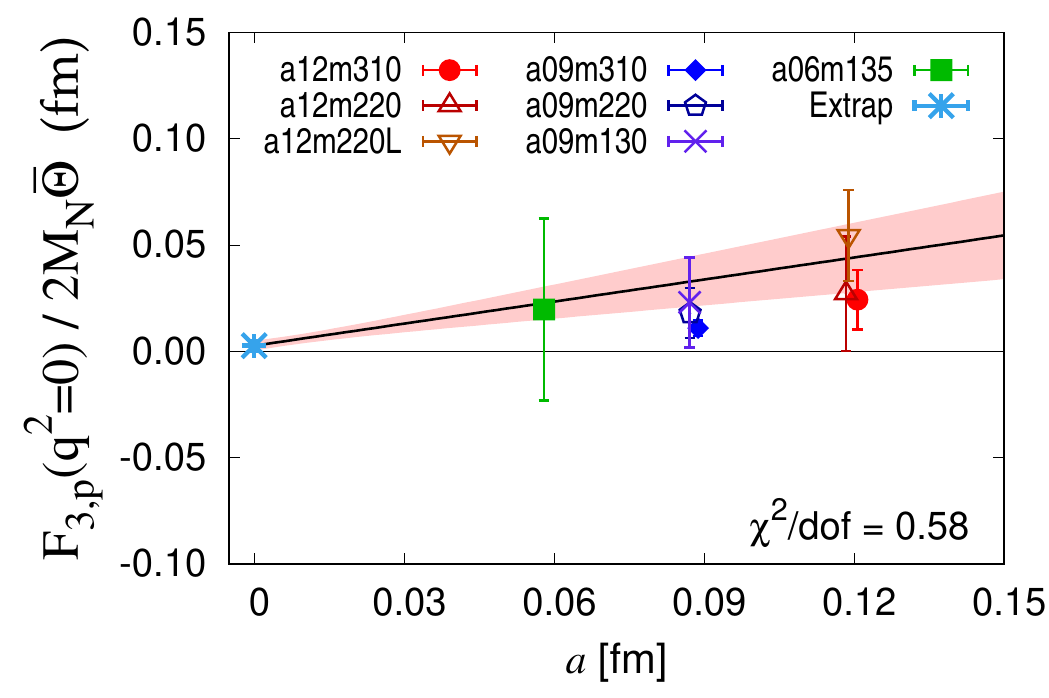}
    \includegraphics[width=0.425\linewidth]{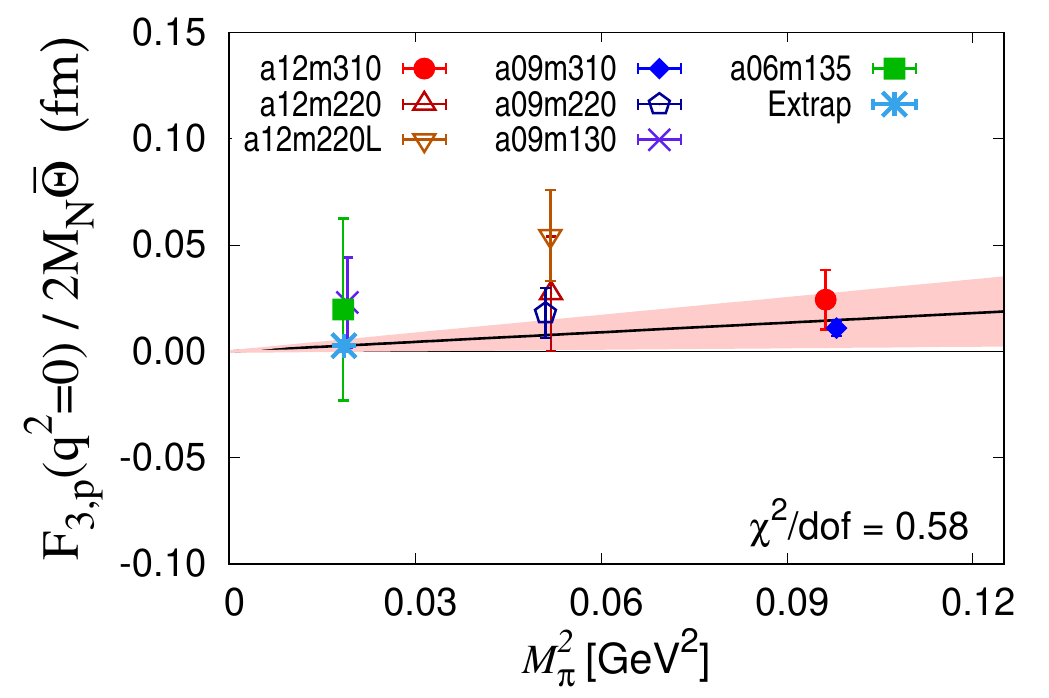}
  }
  \subfigure{
    \includegraphics[width=0.425\linewidth]{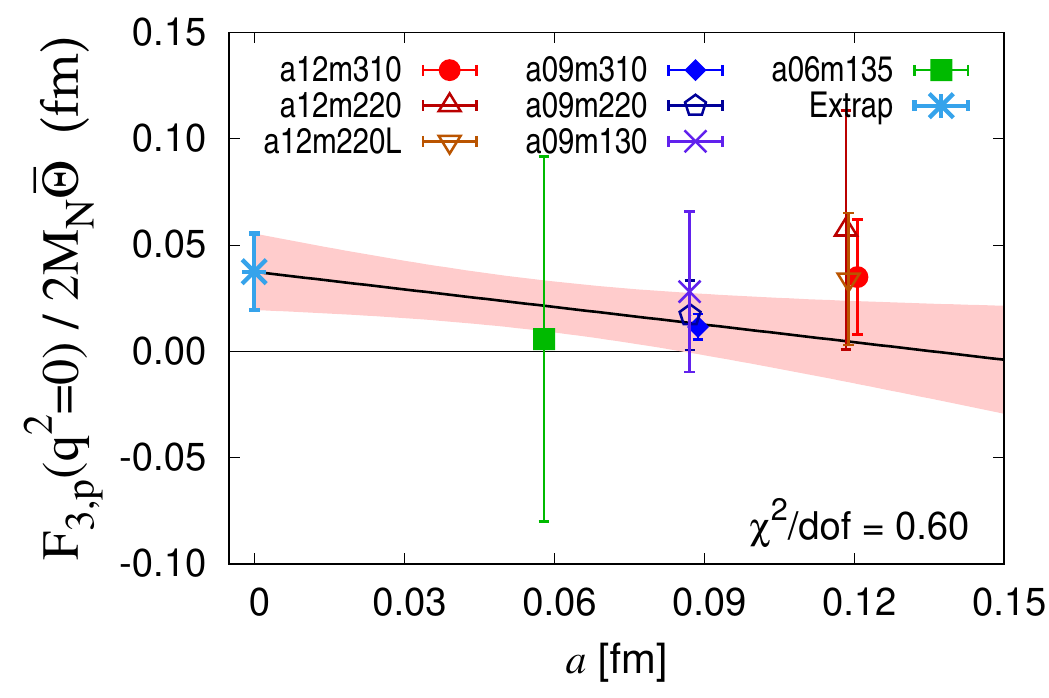}
    \includegraphics[width=0.425\linewidth]{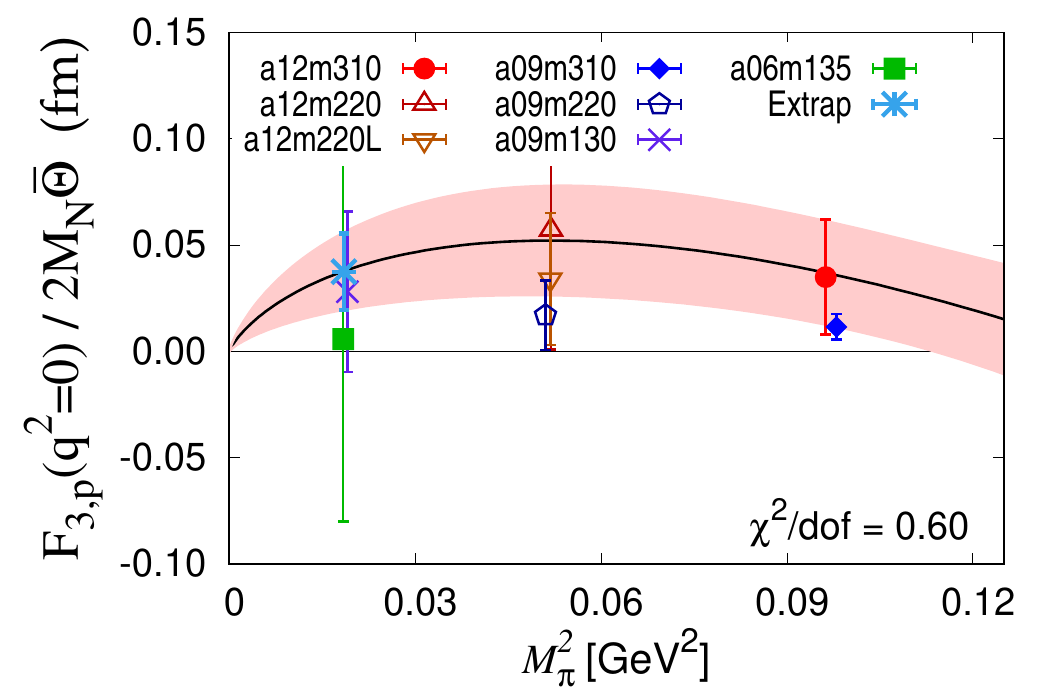}
  }
  \subfigure{
    \includegraphics[width=0.425\linewidth]{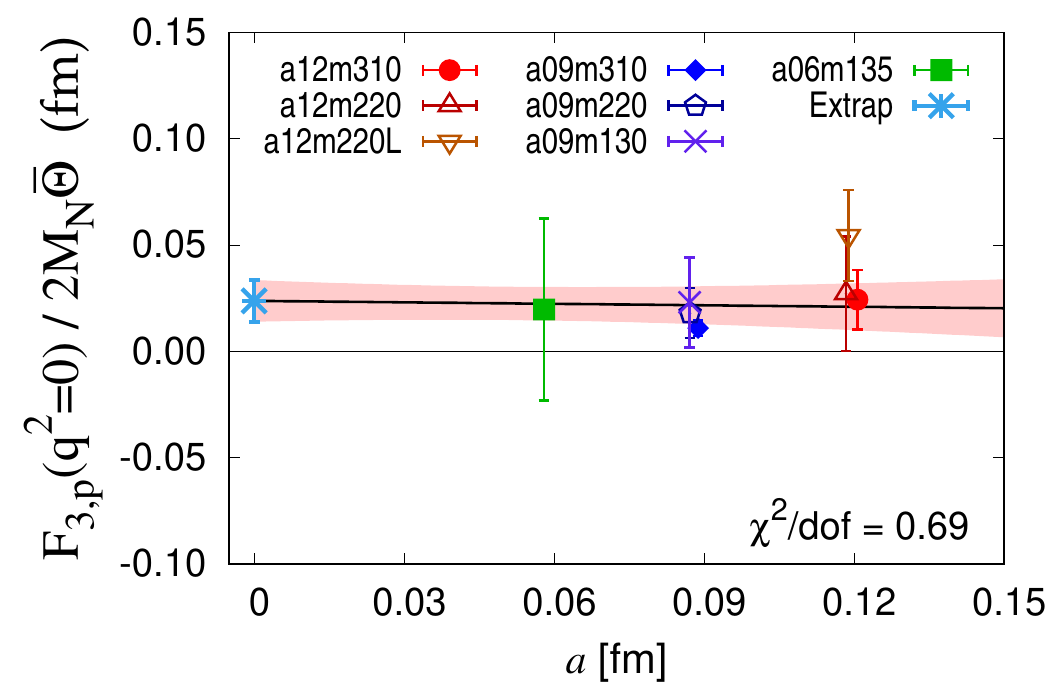}
    \includegraphics[width=0.425\linewidth]{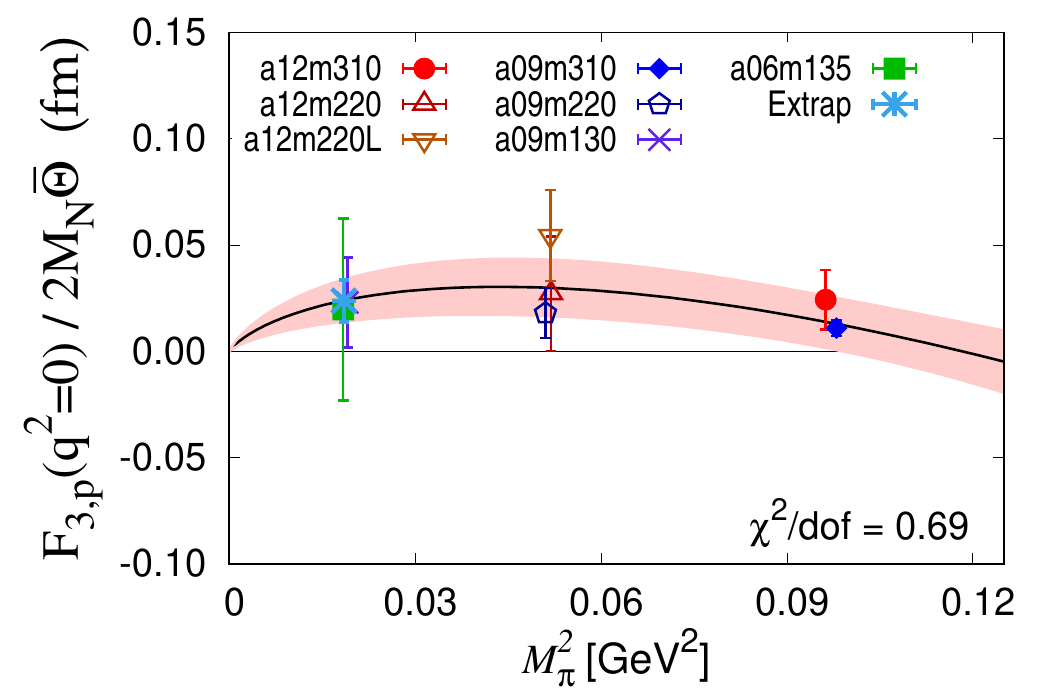}
  }
\caption{The chiral-continuum extrapolation of $d_p$ using the ansatz given in Eq.~\eqref{eq:CCfit}. 
The rest is the same as in Fig.~\protect\ref{fig:CCn}. 
\label{fig:CCp}}
\end{figure*}

In this section, we present the chiral-continuum (CC) extrapolation of data
for $d_n$ (and, similarly, $d_p$) obtained on the seven ensembles. For each, we examine 
four cases. These consist of two CC fits, Linear and $\chi$PT,  using the leading order 
terms 
\begin{eqnarray}
d_n(a,M_\pi) &=& c_1 M_\pi^2 + c_2 a M_\pi^2 + c_3 a  
\label{eq:CCfit0}
\\
d_n(a,M_\pi) &=& c_1 M_\pi^2 + c_{2L}  M_\pi^2 \ln\left(\frac{M_\pi^2}{M_N^2}\right) + c_3 a   \,,
\label{eq:CCfit}
\end{eqnarray}
where the term $c_3 a $ is the $O(a)$ effect discussed in Sec.~\ref{sec:Oa},
because of which $d_{n,p}$ do not vanish in the chiral limit at finite
$a$. The ansatz are distinguished by the terms proportional to $c_2$
(Linear) and $c_{2L}$ ($\chi$PT).  In these fits, $M_N$ is set to its
physical value $940$~MeV. We make these two fits to the data for
$d_{n,p}$ obtained using (i) the linear and (ii) $\chi$PT
extrapolation in $q^2$, which leads to four estimates. 
These four CC fits for the neutron and the proton are shown in
Figs.~\ref{fig:CCn} and~\ref{fig:CCp}.  The results and the fit coefficients $c_i$ are
given in Table~\ref{tab:CCresults}. 

As discussed in Appendix~\ref{chiral}, at NLO in $\chi$PT 
the coefficient of the chiral logarithm $c_{2L}$ is fixed in terms of the isovector scalar charge, the quark condensate and the pion decay constant, leading to $(c_{2L})_n = - (c_{2L})_p = 0.033$ fm-GeV$^2$. 
Although the central values of the fits are approximately one order of magnitude larger, our results are compatible with this estimate at the 1$\sigma$--2$\sigma$ level. 
 
For the central value we take the
$\chi$PT($q^2$)$|\chi$PT(CC) result and the full spread between the four for the error. 
The final results, using the definition in Eq.~\eqref{eq:dndef},  are 
\begin{eqnarray}
d_n  &= -0.003(7)(20) {\overline \Theta} \  e \cdot {\rm fm} \\
d_p  &= 0.024(10)(30) {\overline \Theta} \ e \cdot {\rm fm} 
\label{eq:Final}
\end{eqnarray}
where the second systematic error is the spread in the four estimates given in Table~\ref{tab:CCresults}.

%

\begin{table*}  
  \begin{ruledtabular}
    \begin{tabular}{r|cc|cccc}
\multicolumn{7}{c}{Neutron}\\
\hline
Fit types                      &  $F_3/2M_N$  & $\chi^2$/dof & $c_1$        & $c_2$       & $c_{2L}$    & $c_3$     \\
                               &  (fm)        &              & fm-GeV${}^2$ & GeV${}^2$   & fm-GeV${}^2$   &           \\
                                                                                                  
Linear($q^2)|$Linear(CC)       &  $-$0.0044(36) & 0.804   &  $-$0.24(20) & 3.1(2.3)    &             &   0.02(16) \\
                                                                                             
Linear($q^2)|$$\chi$PT(CC)     &  $-$0.018(13)  & 0.782   &     0.76(62) &             & 0.45(33)    &   0.31(18) \\
                                                                                             
$\chi$PT($q^2)|$Linear(CC)     &   0.0005(17) & 1.213   &    0.028(92) &  0.8(1.2)   &             & $-$0.06(11) \\
                                                                                             
$\chi$PT($q^2)|$$\chi$PT(CC)   &  $-$0.0032(66) & 1.212   &     0.30(38) &             & 0.12(19)    &   0.016(81) \\
\hline
\multicolumn{7}{c}{Proton}\\
\hline
Linear($q^2)|$Linear(CC)       &  0.0076(46) & 0.455    &    0.42(25)  & $-$7.6(3.4) &             &    0.42(26) \\
                                                                                                   
Linear($q^2)|$$\chi$PT(CC)     &   0.037(18) & 0.597    & $-$1.84(97)  &             & $-$1.01(49) & $-$0.28(24) \\
                                                                                                   
$\chi$PT($q^2)|$Linear(CC)     &  0.0027(23) & 0.578    &    0.15(13)  & $-$4.8(1.9) &             &    0.43(17) \\
                                                                                                   
$\chi$PT($q^2)|$$\chi$PT(CC)   &  0.0238(98) & 0.687    & $-$1.40(58)  &             & $-$0.70(28) & $-$0.02(11) \\
\hline
\multicolumn{7}{c}{Neutron (with $N\pi$ excited state)}\\
\hline
Linear($q^2)|$Linear(CC)       & $-$0.0046(87) & 1.402    &  $-$0.25(48)   & 10.6(7.8)   &             & $-$0.79(70) \\
                                                                                     
Linear($q^2)|$$\chi$PT(CC)     & $-$0.054(37)  & 1.323    &   3.2(2.2)   &             &  1.6(1.1)   &  0.27(45) \\
                                                                                     
$\chi$PT($q^2)|$Linear(CC)     &  0.0039(42) & 2.246    &   0.22(23)   &  8.4(3.8)   &             &  $-$1.07(37) \\
                                                                                     
$\chi$PT($q^2)|$$\chi$PT(CC)   & $-$0.028(18)  & 2.430    &   2.5(1.1)   &             &  1.04(52)   & $-$0.26(20) \\
\hline
\multicolumn{7}{c}{Proton (with $N\pi$ excited state)}\\
\hline
Linear($q^2)|$Linear(CC)       &  0.019(12)  & 0.347    &  1.04(66)    & $-$29(12)     &             &   2.2(1.0) \\
                                                                                     
Linear($q^2)|$$\chi$PT(CC)     &  0.140(54)  & 0.358    & $-$7.7(3.2)    &             &   $-$4.0(1.6) & $-$0.70(66) \\
                                                                                     
$\chi$PT($q^2)|$Linear(CC)     &  0.0040(50) & 0.398    &  0.22(27)    & $-$15.7(5.4)  &             &  1.51(52) \\
                                                                                     
$\chi$PT($q^2)|$$\chi$PT(CC)   &  0.068(25)  & 0.522    & $-$4.4(1.6)    &             &  $-$2.09(75)  & $-$0.02(27) \\
\end{tabular} \vspace{5mm}
  \end{ruledtabular} \caption{Results for the contribution of the
  $\Theta$-term to $d_n$ and $d_p$ for the four fit strategies defined
  in the text. Also given are the fit parameters $c_i$ defined in
  Eqs.~\protect\eqref{eq:CCfit0}-\protect\eqref{eq:CCfit} 
  and the $\chi^2$/dof of the fit. Results are 
  given for two choices of the first excited state energy: (top) from a 
  three-state fit to the two-point function, and  (bottom) the noninteracting $N\pi$ 
  state.  \label{tab:CCresults}}
\end{table*}

\section{Analysis including the \texorpdfstring{$N\pi$}{N\textpi} excited state}
\label{sec:Npi}

In this section, we describe how all ground state quantities change
when the $N\pi$ excited state is included. This analysis should be
considered exploratory because (i) the extrapolations in the fits to
remove ESC (see Fig.~\ref{fig:ESCV4}), (ii) the errors, and (iii) the
cancellations when combining different terms to get $F_3$ using 
Eqs.~\eqref{eq:Vinear} are all large.

\begin{figure*}[tbp]
  \subfigure{
    \includegraphics[width=0.42\linewidth]{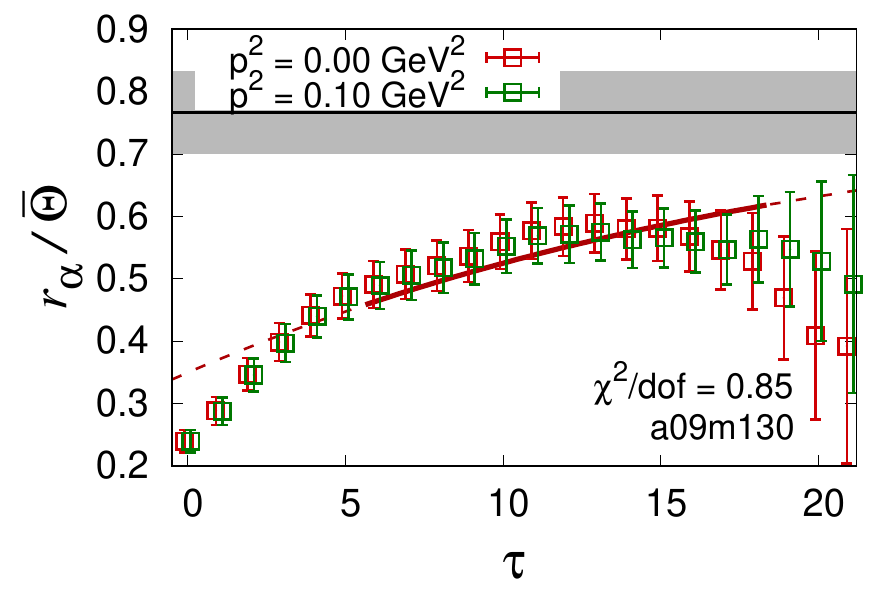}\qquad
    \includegraphics[width=0.42\linewidth]{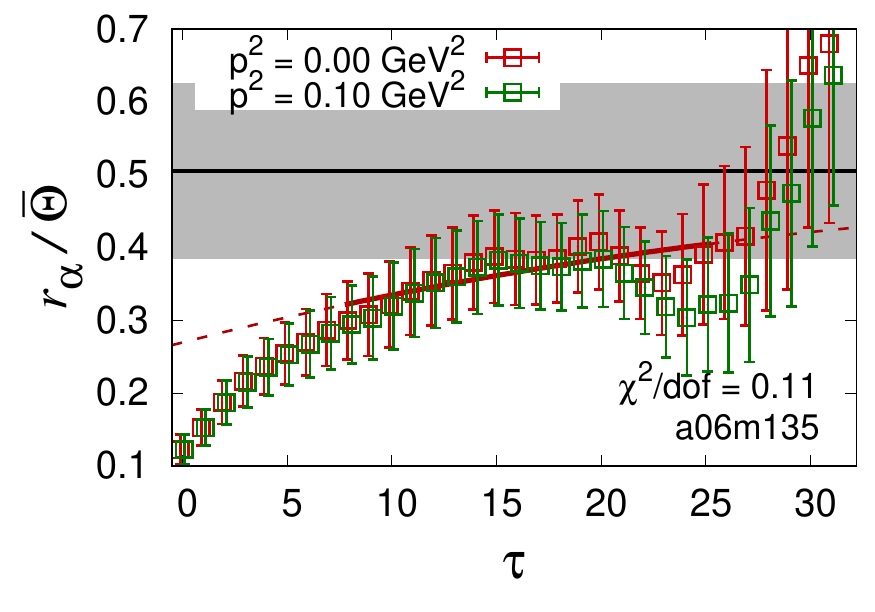}
  }
\caption{The increase in the value of $\alpha$ when fits to the two-point functions are made 
including a $N\pi$ excited state as compared to data in
Fig.~\protect\ref{fig:alpha}. }
\label{fig:Npialpha}
\end{figure*}

\begin{figure*}[tbp]
  \subfigure{
    \includegraphics[width=0.425\linewidth]{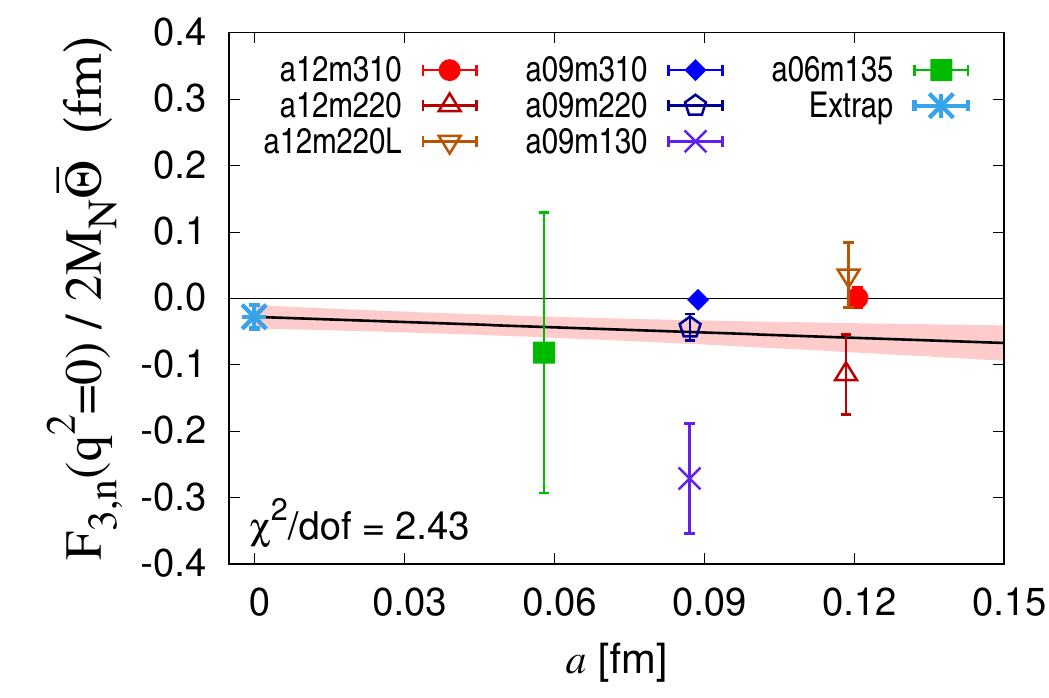}
    \includegraphics[width=0.425\linewidth]{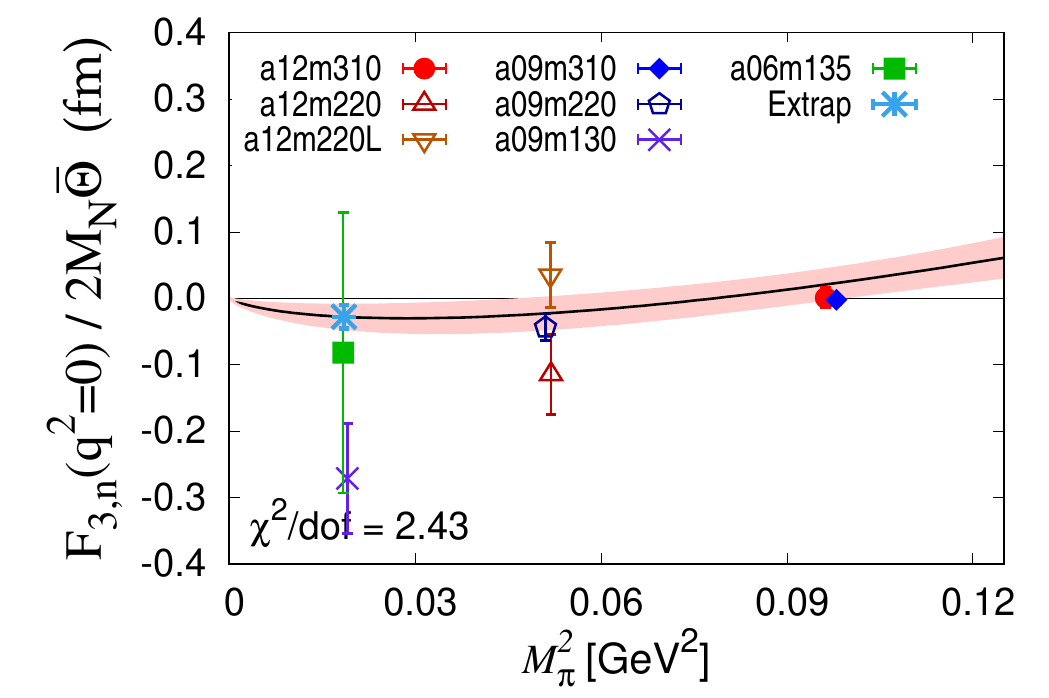}
  }
  \subfigure{
    \includegraphics[width=0.425\linewidth]{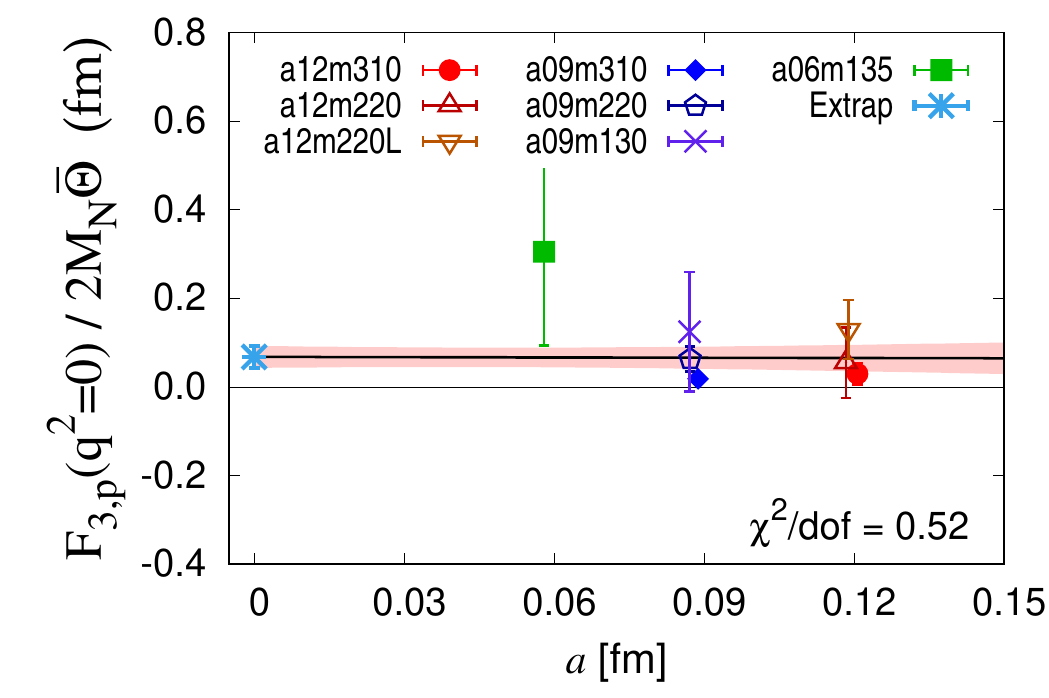}
    \includegraphics[width=0.425\linewidth]{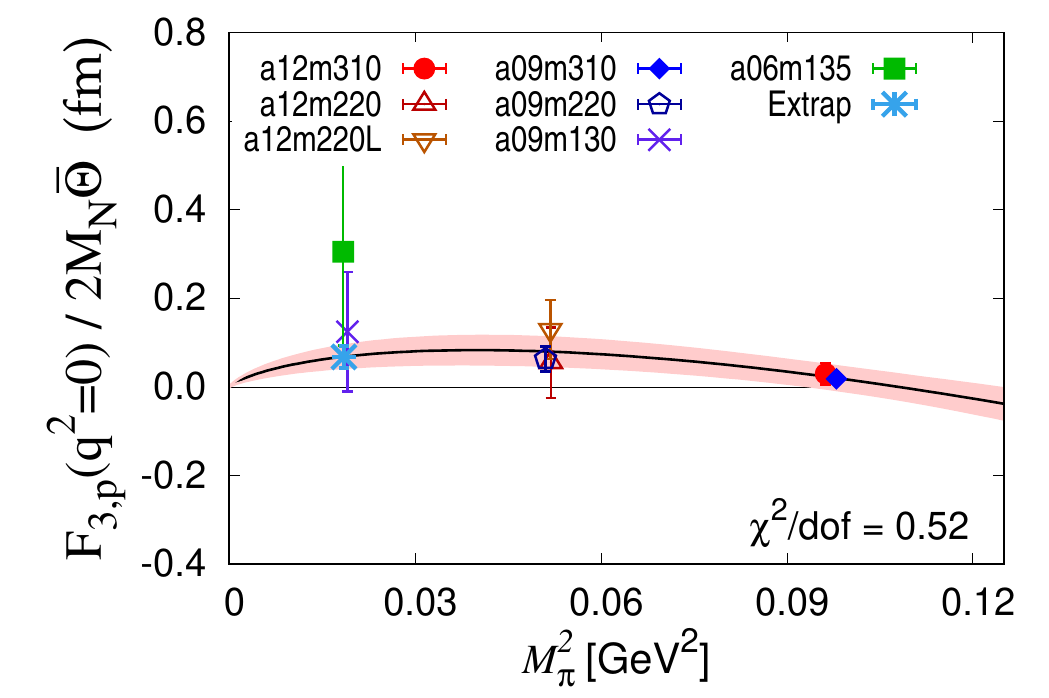}
  }
\caption{The chiral-continuum extrapolation of $d_n$ (top) and $d_p$ (bottom) 
using the ansatz given in Eq.~\eqref{eq:CCfit}, 
using $N \pi$ as the excited state fits, and with the
$\chi$PT($q^2)|$$\chi$PT(CC) strategy. All data are with
${\overline \Theta}=0.2$.
\label{fig:NpiCC}}
\end{figure*}

In Fig.~\ref{fig:Npialpha}, we show the increase in the value of $\alpha$ for the 
two physical mass ensembles as compared to the data presented in Fig.~\ref{fig:alpha}. 
The $q^2$ behavior is similar to that shown in Figs.~\ref{fig:qsqN} and~\ref{fig:qsqP}, 
and the final results for the four strategies are given in Table~\ref{tab:CCresults}. 
The CC fits for the neutron and the proton using the $\chi$PT($q^2)|$$\chi$PT(CC) strategy 
are shown in Fig.~\ref{fig:NpiCC}

For the central value we again take the
$\chi$PT($q^2$)$|\allowbreak\chi$PT(CC) result and the full spread for the error. 
This gives 
\begin{eqnarray}
d_n|_{N\pi}   &= -0.028(18)(54) {\overline \Theta} \  e \cdot {\rm fm} \\
d_p|_{N\pi}   &= 0.068(25)(120) {\overline \Theta} \ e \cdot {\rm fm} 
\label{eq:FinalNpi}
\end{eqnarray}
where the second systematic error is the spread in the four estimates given in Table~\ref{tab:CCresults}.

\section{Comparison to previous work}
\label{sec:comp}

There are two 
estimates~\cite{Dragos:2019oxn,Alexandrou:2020mds}
of the contribution of the $\Theta$-term to the nEDM since the
clarification of the impact of the phase $\alpha$ that arises in the nucleon spinor in a theory
with \CPV\ in Ref.~\cite{Abramczyk:2017oxr}. That work also contains a review of 
previous results, which after correction were consistent with zero. 
No estimate is given in  Ref.~\cite{Abramczyk:2017oxr}, but there is a preliminary value 
in a subsequent conference  proceedings, Ref.~\cite{Syritsyn:2019vvt}.
All three of these calculations use the small $\Theta$ expansion and gradient flow 
method for topological charge renormalization as in this work. All results are summarized in Table~\ref{tab:dnSummary}.

The ETM collaboration~\cite{Alexandrou:2020mds} has performed the
calculation on one 2+1+1-flavor twisted mass clover-improved ensemble
with $a = 0.0801(4)$~fm, $M_\pi = 139(1)$~MeV, $M_\pi L= 3.62$.  Data
are presented for a single value of $\tau=12$ so there is no
information on excited state effects, continuum extrapolation, chiral
behavior, or finite-size effects. They also implicitly implement the
$\overline{\Theta}=0$ subtraction (see Eqs.~\eqref{eq:Vinear}) that we find
reduces the statistical noise by using the spin projector
$(1+\gamma_4)i \gamma_5 \gamma_k/4$. They determine $F_3(0)$ by making a 
constant fit to the lowest three $q^2$ points. Their final result is taken using
the spectral projectors method, which they find reduces the errors by a
factor of about two compared to the field-theoretic definition of the
topological charge used in this work. They do not, however, assess a systematic error 
associated with excited-state effects, extrapolation in $q^2$, or the 
chiral-continuum fit. 

The calculation presented in Ref.~\cite{Dragos:2019oxn} uses six
2+1-flavor Wilson-Clover ensembles but only one below $M_\pi = 567
$~MeV, with $M_\pi = 410$~MeV. The values of lattice spacings range
between $ 0.068 < a < 0.11$~fm.  
A linear fit in $q^2$ is made to obtain $F_3(0)$. 
 Also, artifacts due to ESC are not analyzed and, in any case, 
data with the heavy pion masses studied, $M_\pi > 410$~MeV,  would not provide 
sensitivity to analyses with or without including a $N \pi$ state.  
This is the only other calculation that has presented a chiral extrapolation
using the $\chi$PT ansatz (Eq.~\eqref{eq:CCfit} but with a \(O(a^2)\)
discretization correction instead of our $c_3
 a$ term). As shown in the bottom right panels in Figs.~\ref{fig:CCn}
and~\ref{fig:CCp}, such chiral fits have an inflection point close to
the smallest $M_\pi$ data point in order to satisfy the constraint $F_3 = 0$ at
$M_\pi = 0$.  In the case of Ref.~\cite{Dragos:2019oxn}, this occurs
around $M_\pi = 400$~MeV, raising questions on the reliability of the
extrapolation.

\begin{table*}  
  \begin{ruledtabular}
    \begin{tabular}{l|l|l}
                                   &  Neutron                         &   Proton      \\
                                   &  ${\overline \Theta}\ {\rm e \cdot fm}$ & ${\overline \Theta}\ {\rm e \cdot fm}$   \\
\hline                                                                                                  
This Work                          & $d_n = -0.003(7)(20)$            &  $d_p = 0.024(10)(30)$                 \\
                                                                               
This Work with $N \pi$             & $d_n = -0.028(18)(54)$           &  $d_p = 0.068(25)(120)$                 \\
                                                                               
ETMC~\cite{Alexandrou:2020mds}     & $|d_n| = 0.0009(24)$             &     --                                  \\
                                                                               
Dragos et al.~\cite{Dragos:2019oxn}       & $ d_n = -0.00152(71)$            &  $ d_p = 0.0011(10)$                 \\
                                                                               
Syritsyn et al.~\cite{Syritsyn:2019vvt}   & $d_n \approx 0.001$              &     --                                    \\
                                                                               
\end{tabular} 
  \end{ruledtabular} \caption{Summary of lattice results for the
  contribution of the $\Theta$-term to the neutron and proton electric
  dipole moment.  \label{tab:dnSummary}}
\end{table*}

\section{Conclusions}
\label{sec:Conclusions}

This paper presents a calculation of the contribution of the
$\Theta$-term to the nucleon electric dipole moment using 2+1+1-flavor
HISQ ensembles and Wilson-clover valence quarks. Two of the seven
ensembles analyzed are at the physical pion mass, which anchor our
chiral fits. The calculation has been done using the small $\Theta$
expansion method. Significant effort has been devoted to getting a
reliable signal in the \CPV\ violating form factor $F_3$.  The
gradient flow scheme has been used to renormalize the $\Theta$-term
and the results are shown to be independent of the flow time. Our
estimate of the topological susceptibility for the 2+1+1 theory is
$\chi_Q = (66(9)(4)~{\rm MeV})^4$ in the continuum limit at $M_\pi =
135$~MeV.

We also present two technical issues. First, in
Appendix~\ref{sec:ESCappendix}, we show that, in chiral perturbation
theory, the $N \pi$ excited state should provide the dominant
contamination.  We have, therefore, used two strategies for removing
excited state contamination. In the first, the mass gaps are taken from
fits to the spectral decomposition of the nucleon two-point function,
and in the second we assume they are given by the non-interacting
energy of the $N(\bm 0) \pi (\bm 0)$ state. We find a very significant
difference between the two as shown in Secs.~\ref{sec:ESC}
and~\ref{sec:Npi}, and by the results summarized in
Tables~\ref{tab:CCresults} and~\ref{tab:dnSummary}.

The second technical issue discussed in Sec.~\ref{sec:Oa} and appendix~\ref{sec:appendix2} is that
lattice artifacts introduce a term proportional to $a m_q^0$, because
of which $d_n$ does not vanish in the chiral limit at finite $a$. Our
chiral-continuum fits have been made including this term.

The analysis of the $q^2$ dependence of $F_3^{\Theta}$ has been
carried out using both a linear and the leading order $\chi$PT expression
as described in Sec.~\ref{sec:qsq}.  The current data do not
distinguish between the two. Similarly, the chiral fit is also carried
out using a linear and the leading order $\chi$PT expression as
described in Sec.~\ref{sec:Results}. The results from these four sets
of fits and the two strategies to remove excited-state contributions
are summarized in Table~\ref{tab:CCresults}.

Our preferred values are obtained using the leading order $\chi$PT expressions. 
The analysis using excited states from fits to the two-point function 
indicate that $d_n^\Theta$ is
small, $|d_n^\Theta| \lesssim 0.01\ {\overline \Theta}\ {\rm e \cdot
fm}$, whereas for the proton we get $|d_p^\Theta| \sim 0.02\ {\overline \Theta}\ {\rm e \cdot fm} $. On the other hand, if the dominant excited-state contribution is
from the $N \pi$ state, then $|d_n^\Theta|$ could be as large as $0.05\
{\overline \Theta}\ {\rm e \cdot fm}$ and  $|d_p^\Theta| \sim 0.07\ {\overline \Theta}\ {\rm e \cdot fm} $. 
Lastly, we find  the sign of $d_p^\Theta$ to be opposite to that of $d_n^\Theta$. 

From the final summary of results presented in
Table~\ref{tab:dnSummary}, which also includes estimates from previous
works, it is clear that, at present, lattice calculations do not  yet provide a reliable estimate. To improve the current 100\%  uncertainty to a $3\sigma$ result will require a factor of at least ten  improvement in statistics.

\begin{acknowledgments}
The calculations used the Chroma software
suite~\cite{Edwards:2004sx}. This research used resources at (i) the
National Energy Research Scientific Computing Center, a DOE Office of
Science User Facility supported by the Office of Science of the
U.S. Department of Energy under Contract No. DE-AC02-05CH11231; (ii)
the Oak Ridge Leadership Computing Facility, which is a DOE Office of
Science User Facility supported under Contract DE-AC05-00OR22725;
(iii) the USQCD Collaboration, which are funded by the Office of
Science of the U.S. Department of Energy, and (iv) Institutional
Computing at Los Alamos National Laboratory.  T. Bhattacharya and
R. Gupta were partly supported by the U.S. Department of Energy,
Office of Science, Office of High Energy Physics under Contract
No.~DE-AC52-06NA25396.  We acknowledge support from the
U.S. Department of Energy, Office of Science, Office of Advanced
Scientific Computing Research and Office of Nuclear Physics,
Scientific Discovery through Advanced Computing (SciDAC) program, and
of the U.S. Department of Energy Exascale Computing Project.
T. Bhattacharya, V. Cirigliano, R. Gupta, E. Mereghetti and B.Yoon
were partly supported by the LANL LDRD program.
\end{acknowledgments}

\appendix
\section{Connection between Minkowski and Euclidean notations}
\label{sec:appendix0}

To make our conventions explicit, we present the connection between Minkowski and Euclidean variables in Table~\ref{tab:E2Mconnect}.

\begin{table*}  
  \begin{ruledtabular}
    \begin{tabular}{l|l|l}
Quantity                       &  Minkowsky $\leftrightarrow $ Euclidean       &   Remarks      \\
\hline                                                                                                  
4-vector $v^\mu$               &  $v_M^0 = v_{M0} = -iv_E^4 = -iv_{E4} $       &  Ensures $v_M \cdot v_M^\prime = -v_E \cdot v_E^\prime $;\\ 
                                                                               
                               &  $v_M^i = -v_{Mi} = v_E^i = iv_{Ei  } $       & In particular, \(v^2_M = - v^2_E\). \\
                                                                               
                               &  $t \equiv x^0_M = -ix_E^4 \equiv -i\tau$     &                 \\
                                                                               
                               &  $p^0_M = -ip_E^4 = E$                        &                 \\
                                                                               
\hline                                                                                                  
Derivatives                    &  $\partial_0^M = i\partial_4^E $       &  $ \partial_\mu = \partial/\partial x^\mu$ and   $ \partial^\mu = \partial/\partial x_\mu$ in both E and M       \\
                                                                               
                               &  $\partial^{Mi} = -\partial^M_i = -\partial_i^E = -\partial^{Ei} $  &  \\
                                                                               
\hline                                                                                                  
Gauge Fields                   &  $A^M_0 = iA_4^E $       & \(D_\mu = \partial_\mu - A_\mu\) transforms homogeneously \\
                                                                               
                               &  $A^{Mi} = -  A^M_i = -A^E_i = -A^{Ei}$  &    \\
                               &  $(G_M)^{0i} = -i (G_E)^{4i}$, $(G_M)_{0i} =  i (G_E)_{4i}$  &               \\
                                                                               
                               &  $(G_M)^{ij} =  (G_E)^{ij}$, $(G_M)_{ij} =  (G_E)_{ij}$  &               \\
                                                                               
\hline                                                                                                  
$\gamma$ matrices              &  $\gamma_E^4 = \gamma_M^0$, $\gamma_E^i = - i \gamma_M^i$   &   We adopt the DeGrand-Rossi basis~\cite{DeGrand:1990dk}. These  \\
                               &  $\gamma_E^5 = \gamma_E^1 \gamma_E^2 \gamma_E^3 \gamma_E^4 = -i \gamma_M^0 \gamma_M^1 \gamma_M^2 \gamma_M^3 = -\gamma_M^5 = - \gamma_c^5$  &           Euclidean gamma matrices are Hermitean.      \\
                               & & \\
                               &  \(\gamma_M^\mu \equiv \gamma_{c1} \gamma_{c3} \gamma_c^\mu \gamma^3_c \gamma^1_c \)  &                                Minkowski gamma matrices are unitarily transformed       \\

                               &               & from the standard chiral basis, \(\gamma^\mu_c\) ~\cite{Peskin:1995ev}      \\
                               &  $\slashed p_M = -i \slashed p_E $       &  \\
                               &  $\slashed D_M =  i \slashed D_E$        &  \(\psi_M = \gamma_{c1}\gamma_{c3}\psi_c\) and \(\bar\psi_M = \bar\psi_c\gamma_c^3\gamma_c^1\).\\ 
\hline                                                                                                  
Charge                         &  $C_M = i \gamma^M_{0}\gamma^M_{2} $       &                 \\
Conjugation                    &  $C_c = i \gamma^c_{0}\gamma^c_{2} $       &                 \\
Matrix                         &  $C_E =   \gamma^E_{2}\gamma^E_{4} $       &                 \\
                                                                               
\end{tabular} 
  \end{ruledtabular} 
  \caption{Connection between Euclidean and Minkowsky variables.
 \label{tab:E2Mconnect}} 
\end{table*}

To connect the Lagrangian density for the $\Theta$ term in Minkowski
and Euclidean spaces, we take the Minkowski action associated with the
QCD $\Theta$ term to be
\begin{eqnarray}
S_\Theta^M &=& 
-    \frac{\Theta}{32 \pi^2}    \int d^4x \  (G_M)^{a \, \mu \nu} (x)  \,  (\tilde G_M)^{a}_{\mu \nu} (x)
\ \ 
\end{eqnarray}
where $ (\tilde G_M)^{a}_{\mu \nu} = (1/2)  \epsilon_M^{\mu \nu \alpha \beta} (G_M)^a_{\alpha \beta}$ 
and  $(\epsilon_M)_{0123}= +1 = -\epsilon_M ^{0123}$. 
Upon rotating  to the Euclidean space one gets  $d^4 x_M = -i d^4x_E$ 
and 
\begin{equation}
\epsilon_M^{\mu \nu \alpha \beta}    (G_M)^{a}_{\mu \nu} \,  (G_M)^{a}_{\alpha \beta}  = 
i (\epsilon_E)_{\mu \nu \alpha \beta}    (G_E)^{a}_{\mu \nu} \,  (G_E)^{a}_{\alpha \beta} .
\end{equation}
The factor of $+i$ arises  from the transformation 
of the field strength and 
because each term in the sum has one 
factor of $G_{0i}$ (or $G_{i0}$) and one factor of $G_{jk}$. 
Moreover, we used
\begin{equation}
\epsilon_M^{0 ijk} \equiv  (\epsilon_E)_{4ijk}  =  (\epsilon_E)^{4ijk} \,,
\end{equation} 
which implies 
\begin{equation}
 (\epsilon_E)_{ijk4}  
= 
- (\epsilon_M)^{0 ijk} 
\end{equation} 
and hence $(\epsilon_E)_{1234} = +1$.

Putting together the change in the measure and the change in the Lagrangian density we have
\begin{equation}
S_\Theta^M =  -   \frac{\Theta}{64 \pi^2}  \epsilon_E^{\mu \nu \alpha \beta}   \int d^4x_E \  (G_E)^{a}_{\mu \nu} (x) \, (G_E)^{a}_{\alpha \beta} (x),  
\end{equation}
and hence  ($i S^M = - S^E$)
\begin{equation}
S_\Theta^E=  +  i   \frac{\Theta}{64 \pi^2}  \epsilon_E^{\mu \nu \alpha \beta}   \int d^4x_E \  (G_E)^{a}_{\mu \nu} (x) \, (G_E)^{a}_{\alpha \beta} (x),
\end{equation}
consistently with Eq.~(\ref{eq:Lcpv}).

\section{Extraction of \texorpdfstring{$F_3$}{F\unichar{"2083}}}
\label{sec:appendix1}

The Euclidean four-vector ${\cal V}^\mu ({\bf q})$  defined in Eq.~\eqref{eq:V1t} can be determined from lattice data  
by taking appropriate ratios of 3-pt function and 2-pt functions. 
This is achieved by defining the projected 2- and 3-point functions as follows, 
\begin{eqnarray}
{\cal C}_{2pt} (t,\bm{p})  &=& \Tr
\left[ {\cal P}_{2pt}
\langle\Omega|N(\bm p,t)\bar N(\bm p,0)|\Omega \rangle 
 \right]
\\
{\cal C}_{3pt}^\mu (\tau, t, \bm{q}) &=&  \Tr
\left[ {\cal P}_{3pt} \langle \Omega | {N(\bm{p}^\prime,\tau) J^{\rm EM \, 
\mu}(t) \bar N(\bm p,0)} | \Omega \rangle \right],\nonumber\\
\label{eq:C3pt}
\end{eqnarray}%
with $\bm{q} = \bm{p}^\prime - \bm p$, $\bm{p}^\prime=0$, 
${\cal P}_{3pt}$ given in Eq.~\eqref{eq:projection}, 
\begin{eqnarray}
{\cal P}_{2pt} &=& \frac12 (1+\gamma_4) \,, 
\end{eqnarray}%
and, neglecting the contributions of heavier quarks,
\begin{equation}
J^{\rm EM}_\mu = e \Big(  (2/3) \bar{u} \gamma_\mu u 
-  (1/3) \bar{d} \gamma_\mu d-  (1/3) \bar{s} \gamma_\mu s \Big)\,.
\end{equation}
The ratio 
\begin{eqnarray}
\tilde R^\mu &\equiv& \frac{{\cal C}^\mu_{3pt} (\tau, t, \bm q)}{{\cal C}_{2pt}(\tau,\bm{p}^\prime) }
               \times {}\nonumber\\
      &&\qquad 
\left( 
\frac{
{\cal C}_{2pt} (t, \bm p')  {\cal C}_{2pt} (\tau, \bm p') {\cal C}_{2pt} (\tau-t, \bm p) 
}{{\cal C}_{2pt} (t, \bm p)  {\cal C}_{2pt} (\tau, \bm p) {\cal C}_{2pt} (\tau-t, \bm p') }
\right)^{1/2}
\label{eq:Rmu}
\end{eqnarray}
becomes independent of $t$ and $\tau$ if 
$t, \tau$ are sufficiently large that excited state effects can be neglected, 
and takes the form
\begin{equation}
\frac{{\cal V}^\mu (\bm{q}) }{\sqrt{E_p E_{p'} (E_p + M_N \cos (2 \alpha_N)) (E_{p'} + M_N \cos (2 \alpha_N)  )}} 
\,.
\label{eq:Rmu2}
\end{equation}
In our plots to demonstrate the signal and excited states, we, therefore, choose to show the quantity
\begin{align}
& R^\mu (\tau, t, \bm{q})  \equiv \frac{\tilde R^\mu}{g_V} \times
\nonumber \\
 & {\sqrt{E_p E_{p'} (E_p + M_N \cos (2 \alpha_N)) (E_{p'} + M_N \cos (2 \alpha_N)  )}} \,,
 \label{Rmu}
\end{align}
where \(g_V \equiv {\cal C}_{3pt}^\mu (\tau, t, {\bm 0})/{\cal C}_{2pt} (t,\bm{0})\), and \(\alpha_N\) is calculated from fits
to the 2-pt functions with momentum \(p\) or \(p'\) as discussed in Section~\ref{sec:alpha}.

The components of ${\cal V}_\mu$  are expressed in terms of 
form factors $F_{1,2,3,A} (q^2)$ defined in Eq.~\eqref{eq:FFdef} 
as follows:
\begin{widetext}
\begin{subequations}
\label{eq:components}
\begin{eqnarray}
  {\cal V}_1  
  &=&
     i c_{\alpha_N} M_N (q_2 +i q_1) F_1(q^2)\nonumber\\
    && {} + \left\{- c_{\alpha_N}  M_N q_2 - \frac {1}{2} [s_{\alpha_N}  q_1q_3 +
                                    i c_{\alpha_N} q_1(E_N-M_N)]\right\} F_2(q^2) \nonumber\\
    &&  {} - 2 i\left[ s_{\alpha_N} q_2 (E_N - m_N) -
       c_{\alpha_N} q_1 q_3\right] F_A(q^2) \nonumber\\
    && {} - \frac12
       \left[c_{\alpha_N} q_1q_3  -i s_{\alpha_N}q_1(E_N-M_N)\right] F_3(q^2)\,,\\
  %
 {\cal V}_2   
  &=&
     c_{\alpha_N} M_N (q_1 + i q_2) F_1(q^2)\nonumber\\
    && {} + \left\{c_{\alpha_N}  M_N q_1  - \frac {1}{2} [s_{\alpha_N}  q_2q_3 +
                                   i  c_{\alpha_N} q_2(E_N-M_N)]\right\} F_2(q^2) \nonumber\\
    &&  {} + 2i \left[ s_{\alpha_N} q_1 (E_N - m_N) +
       c_{\alpha_N} q_2 q_3\right] F_A(q^2) \nonumber\\
    && {} - \frac12
       \left[c_{\alpha_N} q_2q_3  - i s_{\alpha_N}q_2(E_N-M_N)\right] F_3(q^2)\,,\\
  %
 {\cal V}_3
&=&
     M_N [i c_{\alpha_N} q_3 + s_{\alpha_N} (E_N-M_N)] F_1(q^2)\nonumber\\
    && {} + \frac{1}{2}\left\{- i c_{\alpha_N}  (E_N-M_N) q_3 - s_{\alpha_N} q_3^2 +
                                    2 s_{\alpha_N} M_N(E_N-M_N)]\right\} F_2(q^2) \nonumber\\
    &&  {} - 2 i c_{\alpha_N} \left[q_1^2 + q_2^2\right] F_A(q^2) \nonumber\\
    && {} - \frac12
       \left[c_{\alpha_N} q_3^2  - i s_{\alpha_N}q_3(E_N-M_N)\right] F_3(q^2)\,,\\
  %
 {\cal V}_4  
&=&
    M_N [c_{\alpha_N} (E_N+M_N) - i s_{\alpha_N} q_3] F_1(q^2)\nonumber\\
    && {} -\frac12\left\{c_{\alpha_N}  (E_N^2-M_N^2) - i s_{\alpha_N} q_3 (E_N-M_N)]\right\} F_2(q^2) \nonumber\\
    && {} + \frac12
       \left[i c_{\alpha_N} q_3 (E_N+M_N)  + s_{\alpha_N}(E_N^2-M_N^2)\right] F_3(q^2)\,,
\end{eqnarray}
\end{subequations}
\end{widetext}
where \(c_{\alpha} \equiv (\cos 2 \Re \alpha + \cosh 2 \Im \alpha)/2\)
and \(s_\alpha \equiv (\sin 2 \Re \alpha + i \sinh 2 \Im
\alpha)/2\). For PT symmetric theories, where \(\alpha\) is real,
these expressions simplify to \(c_\alpha = \cos^2\alpha \)  and
\(s_\alpha = \cos\alpha\sin\alpha\). 

From the above expressions we want to extract $F_{3} (q^2)$, that gives the neutron EDM.  It turns out that the RHS of Eqs.~\eqref{eq:components} is most naturally expressed in terms of $G_{1,2,3}$  given by
\begin{subequations}
\begin{eqnarray}
G_1 &=& F_1 + F_2
\\
G_2 &=&  F_1 -  \frac{q_E^2}{4 m^2} F_2  +  \frac{s}{c}  \frac{q_E^2}{4 m^2} F_3\,,
\\
G_3 &=& F_3 + \frac{s}{c} F_2
\end{eqnarray}
\end{subequations}
where $q_E^2  = {\bm q}^2  + q_4^2$ and $s\equiv \sin \alpha \cos \alpha$, $c \equiv \cos^2 \alpha$.

For a given momentum transfer $\bm{q} = (q_1,q_2,q_3)$,  Eqs.~\eqref{eq:components} thus represents eight equations for $G_{1,2,3}$. 
They can be written in a compact form as follows:
\begin{equation}
K(q)    \, 
\left(
\begin{array}{c} 
G_1\\
G_2\\
G_3
\end{array}
\right) - V (q)  = 0\,,
\end{equation}
where $K(q)$ is an $8 \times 3$ matrix given in block form by 
\begin{subequations}
\label{eq:Kmat}
\begin{eqnarray}
K (q) &=& 
\left(
\begin{array}{ccc}
X_1 (q)  & 0 & X_3 (q) 
\\
0 & Y_1 (q) & 0
\end{array}
\right)
\\
X_1 (q) &=&
m \, \left(
\begin{array}{c}
- c q_2
\\
c q_1
\\
s (E - m) 
\\
- i s q_3
\end{array}
\right)
\\
X_3 (q) &=&
- \frac{c}{2}  \, q_3   \, \left(
\begin{array}{c}
q_1
\\
q_2
\\
q_3
\\
- i  (E + m)
\end{array}
\right)
\\
Y_  1 (q) &=&
m \, c    \, \left(
\begin{array}{c}
q_1
\\
q_2
\\
q_3
\\
- i  (E + m)
\end{array}
\right)\,, 
\end{eqnarray} 
\end{subequations}
and $V(q)$ is an eight-dimensional array  given by 
\begin{subequations}
\begin{align}
V (q) &= 
\left(
\begin{array}{c} 
V_R (q)\\
V_I (q) 
\end{array}
\right)\,,
\quad  \\
V_R (q) &= 
\left(
\begin{array}{c} 
{\rm Re}  \vec{\cal V} (q)
\\
i {\rm Im}  {\cal V}_4 (q)
\end{array}
\right)\,,
\quad  \\
V_I (q) &= 
\left(
\begin{array}{c} 
{\rm Im}  \vec{\cal V} (q)
\\
- i {\rm Re}  {\cal V}_4 (q)
\end{array}
\right)\,.
\end{align} 
\end{subequations}

To solve for $G_{1,2,3}(q^2)$, for a given three-momentum transfer  $\bm{q} = (q_1,q_2,q_3)$  we 
can use a least squares estimator.  
Namely,  we   minimize the function 
\begin{equation}
F (G_{1,2,3})=   \sum_{\vec q  \in P (\vec q)}  \sum_{i,j=1}^{8} \ w_{ij} (q) \  E_i (q)  \  E_j (q) 
\label{eq:F}
\end{equation}
where 
\begin{eqnarray} 
E_i (q) &=&  \sum_{\beta = 1}^{3}  K_{i \beta} (q)  \, G_\beta  - V_i (q)
\\
w_{ij} (q) &=&   \left[ C_V^{-1} (q) \right]_{ij}
\end{eqnarray}
where the weights matrix is the inverse of the covariance matrix of  lattice ``measurements" $V_i (q)$:
\begin{equation}
\left[ C_V (q) \right]_{ij}  =   {\rm Cov} \left( V_i(q), V_j (q)  \right)\,.
\end{equation} 
For independent variables $V_i (q)$  the covariance matrix $C_V$ and it inverse are positive definite.\footnote{For ease of notation, we are ignoring current conservation, which relates the various components $V_i (q)$. Strictly speaking, we need to eliminate the dependent components of $V_i (q)$ when using a conserved current to get an invertible covariance matrix.} 
This guarantees that  $F (G_{1,2,3})$ is minimized if and only if  $E_i (q) = 0$  for all $i$. 
The sum over momenta runs over the six permutations 
$(q_1,q_2,q_3)$, 
$(q_1,q_3,q_2)$, 
$(q_2,q_1,q_3)$, 
$(q_3,q_1,q_2)$, 
$(q_2,q_3,q_1)$, 
$(q_3,q_2,q_1)$.

The function $F (G_{1,2,3})$ is stationary for 
\begin{equation}
\frac{\partial F}{\partial G_\alpha} = 0  \qquad \alpha = 1,2,3\,.
\end{equation}
Explicitly, since $\partial E_j/\partial G_\alpha =  K_{j  \alpha}$,  one finds
\begin{equation}
2  \sum_{\vec q  \in P (\vec q)}  \sum_{i,j=1}^{8} \ w_{ji}  (q) \  E_i (q)  \ K_{j \alpha} (q)  = 0\,, \qquad \alpha = 1,2,3\,.
\end{equation}
or even more explicitly
\begin{eqnarray}
  \sum_{\vec q  \in P (\vec q)}  \sum_{i,j=1}^{8} \ w_{ji} (q) \ 
\left(
\sum_\beta K_{i \beta} (q)  \, G_\beta  - V_i (q)
\right)   
\ K_{j \alpha} (q) \span\omit\span\nonumber\\
 \qquad\qquad\qquad\qquad&=& 0\,, \qquad \alpha = 1,2,3\,, 
\end{eqnarray}
which is a system of three equations for $G_{1,2,3}$. 
The extremum condition for $F (G_{1,2,3})$  implies the following linear equation for $G_{1,2,3} (q^2)$: 
\begin{equation}
A_{\alpha \beta}  G_\beta =    B_\alpha 
\end{equation}
where the $3\times 3$ matrix $A$ and the three dimensional array $B$ are given by
\begin{subequations}
\begin{eqnarray}
A_{\alpha \beta} &=&  \sum_{\vec q  \in P (\vec q)}  \sum_{i,j=1}^{8}   
\ K_{j \alpha} (q) \ w_{ji} (q) \  K_{i \beta} (q) 
\\
B_{\alpha} &=&  \sum_{\vec q  \in P (\vec q)}  \sum_{i,j=1}^{8}   
\ K_{j \alpha} (q) \ w_{ji} (q) \ V_i (q)  \,.  
\end{eqnarray}
\end{subequations}

So from the lattice data on $V_{i}(q)$, their covariance matrix,  and the explicit form of the matrix $K_{i \alpha} (q)$ 
given in Eq.~\eqref{eq:Kmat}
one  can construct $A_{\alpha \beta}$ and $B_\alpha$ and solve for $G_{1,2,3}$. 
Error on $G_{1,2,3}$ can be assigned with the bootstrap method. 

\section{Chiral extrapolation formulae}\label{chiral}

We can express the electric dipole form factor as 
\begin{equation}
\frac{F^i_3(q^2)}{2 M_N}= d_i - S'_i \; q^2 + H_i(q^2),
\label{Hdef}
\end{equation}
where $d_i$ is the EDM, $S'_i$ the Schiff moment (with some abuse of notation), and $H_i(q^2)$ 
account for the higher order dependence on $q^2$.  Here, $i$ is an isospin label, and the results are more conveniently expressed in terms of an isoscalar ($i=0$) and isovector ($i=1$) component.
The neutron and proton form factors are 
\begin{eqnarray}
F_{3,\, p}(q^2) &=& F^0_3(q^2) + F^1_3(q^2), \nonumber \\
F_{3,\, n}(q^2) &=& F^0_3(q^2) - F^1_3(q^2).
\end{eqnarray}

At NLO in $\chi$PT, the EDMs are given by \cite{Crewther:1979pi,Hockings:2005cn,Ottnad:2009jw,Mereghetti:2010kp},
\begin{eqnarray}
d_0 &=& e  \bar{d}_0
        +\frac{eg_A\bar{g}_0}{(4\pi F_{\pi})^2}\; 
        \left[\frac{3 \pi M_{\pi}}{4 M_N}
        \right],
\label{d0}\\
d_1 &=& e  \bar{d}_1(\mu) 
        +\frac{eg_{A}\bar{g}_{0}}{(4\pi F_{\pi})^2}
        \left[-\ln\frac{M_{\pi}^2}{\mu^2}
        +\frac{5\pi}{4}\frac{M_\pi}{M_N}\right],
\label{d1}
\end{eqnarray}
where the renormalization scale dependence of the LEC $\bar d_1$ cancels the $\mu$ in the logarithm. 
Here $g_A = 1.27$, $F_\pi= 92.4$ MeV.  $\bar g_0$ is a  \CPV\, pion-nucleon coupling, defined as 
\begin{eqnarray}
\mathcal L = - \frac{\bar g_0}{2 F_\pi} \bar N \boldsymbol{\pi} \cdot \boldsymbol{\tau} N,
\end{eqnarray}
which is related by chiral symmetry to the neutron-proton mass splitting \cite{Crewther:1979pi}
\begin{equation}\label{g0}
\bar g_0 = \left( \frac{M_n - M_p}{\bar m \varepsilon}  + \mathcal O\left( \frac{M_\pi^2}{\Lambda_\chi^2}\right) \right)\, m_* \bar\Theta   =   g_S \bar m\,   \bar\Theta,
\end{equation}
where $m_*^{-1} = m_u^{-1} + m^{-1}_d $, $2\bar m = m_u + m_d$, and $\Lambda_\chi \sim 1$ GeV is the scale at which the $\chi$PT expansion breaks down. $g_S$ is the isovector scalar charge, and the last equality holds in the isospin limit.
At the physical pion mass,  one obtains \cite{deVries:2015una}
\begin{equation}
\frac{\bar g_0}{2 F_\pi}  = \left( 15.5 \pm 2.6 \right) \cdot 10^{-3} \bar\Theta,
\label{eq:g0def}
\end{equation}
but the last term in Eq.~\eqref{g0} allows to extend the relation to arbitrary masses in the regime of validity of $\chi$PT. 
In particular, in the $\chi$PT fits to $F_3(q^2)$ we use
\begin{equation}
\bar g_0  = \frac{g_S}{2 B} M^2_\pi \, \bar\Theta,
\label{eq:g0def2}
\end{equation}
with $g_S = 1.0$ and $B=2.8$ GeV.
$\bar d_{0,1}$ are two low-energy constants, which, by naive-dimensional-analysis, scale as
\begin{eqnarray}
  \bar d_{0,1} = \mathcal O\left(\frac{M_\pi^2}{(4\pi F_\pi)^3}\right)
\end{eqnarray}

The first derivative of the form factor is \cite{Hockings:2005cn,Ottnad:2009jw,Mereghetti:2010kp}
\begin{eqnarray}
S'_0&=& 0,
\\
S'_1&=&\frac{eg_A\bar{g}_0}{6 (4\pi F_{\pi})^2 M_{\pi}^2}
       \left[1-\frac{5\pi}{4}\frac{M_{\pi}}{M_N} \right].
\label{radius}
\end{eqnarray}
At N$^2$LO there are additional long- and short-distance contributions to both isoscalar and isovector components.

The remaining momentum dependence of the EDFF is given 
by the functions $H_i(q^2)$ introduced in Eq.~(\ref{Hdef}),
\begin{eqnarray}
H_{0}(q^{2})&=& 0,
\\
H_{1}(q^{2})&=&\frac{4eg_{A}\bar{g}_{0}}{15(4\pi F_{\pi})^{2}}
               \left[h_a(x)
       -\frac{7\pi}{8}\frac{M_{\pi}}{M_N}
                      \, h_b(x)\right], 
                      \nonumber \\
\label{FD}
\end{eqnarray}
with $x \equiv q^2 / 4M_\pi^2$. $h_a$ appears at leading order, 
\begin{eqnarray}
h_a(x)&=&-\frac{15}{4}\left[
        \sqrt{1+\frac{1}{x}} \;
        \ln{\left(\frac{\sqrt{1+1/x}+1}{\sqrt{1+1/x}-1}\right)}\right.\nonumber\\
      &&\qquad\left.{}-2\left(1+\frac{x}{3}\right)\right],
\label{f0}
\end{eqnarray}
while $h_b$ is generated at NLO
\begin{eqnarray}
h_b(x)
&=&-\frac{1}{7}\Bigg[3(1+2x)\, \left( 5\left(
\frac{1}{\sqrt{x}} \arctan \sqrt{x} -1 +\frac{x}{3}\right) \right) \nonumber \\
& & \qquad{}-10x^2\Bigg].
\label{f1}
\end{eqnarray}
Since these behave as $h_i^{(n)}(x)=x^2 +{\cal O}(x^3)$ for
$x\ll 1$, the leading, $O(q^4)$, dependence of $H_i$ is consistent
with the definition in Eq.~\eqref{Hdef}.

\section{Excited state contamination in chiral perturbation theory}
\label{sec:ESCappendix}

\begin{figure}[tbp]
    \centering
  \subfigure{    \includegraphics[width=0.96\linewidth]{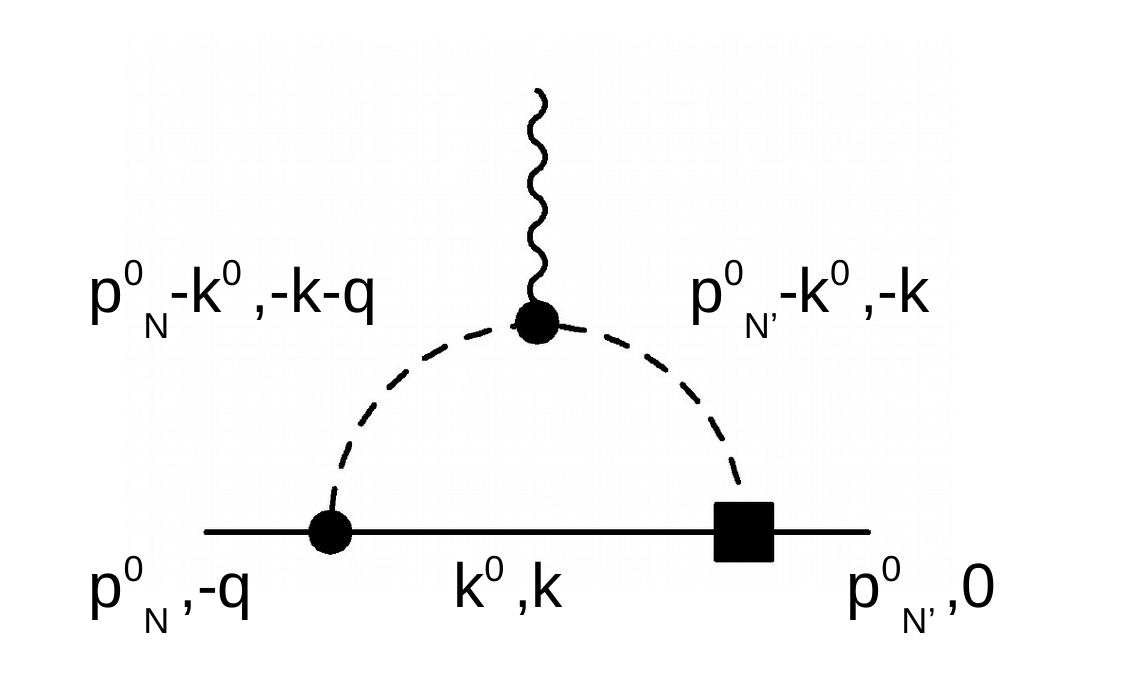}}
    \caption{Leading order diagram for the excited states contribution to the three-point function ${\cal C}_{3pt}^\mu$ in chiral perturbation theory. A black square denotes an insertion of the CP-odd pion-nucleon couplings $\bar g_0$. Filled circles denote CP-even pion-nucleon and pion-photon couplings.  }
    \label{fig:pionloop}
\end{figure}

In this appendix, we show that, in $\chi$PT, the gap between the ground state and excited state contributions to the CP-odd components of the three-point function ${\cal C}_{3pt}^\mu$ is expected to be of order of the pion mass $M_\pi$. This can be intuitively understood from the fact that the nucleon EDM induced by the QCD $\bar\Theta$ term receives a LO
contribution from a long-range pion loop \cite{Crewther:1979pi}, shown in Figure \ref{fig:pionloop}. In Minkowski space, this diagram has a branch cut when the intermediate pions and nucleon go on-shell. In Euclidean space, this translates into a $N\pi$ excited state, whose amplitude is of the same size as the ground state contribution.  
For simplicity, we focus only on the diagram shown in Figure \ref{fig:pionloop}, and assume that the nucleon interpolating field does not couple to nucleon plus pions. 

We start from the 4$^{\rm th}$ component of the three-point function. 
Carrying out the Dirac traces in Eq.~\eqref{eq:V1t}, in the limit $M_N \gg q$, we find 
\begin{eqnarray}
\mathcal C_{3pt}^4 &=& q_3  \tau_3 \frac{\bar g_0 g_A}{(4 \pi F_\pi)^2} e^{- M_N t_B - E_N t } \Bigg\{ 
f_0(M_\pi, q,  L) \nonumber \\ & & +  \frac{(4 \pi)^2}{L^3 M_\pi E_\pi^2}
\Bigg( e^{- M_\pi t} + e^{- M_\pi t_B}     \nonumber\\
      && \quad {}+\frac{E_\pi}{M_\pi} \left(e^{- E_\pi t}+
e^{- E_\pi t_B}\right)  \nonumber \\ & & 
\quad {}- \frac{M_\pi + E_\pi}{2 M_\pi} \left(e^{- E_\pi t - M_\pi t_B } +s
e^{- M_\pi t - E_\pi t_B} \right) \nonumber \\
& & \quad{}+ \frac{(E_\pi - m_\pi)^2}{2 M_\pi (E_\pi + M_\pi)  } \nonumber\\
&& \qquad\qquad{}\times\left(e^{- (M_\pi + E_\pi) t} + e^{- (M_\pi + E_\pi) t_B}\right)
 \Bigg) \nonumber \\ & & + \ldots
\Bigg\}, \label{V4excited}
\end{eqnarray}
where $t_B = \tau - t$, 
$E_N = \sqrt{M_N^2 + q^2} \sim M_N$, $E_\pi = \sqrt{M_\pi^2 + q^2}$ and $\ldots$ denotes terms with a gap with two or more units of momentum.
$f_0(M_\pi, q, L)$ denotes the ground state loop function, which we write as an infinite volume term $f^{\infty}_0$ and a correction $\Delta$
\begin{eqnarray}
f_0(M_\pi, q, L) &=&  f^{\infty}_0(M_\pi, q) + \Delta(M_\pi, q, L ).
\end{eqnarray}
In the non-relativistic limit,  $f^{\infty}_0(M_\pi, q)$ is given by 
\begin{eqnarray}
f^{\infty}_0(M_\pi, q) &=& (4\pi)^2 \Bigg( \int \frac{d^4 k}{(2\pi)^4} \frac{1}{k_0^2 + \vec k^2 + M_\pi^2} \nonumber  \\ & & \frac{1}{k_0^2 + (\vec k+ \vec q)^2 + M_\pi^2} \Bigg),
\end{eqnarray}
and is ultraviolet divergent. In dimensional regularization and in the $\overline{\rm MS}$ scheme
\begin{eqnarray}\label{D4}
f^{\infty}_0(M_\pi, q) = \log \frac{\mu^2}{M^2_\pi} + 2 - \sqrt{1 + \frac{1}{x}} \ln \frac{\sqrt{1+ \frac{1}{x}}+1}{\sqrt{1+ \frac{1}{x}}-1},\nonumber\\
\end{eqnarray}
with $x= q^2/(4 M^2_\pi)$, which is of course the same function as in Section \ref{chiral}. 
The finite volume correction is given by
\begin{eqnarray}
\Delta(M_\pi, q,  L )  &=& (4\pi)^2 \int \frac{d k_0}{2\pi} \left( \frac{1}{L^3} \sum_{\vec k} - \int \frac{d^3 k}{(2\pi)^3}\right)\nonumber \\ & & \frac{1}{k_0^2 + \vec k^2 + M_\pi^2} \frac{1}{k_0^2 + (\vec k+ \vec q)^2 + M_\pi^2},\nonumber\\
\end{eqnarray}
which can be written in terms of Bessel functions as \cite{Beane:2004tw}
\begin{eqnarray}
\Delta(M_\pi, q,  L ) \span\omit=\span\nonumber\\
&&\qquad\qquad2  \sum_{\vec n \neq 0}\int_0^1 d x K_0\left(L \sqrt{M_\pi^2 + q^2 x (1-x)} |\vec n|\right).\nonumber \\ 
\end{eqnarray}
At $q=0$, for $M_\pi L \sim 4$, $\Delta$ amounts to a $0.1\%$ correction.
Eqs.~\eqref{V4excited} and \eqref{D4} thus show that the excited states have a gap of $\mathcal O(M_\pi)$. The ratio of the ground and excited state contributions is determined by the quantity $(4\pi)^2/(L M_\pi)^3$, which is a number of order 1 for $L M_\pi = 4$.  
We thus do not expect a significant suppression of the excited states.
A similar calculation can be performed for the spatial components $\mathcal C_{3pt}^i$, yielding a result similar to Eq.~\eqref{V4excited}, but   with a $\sinh$ rather than $\cosh$ behavior.
\section{\texorpdfstring{\(O(a)\)}{O(a)} corrections in the Wilson-Clover theory}
\label{sec:appendix2}
In this appendix, we analyze CP violation due to the topological charge in the Wilson-Clover theory at \(O(a)\).
We will  denote by $O_n^{(d)}$,   $\tilde{O}_n^{(d)}$,  $O_n^{(d),{\rm ren}}$,  
 the set of bare, subtracted, and renormalized  operators of dimension $d$, respectively. 
Subtracted operators, i.e., operators free of power divergences,  are defined by 
\begin{equation}
\tilde O_{n'}^{(d)} =  O_{n'}^{(d)}  -  \sum_{d'<d}\sum_k   \frac{\beta_{n' k}^{(d)}}{a^{d- d'}} \, \tilde O_k^{(d')}
\label{eq:Osubs}
\end{equation}
while finite (renormalized) operators are given by
\begin{equation}
O_{n}^{(d),{\rm ren}} =   Z_{n n'}   \  \tilde O_{n'}^{(d)}  \,.
\label{eq:Zds}
\end{equation}
The presence of $\tilde O_k^{(d')}$ and not  $O_k^{(d')}$ in Eq.~\eqref{eq:Osubs}
 is needed to avoid ambiguities in the definition of lower-dimensional coefficients 
$\beta_{n' k}^{(d)}$. Note, however, that like all operators, the subtracted operators allow any amount of admixture of \(a^{d'-d}\tilde O^{(d')}\) for \(d'\geq d\).

We use the Wilson-Clover quark action, in which the Dirac operator reads: 
\begin{eqnarray}
O_D &=&  D_L  + m_W    
\\
D_L &=&  \slashed{D} - a \,  \left( \frac{r}{2}  D^2  +
\frac{r c_{SW} }{4} \sigma \cdot G \right),  
\label{eq:DO1s}
\end{eqnarray}
with $c_{SW} = 1 + O(g^2)$.\footnote{Throughout, we use 
$D_\mu = \partial_\mu + i  A_\mu$, $G_{\mu \nu} = \partial_\mu A_\nu - \partial_\nu A_\mu + i  [A_\mu, A_\nu]$,  so that $[D_\mu , D_\nu] = i  G_{\mu \nu}$ and $\slashed{D} \slashed{D} = D^2 + (1/2) \sigma \cdot G$.}
To simplify the analysis, in the following discussion we will
first assume that the quark mass matrix is proportional to the identity, 
pointing out the minor modifications at the end. 

The starting point of our analysis is the singlet axial Ward Identity (AWI) obtained by considering the 
axial transformation on the quark fields $\psi^T = (u,d,s)$:  
\begin{eqnarray}
\psi (x) &\to& (1 + i \alpha (x)   \gamma_5 ) \psi (x) 
\nonumber \\
\bar \psi (x)  &\to & \bar \psi  (x) (1 + i \alpha (x)   \gamma_5)\,,  
\end{eqnarray} 
where $\alpha (x)$ is the local transformation parameter. 
Denoting by $O(x_1, ..., x_n)$  any product of local operators, the   singlet AWI reads 
\begin{eqnarray}
\left\langle O(x_1,..., x_n)  \Big(   \partial_x^\mu A_\mu(x) 
- 2 m_W   \bar{\psi} (x)   \gamma_5 \psi  (x)  - X  (x)  \Big) \right \rangle
\span\omit\span\nonumber \\ 
\qquad\qquad\qquad\qquad& = & - \left \langle 
\frac{\delta O (x_1, ..., x_n)}{\delta (i \alpha (x))} 
\right \rangle \,,
\label{eq:AWI1s}
\end{eqnarray}
where  
\begin{equation}
A_\mu (x) =  \bar{\psi}  (x)  \gamma_\mu \gamma_5 \psi  (x)  
\end{equation}
and $X (x)$ is given by the variation of the Wilson-Clover term~\cite{Karsten:1980wd,Bochicchio:1985xa,Guadagnoli:2003zq}.   
\begin{equation}
\frac{X}{2} =  -  a   \bar{\psi}\,  \Big( \frac{r}{2}  D^2 + \frac{r c_{SW} }{4}  \sigma \cdot G\Big) \gamma_5 \,  \psi  \,.
\label{eq:id2s}
\end{equation}
Insertions of $X (x)$ vanish at tree level in the continuum limit, but quantum effects induce power-divergent mixing with lower dimensional operators, 
that have to be taken into account when taking the continuum limit.    
This is done by writing ~\cite{Karsten:1980wd,Bochicchio:1985xa,Guadagnoli:2003zq}
\begin{eqnarray}
X (x) &=&  a   \tilde{X}  (x)  -  2 \bar m  \bar \psi (x)   \gamma_5 \psi (x)  - ( Z_A-1)   \partial_x^\mu A_\mu (x)   
\nonumber \\
&+& Z_{G \tilde G} \,  \frac{2 N_F}{32 \pi^2} \, (G \tilde G)_{\rm sub} \,, 
\label{eq:X1s}
\end{eqnarray}
where $N_F$ is the number of quark flavors and $\tilde{X}(x) $ 
is a  `subtracted'  dimension-five operator,  i.e., it is free  of power divergences, expanded according to Eq.~\eqref{eq:Osubs}. 
The operator $a \tilde{X}(x) $ has no impact on the analysis of the axial WI with elementary fields, 
 while  it induces contact terms in the continuum limit of axial WIs involving composite fields~\cite{Bochicchio:1985xa, Testa:1998ez}.
It is, however, essential in order to identify the  $O(a)$ corrections to $d_n ({\overline \Theta})$.
Using the above expression in \eqref{eq:AWI1s}, 
and taking into account the mixing between $(G \tilde G)$ and  $\partial_\mu A^\mu$ (which involves 
the renormalization constant $Z_C$)  one arrives at~\cite{Testa:1998ez,Guadagnoli:2003zq}
\begin{eqnarray}
\Big \langle O(x_1,..., x_n)  \Big(  Z_A (1 - Z_C)  \partial_x^\mu A_\mu (x) 
-  2 m  \bar{\psi} (x)  \gamma_5 \psi  (x)  \span\omit\span
\nonumber \\ 
\qquad\qquad\qquad{} -   \frac{2 N_F}{32 \pi^2} \, (G \tilde G)_{\rm ren}  
- a \tilde{X} (x) 
 \Big) \Big\rangle \span\omit\span\nonumber\\
\qquad\qquad\qquad\qquad&=& -   \left \langle   \frac{\delta O (x_1, ..., x_n)}{\delta (i \alpha (x))}  
 \right\rangle \,, \quad 
\label{eq:AWI1.5s}
\end{eqnarray}
where  
\begin{equation}
m = m_W - \bar{m}
\end{equation}
is the quark mass free of power divergences as we take the continuum limit. Here, and henceforth, the \(O(ma)\) dependence of the coefficients of the operators are suppressed.
Finally, upon integrating over $\int d^4x$ we arrive at
%
\begin{equation}
\begin{array}{rl}
 \int d^4x  \Big\langle O(x_1,..., x_n)  
  &\Big(-  2 m \bar{\psi} (x)  \gamma_5 \psi  (x)  \\
  &{}-   \frac{2 N_F}{32 \pi^2} \, (G \tilde G)_{\rm ren}    
 - a \tilde{X}(x)
\Big)
\Big \rangle \\
 =  - \int d^4x \Big \langle   \frac{\delta O (x_1, ..., x_n)}{\delta (i \alpha (x))}  
\Big \rangle & \,. 
\end{array}
\label{eq:AWI3ns}
\end{equation}
%
Ref.~\cite{Guadagnoli:2002nm} performed a detailed diagrammatic analysis of 
Eq.~\eqref{eq:AWI3ns}, with  $O(x_1, x_2. x_3) = N(x_1) \, J^{\rm EM}_\mu (x_2) \, \bar N(x_3)$ in the $a \to 0$ case, 
showing that the $\delta O$ terms cancel the connected insertions of $2 m \bar \psi \gamma_5 \psi$.
Their analysis shows that  insertions of the operator $G \tilde G$  can be replaced by $2 m$ times the disconnected insertions of the isosinglet pseudoscalar 
density $\bar \psi \gamma_5 \psi$. Since the disconnected matrix elements of the isoscalar density do not diverge in the chiral limit, this implies as a corollary that the neutron EDM should vanish as $m \to 0$. 
$O(a)$ effects would modify the result of Ref.~\cite{Guadagnoli:2002nm} by modifying the RHS of their Eqs.~{(2.11)} and {(3.5)}. 
In the context of our analysis,   the term proportional to $a \tilde X$ in Eq.~\eqref{eq:AWI3ns}   provides  $O(a)$ effects, which we  discuss next. 

First, we  project the subtracted operator $\tilde X$  on the basis of (subtracted) dim-5 operators, 
given in Ref.~\cite{Bhattacharya:2015rsa}, 
\begin{equation}
\tilde X  = 
\sum_n
K_{Xn}
\tilde O_n^{(5)} 
\label{eq:Xtilde1s}
\end{equation}
and  analyze the consequences of Eq.~\eqref{eq:Xtilde1s}   for Eq.~\eqref{eq:AWI3ns}. 
The basis of dimension-5 operators  $O_n^{(5)}$   appearing on the RHS of Eq.~\eqref{eq:Xtilde1s} 
is given in \cite{Bhattacharya:2015rsa} assuming  generic diagonal quark mass $\hat m$, 
and we repeat it here for completeness: 
\begin{eqnarray}
O^{(5)}_1 &= &  i\,  \bar\psi\tilde\sigma^{\mu\nu}G_{\mu\nu} \psi 
 \label{eq:CEDMdef} \\
O^{(5)}_2 &=&    \partial^2  \left(   \bar\psi i\gamma_5  \psi \right) 
\\
 O^{(5)}_3 &=& i e \,  \bar\psi\tilde\sigma^{\mu\nu} Q F_{\mu\nu}  \psi 
\\
O^{(5)}_4  &=&     
 \textrm{Tr} \left[ \hat m Q^2  \right] \,
  \frac{1}{2} \epsilon^{\mu \nu \alpha \beta}  F_{\mu \nu}  F_{\alpha \beta}
\\
O^{(5)}_5  &= &    
 \textrm{Tr} \left[  \hat m  \right] \,
  \frac{1}{2} \epsilon^{\mu \nu \alpha \beta}  G^b_{\mu \nu}  G^b_{\alpha \beta}
\\
O^{(5)}_6  &=&  
\textrm{Tr}\left[\hat m \right]  \partial_\mu  \left( \bar\psi\gamma^\mu\gamma_5   \psi \right)
\\
O^{(5)}_7  &=&  
\partial_\mu  \left(  \bar\psi\gamma^\mu\gamma_5   \hat m  \psi \right) 
- \frac{1}{3} \textrm{Tr}\left[\hat m  \right]  \partial_\mu  \left( \bar\psi\gamma^\mu\gamma_5   \psi \right)
\\
O^{(5)}_8   &= &    
 \bar\psi i\gamma_5   \hat m^2 \  \psi
\\
O^{(5)}_9   &= &  
\textrm{Tr} \left[\hat m^2 \right]  \ \bar\psi i\gamma_5   \psi
\\
O^{(5)}_{10}   &= &    
\textrm{Tr} \left[\hat m  \right]  \ \bar\psi i\gamma_5  \hat m  \psi
\\
O^{(5)}_{11} &\equiv&    P_{EE}  =  i\bar\psi_E\gamma_5  \psi_E 
\\
O^{(5)}_{12} &\equiv&    \partial \cdot A_E  
=  \partial_\mu[\bar\psi_E\gamma^\mu\gamma_5   \psi+\bar\psi\gamma^\mu\gamma_5   \psi_E]
\\ 
O^{(5)}_{13}  &\equiv&  A_\partial  = 
  \bar\psi  \gamma_5  \slashed{\partial}    \psi_E   \ -   \bar {\psi}_E   \overleftarrow{\slashed{\partial}}  \gamma_5     \psi  
  \\ 
O^{(5)}_{14} &\equiv &   A_{A^{(\gamma)}}  =
 i  e \left( \bar\psi   Q  \slashed{A} ^{(\gamma)}  \gamma_5\psi_E - \bar\psi_E Q \slashed{A}^{(\gamma)}  \gamma_5\psi \right)\,,\nonumber\\
\end{eqnarray}
where \(\tilde{\sigma}^{\mu \nu} \equiv \frac{1}{2} \left( \sigma^{\mu \nu} \gamma_5
 + \gamma_5 \sigma^{\mu \nu} \right) \) and $\psi_E = (\slashed{D} + \hat m) \psi$.

Keeping in mind that $O(x_1, ..., x_n)$ has the structure $N(x_1) \, J^{\rm EM}_\mu (x_2) \, \bar N(x_3)$, 
in terms of the neutron source and sink operator and the electromagnetic current, 
the various  $O_n^{(5)}$  contribute to Eq.~\eqref{eq:AWI3ns} as follows:

\begin{itemize}
\item  $O^{(5)}_1$ is the isoscalar chromo-EDM operator and contributes an $O(a)$ term  to the LHS of Eq.~\eqref{eq:AWI3ns}. 
In fact, as shown below, this is the leading $O(a)$ contribution, thus proving a linear relation between 
isovector insertions of the pseudoscalar density and the chromo-EDM.  
\item $O^{(5)}_{2,6,7}$ are total derivatives and their insertion  in  Eq.~\eqref{eq:AWI3ns} vanish upon integration over $\int d^4x$.  
\item  $O^{(5)}_{3,4}$ involve one and two powers of the electromagnetic field strength. In order to eliminate the photon field 
in the correlation functions in Eq.~\eqref{eq:AWI3ns}, one needs electromagnetic loops, making the 
contribution of $O^{(5)}_{3,4}$ to Eq.~\eqref{eq:AWI3ns} of $O(a \,  \alpha_{\rm EM}/\pi)$, and thus negligible to the order we are working.

\item $O^{(5)}_5$  
provides a  correction  of $O(am)$   proportional to $(G \tilde G)$ 
in the  LHS of Eq.~\eqref{eq:AWI3ns}.


\item  $O^{(5)}_{8,9,10}$  become $\hat m^2  \bar\psi i\gamma_5   \psi $ when  $\hat m \propto I$.  Therefore, their 
contributions have the same form of  the pseudoscalar insertion in  Eq.~\eqref{eq:AWI3ns}, but suppressed  by  $O(a m)$. 

\item The operators $O^{(5)}_{11,12,13,14}$ vanish by using the  quark equations of motion and can contribute contact terms to the 
LHS of Eq.~\eqref{eq:AWI3ns}. 
However, it turns out that none of them actually contributes {at this order}.  
$O^{(5)}_{11}  $ contains two equation of motion operators. 
Therefore, when inserted in Eq.~\eqref{eq:AWI3ns}, it will always involve a contraction with a quark field in the neutron source or sink operator, 
and thus it will not contribute to the residue of the neutron pole. 
$O^{(5)}_{12}$ is a total derivative and drops out of  Eq.~\eqref{eq:AWI3ns}. 
$O^{(5)}_{13}$ is gauge-variant operator and drops out of  Eq.~\eqref{eq:AWI3ns} as long as $O (x_1, ... , x_n)$ is a gauge singlet, 
which is the case for  $O(x_1,x_2,x_4)  \propto N(x_1) \, J^{\rm EM}_\mu (x_2) \, \bar N(x_3)$.
$O^{(5)}_{14}$ involves the photon field and therefore can contribute to  Eq.~\eqref{eq:AWI3ns} only to $O(a \alpha_{\rm EM}/\pi)$.

\end{itemize}

So in summary, for \(\hat m\propto I\),  Eq.~\eqref{eq:AWI3ns} becomes  
\begin{widetext}
\begin{eqnarray}
\int d^4x  \left\langle O(x_1,..., x_n)  
  \Big(-  2 m \bar{\psi} (x)  \gamma_5 \psi  (x)  \Big(1 + O(a m) \Big)
  -   \frac{2 N_F }{32 \pi^2} \, (G \tilde G)_{\rm ren}     \Big(1 + O(a m) \Big)
 - a K_{X1} \tilde O_1^{(5)} 
\Big)
\right \rangle\span\omit\span
\nonumber \\
\qquad\qquad\qquad\qquad\qquad\qquad&=&
  - \int d^4x \left \langle   \frac{\delta O (x_1, ..., x_n)}{\delta (i \alpha (x))}  
\right \rangle \,. 
\label{eq:AWI3ns2}
\end{eqnarray}
\end{widetext}

If $\hat m \neq I$, the singlet AWI,  Eq.~\eqref{eq:AWI3ns}, 
involves $\bar \psi \hat m \gamma_5 \psi$.
All the arguments above go through, except for the effect of 
$O_{8,9,10}^{(5)}$.
$O_{10}^{(5)}$ gives a correction of 
$O(am)$ proportional to $\bar \psi \hat m \gamma_5 \psi$, 
while  $O_{8,9}^{(5)}$ contribute nonmultiplicative  
terms involving the nonsinglet pseudoscalar densities of $O(a \hat m^2)$ in Eq.~\eqref{eq:AWI3ns2}. 
The presence of these additional terms does not affect our conclusion about the 
existence of $O(a m_q^0)$ corrections.

\onecolumngrid\twocolumngrid
\hrule width 0pt
%
\bibliographystyle{apsrev4-2} 
{%
\let\origem\em
\def\em{\sfcode`.=1000\relax\origem}
\newcommand\Eprint[2]{\href{#1}{#2}\afterassignment\temp\let\temp}
\bibliography{ref} 
}

\end{document}